\documentclass[final]{jpp}
\usepackage{graphicx}
\usepackage[T1]{fontenc}
\usepackage{amsmath}
\usepackage{bm}
\usepackage{bbold}
\usepackage{natbib}
\usepackage{upgreek}
\usepackage{float}
\usepackage{mathtools}
\usepackage[section]{placeins}
\usepackage{physics}
\usepackage{siunitx}
\usepackage{verbatim}
\usepackage{comment}
\usepackage[colorlinks=true,
  linkcolor=red, citecolor=blue, urlcolor=blue]{hyperref}   
\usepackage[dvipsnames,svgnames,x11names,hyperref]{xcolor}  
\usepackage[capitalize]{cleveref}
\renewcommand{\vec}[1]{\bm{#1}}

\newcommand{\B}{\vec{B}}

\renewcommand{\r}{\vec{r}}

\newcommand{\vi}{\bm{v}}

\newcommand{\R}{\bm{R}}

\renewcommand{\b}{\vec{b}}

\newcommand{\g}{\bm{g}}

\newcommand{\kernel}[1]{\mathcal{K}_{#1}}
\newcommand{\gyaver}[1]{\left<  #1 \right>}

\newcommand{\vparallel}{v_{\parallel}}

\newcommand{\corr}[1]{\textbf{\color[red]{}}}
\usepackage[colorinlistoftodos]{todonotes}

\newcommand{\C}{\mathcal{C}}
\renewcommand{\g}{\textsl{g}}
\usepackage[capitalize]{cleveref}
\crefname{subsection}{Sec.}{subsections}
\crefname{section}{Sec.}{sections}
\crefformat{subsection}{\S#1}

\shorttitle{Collisional Theory of Ion-Temperature Gradient Mode}
\shortauthor{B. J. Frei et al.}

\title{Local Gyrokinetic Collisional Theory of the Ion-Temperature Gradient Mode}

\author{B. J. Frei\aff{1}
  \corresp{\email{baptiste.frei@epfl.ch}},
  A. C. R. Hoffmann \aff{1},
 \and P. Ricci\aff{1}}
 
\affiliation{\aff{1} \'Ecole Polytechnique F\'ed\'erale de Lausanne (EPFL),
  Swiss Plasma Center (SPC),
CH-1015 Lausanne, Switzerland}

\begin{document}
\maketitle
\begin{abstract}
We present a study of the linear properties of ion temperature gradient (ITG) modes with collisions modelled by the linearized gyrokinetic (GK) Coulomb collision operator \citep{frei2021} in the local limit. The study is based on a Hermite-Laguerre polynomial expansion of the perturbed ion distribution function applied to the linearized GK Boltzmann equation, yielding a hierarchy of coupled equations for the expansion coefficients, referred to as gyro-moments. We explore analytically the collisionless and high-collisional limits of the gyro-moment hierarchy. Parameter scans revealing the dependence of the ITG growth rate on the collisionality are reported, showing strong damping at small scales as the collisionality increases. These properties are compared with the predictions based on the Sugama, the momentum-conserving pitch-angle scattering, the Hirshman-Sigmar-Clarke, and the Daugherty collision operators. The importance of finite Larmor radius (FLR) terms in the collision operators is pointed out by the appearance of a short wavelength (SW) ITG branch when collisional FLR terms are neglected, this branch being completely suppressed by collisional FLR effects. We demonstrate that energy diffusion is important at high collisionality and small scale lengths and that, among the collision operators considered in this work, the GK Sugama collision operator yields, in general, the smallest deviation on the ITG growth rate compared to the GK Coulomb collision operator. Convergence studies of the gyro-moment method are reported.

\end{abstract}

\section{Introduction}

Belonging to the class of instabilities that develop on the ion-gyroscale, the ion temperature gradient (ITG) modes are widely recognised as the main candidate to explain the experimental observations of anomalous ion heat turbulent transport in the tokamak core \citep{Garbet2004}. The ITG modes can be driven unstable by the parallel plasma compression (slab ITG) and by the presence of magnetic drifts (toroidal ITG). In the core region, since the plasma collision frequency is considerably smaller than the typical ITG mode frequency, it is usually assumed that collisions do not affect significantly the ITG dynamics. In fact, the GK collisionless theory of the ITG modes is well documented in the literature \citep[see, e.g.,][]{Romanelli1989,Hahm1989,Romanelli1990,Chen1991}. 
Because of the temperature drop, the effects of collisions are expected to be important in the tokamak boundary, i.e. the region that encompasses the edge and the scrape-off layer (SOL). Collisions are known to modify the linear and nonlinear properties of micro-instabilities. For instance, collisions contribute to smear out small scale structures in velocity space \citep{watanabe2004,ajay2021}, to affect the linear properties of the ITG mode and the transition to the trapped electron (TEM) mode \citep{manas2015,Belli2017}, to suppress short wavelength structures \citep{Barnes2009}, and can ultimately modify the damping of the zonal flow and, hence, the saturation mechanism of turbulent fluxes \citep{Pan2020,frei2021}. Also, the collisionality dependence of the ITG critical gradient is important in the prediction of the transition from resistive ballooning mode and ITG in the edge region \citep{zeiler1998transition}. In general, despite the fact that the boundary region is particularly important since it sets the heat exhaust and the thermal load on the vessel walls, determines the overall confinement capabilities of the device, controls the level of impurities, and regulates the removal of helium ash, the role of the ITG mode is less clear in this region than in the core. 

The effects of collisions on the ITG linear properties have been introduced in kinetic models using simplified collision operators such as, e.g., an energy-conserving Krook and a pitch-angle operator models \citep{Romanelli1990,Romanelli1991,Pusztai2009}. Among the documented studies of discrepancies produced by the collision operator models \citep{Catto2009,Belli2017,Pan2019,Pan2020} with respect to the linearized Fokker-Planck collision operator \citep{Rosenbluth1957,Hazeltine2003} which we refer to as Coulomb collision operator in the present study, only a few have reported parameter scans and direct operator comparisons when applied to ITG modes. In general, the limited number of investigations of the impact of collisions on the ITG mode, and the related limitations of the understanding of the role of ITG mode on the boundary region stems from the frequent use of drift-reduced Braginskii fluid models  \citep{Zeiler1997} for the simulation of this region \citep{dudson2009,tamain2016,Halpern2016,Paruta2018,zhu2018,Stegmeir2019,Giacomin2020}. These fluid models are based on the assumptions that $k_\parallel \lambda_{mfp} \ll 1$, i.e. that the particle mean-free path $\lambda_{mpf}$ is much shorter than the typical parallel length scales, described by the parallel wavenumber $ k_\parallel$, and that $k_\perp \rho_i \ll 1$, i.e. that the modes develop a perpendicular wavenumber $k_\perp$ associated with a scale length much larger than the ion gyroradius $\rho_i$. Therefore, while the drift-reduced Braginskii models retain finite ion temperature gradient effects \citep{Hallatschek2000,Mosetto2015}, their applicability to the study of ITG modes is questionable since, for instance, these modes develop on a $k_\perp \rho_i \sim 1$ scale. In addition, while the $k_\parallel \lambda_{mfp}  \ll 1$ assumption is well established at the separatrix and in the SOL (with temperature $T \lesssim  10$ eV and density $n \sim 10^{18} $ m$^{-3}$ yielding typically $k_\parallel \lambda_{mfp} \lesssim 10^{-2}$ in, e.g., DIII-D and ITER), the mean free path is larger inside the separatrix where the temperature increases. Hence, for typical H-mode pedestal values ($T \sim  10^{2}-10^{3}$ eV and $n \sim 10^{20} $ m$^{-3}$), an estimate of  $k_\parallel \lambda_{mfp}  \sim 1$ can be obtained. Thus, one expects a level of collisionality sufficient to justify the use of the drift-reduced Braginskii fluid models only in the SOL and, possibly, in the outermost region of the edge in typical tokamak conditions.

 To retain collisionless kinetic and finite Larmor radius (FLR) effects, gyrofluid models have been developed by taking explicitly a finite number of velocity space moments of the GK equation \citep{hammett1992,Dorland1993,Beer1996,Snyder2001}. Collisionless kinetic effects are then introduced in these models by designing \textit{ad hoc} closures for the high-order moments that mimic the kinetic linear response in the absence of collisions. While usually neglected, collisional effects are introduced in, e.g., the gyrofluid model proposed by \citet{Snyder2001} by considering a particle, momentum, and energy-conserving BGK-like operator for ion collisions and a pitch-angle scattering operator for the electrons collisions. However, BGK-like operators and the absence of energy diffusion in the pitch-angle scattering operator can potentially lead to significant deviations when compared to the Coulomb collision operator.
 
 In this work, we aim to study the linear properties of the ITG mode within a GK framework able to capture accurately the collisionless kinetic physics and, at the same time, the collisional effects described through the Coulomb collision operator. This is done by leveraging previous works that led to a formulation of the GK model based on a Hermite-Laguerre expansion of the full-F gyrocenter distribution function \citep{Frei2020}, and the development of a full-F GK Coulomb collision operator derived using the same technique expansion, as presented in \citet{Jorge2019}. More precisely, the ITG model we use in this work is based on the linearized and local limit of this formulation of the GK model, coupled with a linearized GK Coulomb collision operator, derived and numerically implemented in \citet{frei2021}. The Hermite-Laguerre expansion allows us to express the self-consistent GK model in terms of moments of the gyrocenter distribution function, that we refer to as gyro-moments. The Coulomb collision operator we use is valid for arbitrary mass and temperature ratios of the colliding species and arbitrary values of the perpendicular wavenumber. By using the gyro-moment approach, we first retrieve the collisionless and the high-collisional ITG limits. Using the linearized GK Coulomb collision operator, wide parameters scans of the ITG growth rate are then performed. We also compare the GK Coulomb collision operator with the drift-kinetic (DK) Coulomb collision operator, a GK/DK momentum-conserving pitch-angle scattering operator \citep{Helander2002}, the zeroth-order DK Hirshman-Sigmar-Clarke (HSC) collision operator \citep{Hirshman1976} and the DK/GK Dougherty collision operator \citep{Dougherty1964}, in addition to the DK/GK Sugama collision operator implemented in \citet{frei2021}. The gyro-moment approach is systematically applied to all the collision operators considered in this work, allowing their numerical implementation. We show the importance of energy diffusion and FLR terms in the collisional damping of small-scale modes. The GK Sugama operator yields the smallest deviation with respect to the GK Coulomb collision, while the largest deviations are found when the GK Dougherty collision operator is considered. We demonstrate the importance of retaining FLR terms in the collision operator models and, in particular, that a short wavelength ITG (SWITG) mode at high collisionality and steep ion temperature gradients can be destabilized when the DK limit of the collision operator is used. We remark that the assessment of the potential deviations between collision operator models is of primary importance since simplified and DK collision operators are often used in turbulent GK codes \citep{bernard2019gyrokinetic,ulbl2021implementation,francisquez2021improved}. Finally, we investigate the convergence properties of the gyro-moment approach as a function of relevant physical parameters in different collisionality regimes. Additionally, we analyse the Hermite-Laguerre spectrum of the ITG distribution function and show that the magnetic drifts broaden significantly the gyro-moment spectrum in the absence of collisions.
 
 The structure of the paper is the following. In \cref{sec:gyrokineticModel}, we introduce the gyro-moment hierarchy equation. Then, in \cref{sec:CollisionlessHermiteLaguerreTheory}, we derive the collisionless limit of the ITG dispersion relation by leveraging the gyro-moment hierarchy equation, and \cref{eq:highcollisionallimit} focuses on the high-collisional limits. In \cref{sec:ITGwithGKCoulomb}, we evaluate the ITG growth rate over a wide range of parameters with collisions modeled by the GK Coulomb collision operator. \Cref{sec:comparisonsbetweencollisionoperatormodels} is focused on the comparison of collision operator models with the GK Coulomb collision operator. Finally, we analyze the Hermite-Laguerre spectrum and report convergence studies in \cref{sec:GMspectrum} and \cref{sec:convergence}, respectively. The conclusion follow in \Cref{sec:conclusion}. The Hermite-Laguerre expansions of the collision operators used in this work are reported in \cref{sec:pitchangle,sec:HSC,sec:dougherty}, while analytical results are detailed in \cref{sec:AppGeneralizedLinearMomentResponse}.


\section{Gyrokinetic Model}
\label{sec:gyrokineticModel}

 Focusing on a shearless slab configuration and adopting a local GK approach where quantities do not vary along the magnetic field line \citep{kadomtsev2012}, we consider the linearized electrostatic GK Boltzmann equation expressed in gyrocenter phase-space coordinates  $(\R, \mu, v_\parallel,\theta)$ and assume that the electrons are adiabatic. Here, $\R = \r - \bm \rho$ is the ion gyrocenter position (with $\r$ being the particle position and $\bm \rho(\R, \mu, \theta) = \b \times \bm v/\Omega$ the ion gyroradius with $\Omega = q B / m$ the ion gyrofrequency, $q$ the ion charge and $\b = \B / B$), $\mu = m v^2/[2 B(\R)]$ is the magnetic moment, $v_\parallel = \bm b \cdot \bm v$ is the component of the velocity parallel to the magnetic field, and $\theta$ is the gyroangle. Since only the evolution of the ion distribution is considered, we drop the ion species label for simplicity. We denote the equilibrium gradient scale lengths of ion density $N$, temperature $T$ and magnetic field strength $B$ by $L_N = \left| \grad_\perp \ln N \right|^{-1}$, $L_{T} = \left| \grad_\perp \ln  T \right|^{-1}$ and $L_B = \left| \grad_\perp \ln B \right|^{-1}$, respectively. We assume that the gyroaveraged ion \textit{gyrocenter} distribution function, $F = F (\R, \mu, v_\parallel)$, is a perturbed Maxwellian, that is $F = F_{M} + \textsl{g}$, being $\g$ the perturbed part ($\g/ F_{M} \ll1$) and $F_{M}$ a local Maxwellian distribution function defined by $F_{M} = N(\R)   e^{- s_\parallel^2 - x} / (\pi^{3/2} v_{T}(\R)^3)$ with $s_\parallel = v_\parallel / v_{T}(\R)$, $x= \mu B(\R) / T(\R)$, and $v_T(\R)^2 = 2 T(\R)/m$ the ion thermal velocity. The linearized electrostatic GK Boltzmann equation describing the time evolution of a single Fourier harmonic $\g(\bm k,t) = \int d \R \g(\R ,t) e ^{- i \bm R \cdot \bm k}$ is then given by

\begin{align} \label{eq:LinToroidalGKB}
        \frac{\partial}{\partial t } \g& + \left(i \omega_{B} + i  k_\parallel v_\parallel  \right) \g  -      i  \omega_{*}^T  J_0(b\sqrt{x}) \frac{e \phi}{T_e} F_{M}   + i \omega_{\grad B} J_0(b\sqrt{x})  \frac{q \phi }{T}   F_{M}\nonumber \\
        & + \frac{q}{T}   i k_\parallel v_\parallel J_0(b\sqrt{x}) \phi F_{M} =  \C.
\end{align}
\\ 
where $e$ is the electron charge and $\tau = T_i / T_e$. In \cref{eq:LinToroidalGKB}, we express the wavevector $\bm k = k_\parallel \b + \bm k_\perp$, and introduce the velocity-dependent resonant frequencies $\omega_{\grad B} = v_{T}^2  \left ( \bm b \times \grad B /B\right) \cdot \bm k /(2 \Omega)(x + 2 s_\parallel^2)$ and $\omega_{*}^T = \omega_*[1 + \eta ( x + s_\parallel^2 - 3/2)]$, with the ion diamagnetic frequency $\omega_* = T_e (\bm b \times \grad \ln N ) \cdot \bm k /(eB) $. We also define the normalized temperature gradient $\eta = L_N / L_{T}$, a measure of the relative strength between the density and temperature gradients. FLR effects are taken into account through the zeroth order Bessel function, $J_0$, with argument $ k_\perp v_\perp  / \Omega = b \sqrt{x}$, being $b = k_\perp v_{T} /\Omega$. Collisional effects are introduced in \cref{eq:LinToroidalGKB} by the GK linearized ion collision operator defined by

\begin{align} \label{eq:gyaverC}
    \C = \gyaver{C}_{\R}  = \frac{1}{2 \pi}\int_0^{2 \pi} d \theta C,
\end{align}
\\
where $C$ is the linearized ion collision operator acting on the ion perturbed particle distribution function, and $\gyaver{\dots}_{\R} =\int_0^{2 \pi} d \theta \dots /(2 \pi) $ denotes the gyro-average operator evaluated at constant $\R$. As the effects of collisions between ions and electrons develop over a time scale that is longer by a factor $O(\sqrt{m_i/ m_e})$ than the ones due to ion-ion collisions, we neglect the former, i.e. we assume $C = C_{ii}$ in \cref{eq:gyaverC}. The linearized GK Boltzmann equation is closed by the GK quasi-neutrality condition that determines the Fourier component of the electrostatic potential $\phi(\bm k, t)  = \int d \r \phi(\r,t) e^{i \bm k \cdot \r}$, that is

\begin{align} \label{eq:Poisson}
\left[ 1 + \frac{q^2 n_{0}}{T} \left( 1 - \Gamma_0(a) \right) \right] \phi = q \gyaver{\delta n},
\end{align}
\\
where  $\gyaver{\delta n} = 2 \pi \int d v_\parallel d \mu B  / m  J_0(b) \g$ is the gyro-averaged ion gyrocenter perturbed density, $a= b^2 / 2 $ and $\Gamma_0(a) = I_0(a) e^{- a}$, with $I_0(a)$ the modified Bessel function. The first term on the left hand-side of \cref{eq:Poisson} is the adiabatic electron response, while the second term constitutes the ion polarization term.

We now normalize \cref{eq:LinToroidalGKB,eq:Poisson}. The physical quantities $t$, $v_\parallel$, $k_\perp$, $k_\parallel$, $\phi$ and $q$, are normalized respectively to $L_N/c_{s}$, $c_s$,  $1/ \rho_s$, $1/ L_N$, $T_{e}/e$ and $e$, with  $c_{s} = \sqrt{T_{e}/m}$ the ion sound speed and $\rho_s = c_s / \Omega$ the ion sound Larmor radius. The ion-ion collision frequency, $\nu =  4  N q^4 \ln \Lambda / [ 3 \sqrt{\pi} \sqrt{m} T^{3/2}]$ that enters in the ion-ion collision operator $\C$, is normalized to $c_s /  (N L_N)$ and the gyrocenter density to $N$. The same dimensionless units are used in the rest of the present paper. The normalized GK Boltzmann equation, \cref{eq:LinToroidalGKB}, then becomes

\begin{align} \label{eq:LinGK}
&\frac{\partial }{\partial t} \g + i \left( \omega_{\grad B} +  k_\parallel \sqrt{2 \tau} s_{\parallel }  \right)    \g  - i \omega_{*}^T  J_0(b \sqrt{x}) \phi F_{M}   +  \frac{q}{\tau}i \omega_{\grad B} J_0(b \sqrt{x})   \phi    F_{M} \nonumber \\
& + \frac{\sqrt{2}}{\sqrt{\tau}} i k_\parallel s_\parallel J_0(b\sqrt{x}) \phi F_{M} =  \C,
\end{align}
\\
where $\omega_{\grad B} =  \tau \omega_{B} ( x + 2 s_\parallel^2) /q $ (with $\omega_{ B} = k_\perp  R_B  $, $\tau = T_i / T_e$) and $R_B = L_N / L_B$ the normalized magnetic gradient), $\omega_*^T = \omega_{*} [1 + \eta (x + s_\parallel^2 - 3/2)] $ (with $\omega_* = k_\perp$ the normalized ion diamagnetic frequency). The argument of the Bessel function, $b \sqrt{x}$, is normalized to $b = k_\perp \sqrt{2 \tau}$. The self-consistent electrostatic potential, $\phi$, is obtained by the normalized linearized GK quasineutrality,

\begin{equation} \label{eq:LinGKPoisson}
\left[ 1 +  \frac{q^2}{\tau}\left( 1 - \Gamma_0(a) \right) \right] \phi = q \gyaver{\delta n}.
\end{equation}
\\
 \Cref{eq:LinGK,eq:LinGKPoisson} constitute a closed system that evolves self-consistently $\g$ and $\phi$, in presence of the temperature and magnetic equilibrium gradients, once the collision operator is specified (see \Cref{sec:CollisionOperatorModels}).

\subsection{ Gyro-Moment Hierarchy Equation}

To approach the solution of the GK model, given by \cref{eq:LinGK,eq:LinGKPoisson}, we use an Hermite-Laguerre moment expansion of the ion distribution function. This allows us to reduce the dimensionality of the linearized GK equation, \cref{eq:LinToroidalGKB}, by projecting it onto a velocity-space Hermite-Laguerre basis polynomials. More precisely, we decompose the perturbed \textit{gyrocenter} distribution function as \citep{Jorge2017,Jorge2019,Frei2020},

\begin{align} \label{eq:fHL}
  \g = \sum_{p = 0}^\infty \sum_{j = 0}^\infty N^{pj} \frac{H_p(s_{\parallel}) L_j(x)}{\sqrt{2^p p!}} F_{M},
 \end{align}
 \\
where we define the Hermite-Laguerre velocity moments of $\g$,  

\begin{equation} \label{eq:Npjdef}
    N^{pj} = \frac{1}{N} \int d \mu  d \vparallel  d \theta\frac{B}{m} \g \frac{H_p(s_\parallel) L_j(x)}{\sqrt{2^p p!}},
\end{equation}
\\
with $N =  \int d \mu  d \vparallel  d \theta B \g / m$ the gyrocenter density. The Hermite-Laguerre coefficients $N^{pj}$ in \cref{eq:fHL} are referred to as the gyro-moments. In \cref{eq:fHL}, we introduce the Hermite and Laguerre polynomials, $H_p$ and $L_j$, via their Rodrigues' formulas  \citep{gradshteyn}, that is 

\begin{align}
H_p(x) & = (-1)^p e^{x^2} \frac{d^p}{d x^p} \left( e^{- x^2} \right),
\end{align}
\\
and 
\begin{align}
L_j(x)  &= \frac{e^{x}}{j!} \frac{d^j }{d x^j} \left ( e^{- x } x^j\right),
\end{align}
\\
 and their orthogonality relations, that is 
 
 \begin{align}
 \label{eq:hermiteLaguerreorthogonality}
 \int_{- \infty}^\infty d x H_p(x) H_{p'}(x) e^{- x^2} & = 2^p p! \sqrt{\pi} \delta_{p}^{p'},
 \end{align}
 \\
 and 
  \begin{align} \label{eq:Laguerreorthogonality}
\int_0^\infty d x L_j(x) L_{j'}(x) e^{-x} & = \delta_j^{j'},
 \end{align}
\\
 respectively. We now project \cref{eq:LinGK,eq:LinGKPoisson} onto the Hermite-Laguerre polynomial basis. For this purpose, we note that the Bessel function $J_m$  appearing in both \cref{eq:LinGK,eq:LinGKPoisson} and in the GK collision operators \citep{frei2021} can be conveniently projected by expanding the Bessel functions, $J_m$, onto associated Laguerre polynomials, $L^m_n(x) = (-1)^m d^m L_{n + m}(x) / d x^m$, 

\begin{align} \label{eq:J02Laguerre}
J_m(b \sqrt{x}) = \left(\frac{b \sqrt{x}}{2}\right)^{m}\sum_{n=0}^\infty \frac{n!\kernel{n}}{(n + m)!} L^m_n(x),
\end{align}
\\
where we introduce the $nth$-order kernel function 

\begin{align} \label{eq:kerneldef}
\kernel{n}  = \frac{1}{n!}\left(\frac{b}{2}\right)^{2n } e^{- b^2 /4},
\end{align}
\\
with argument $b = \sqrt{2 \tau} k_\perp$ \citep{Frei2020}. The projections of \cref{eq:LinGK,eq:LinGKPoisson} onto the Hermite-Laguerre basis yields the linearized gyro-moment hierarchy equation that states the time evolution of the gyro-moments, $N^{pj}$, that is

\begin{align} \label{eq:dNipjdt}
     \frac{\partial}{\partial t} N^{pj}  + &i k_\parallel \sqrt{\tau} \sum_{p'} \mathcal{H}^{\parallel p}_{p'} N^{p'j} + i \frac{\tau}{q}\omega_{ B } \sum_{p'j'} \left( \mathcal{H}_{B p'j'}^{\parallel pj} + \mathcal{H}_{B p'j'}^{\perp pj} \right)  N^{p'j'}  \nonumber \\
     &=  \left( \mathcal{F}^{pj}  - \mathcal{F}_{B}^{\parallel pj} \right) \phi + \C^{pj} ,
     \end{align}
\\
where we introduce the phase-mixing operators, associated with the parallel and perpendicular drifts,

\begin{subequations}
\begin{align}
    \mathcal{H}^{\parallel p}_{p'} & = \sqrt{p +1 } \delta_{p'}^{p+1}+ \sqrt{p} \delta_{p'}^{p-1}\\
    \mathcal{H}_{B p'j'}^{\parallel pj} & = \delta_{j'}^j \left( \sqrt{(p+1)(p+2)} \delta_{p'}^{p+2}  + \sqrt{2p+1} \delta_{p'}^p  + \sqrt{p(p-1)} \delta_{p'}^{p-2}\right)\\
    \mathcal{H}_{B p'j'}^{\perp pj} & = \delta_{p'}^p \left( (2j+1)\delta_{j'}^j - j \delta_{j'}^{j-1} - (j+1) \delta_{j'}^{j+1}\right), \label{eq:HBpjperppj}
\end{align}
\end{subequations}
\\
and the operators that provide the instability drive and are  proportional to the equilibrium gradients

\begin{subequations} \label{eq:forcemixingoperator}
\begin{align}
    \mathcal{F}^{pj} & =  i \omega_*  \left[ \kernel{j}  \delta_p^0 + \eta \left( \kernel{j} \frac{\delta_p^2}{\sqrt{2}}  + \delta_p^0 \left(2 j\kernel{j}   - j \kernel{j-1} - (j+1)\kernel{j+1} \right) \right) \right] ,\\
    \mathcal{F}_{B}^{\parallel pj} & =\frac{\delta_p^1}{\sqrt{\tau} } i k_\parallel \kernel{j}  + i \omega_{ B} \delta_p^0  \left[ 2 ( j +1) \kernel{j} - j \kernel{j-1} - (j+1)\kernel{j+1} \right]+   i \omega_B \kernel{j} \sqrt{2} \delta_p^2 .
\end{align}
\end{subequations}
\\
 In \cref{eq:dNipjdt}, the Hermite-Laguerre projection of the collision operator, $\C$, is defined by

\begin{align} \label{eq:Cpj}
\C^{pj} = \int d \theta d \mu d v_\parallel \frac{B}{m} \frac{H_p(s_\parallel) L_j(x)}{ \sqrt{2^p p!}} \C.
\end{align}
\\
In this work, different collision operator models for $\C$ are considered and detailed in \cref{sec:CollisionOperatorModels}. The linearized gyro-moment hierarchy equation is coupled with the self-consistent GK quasineutrality condition that, projected onto the Hermite-Laguerre basis, is given by

\begin{equation} \label{eq:GKPoissonNpj}
\left[ 1 + \frac{q^2}{\tau}\left( 1 - \sum_{n=0}^{\infty} \kernel{n}^2 \right)\right] \phi = q \sum_{n=0}^{\infty}  \kernel{n} N^{0n}.
\end{equation}
\\
\Cref{eq:dNipjdt,eq:GKPoissonNpj} constitute an infinite set of linear coupled fluid equations for the gyro-moments $N^{pj}$, and depend only on the Fourier mode wavevector $\bm k$ and time $t$. In order to solve numerically \cref{eq:dNipjdt} together with \cref{eq:GKPoissonNpj}, we apply a simple closure by truncation, i.e. we solve \cref{eq:dNipjdt,eq:GKPoissonNpj} up to $(p,j) \leq (P,J)$ and set $N^{pj} = 0$ for all $ (p,j) > (P,J)$, with $0 < P, J < \infty $. We remark that a high-collisional closure of the gyro-moment hierarchy is derived in \cref{sec:64GM}.

\subsection{Gyrokinetic Linearized Collision Operator Models}
\label{sec:CollisionOperatorModels}

 While numerous collision operator models are available in the literature \citep[see, e.g.,][]{Hirshman1976,Abel2008,Sugama2009,Li2011,Madsen2013a,Sugama2019}, the main focus of this work is a linearized GK Coulomb collision operator, and we leverage its formulation derived and implemented in \citet{frei2021}. This advanced linearized GK collision operator is obtained from the nonlinear full-F GK Coulomb operator derived in \citet{Jorge2019}. To highlight the importance of FLR terms in the Coulomb collision operator, the DK limit of the GK Coulomb collision operator is also considered. In addition, collision operators found in the literature and widely-used in numerical codes are also considered for comparison. In particular, we consider the GK and DK Sugama collision operator \citep{Sugama2009}, the GK and DK momentum-conserving pitch-angle scattering operator \citep{Helander2002}, the zeroth-order DK Hirshman-Sigmar-Clarke collision operator \citep{Hirshman1976}, and the GK and DK Dougherty collision operator \citep{Dougherty1964}. The gyro-moment expansion of the linearized GK/DK Coulomb and the GK/DK Sugama collision operators are reported in \citet{frei2021} where the linearized DK Coulomb and GK/DK Sugama operators are successfully benchmarked against the GK continuum code GENE \citep{Jenko2000}. The Hermite-Laguerre expansion of the other collision operators are reported in \cref{sec:pitchangle,sec:HSC,sec:dougherty}, respectively. We remark that the gyro-moment approach offers an ideal framework to study the impact of collisions and collision operator models on ITG modes thanks to its capability of describing the distribution function at different levels of fidelity, as needed, and as a function of the collisionality.
 
Before turning to the investigation of the impact of collisions on the ITG mode, here we recall some important conservation properties of linearized like-species collision operators. A model of linearized collision operator for ion-ion collisions, $C$, is obtained by linearizing a nonlinear collision operator, that we denote by $C^{NL}$. Since collisions act at the particle position $\r$ (rather than at the gyrocenter position $\R$), the nonlinear collision operator is most often derived as acting on the full \textit{particle} distribution function and is expressed in the particle phase-space $(\r, \bm v)$. While the functional form of $C^{NL}$ depends on the choice of collision model, the linearized collision operator, $C$, can always be expressed as

\begin{align}
    C  = C^T + C^F,    
\end{align}
\\
where we introduce the \textit{test} component of the linearized collision operator, $C^T =  C^{NL}(f,f_{M}) $, and the \textit{field} component, $ C^F =   C^{NL}(f_{M},f) $, with $f_M = f_M(\bm r, \bm v)$ the particle Maxwellian distribution function and $f = f (\r,\bm v)$ a small perturbation. 
Conservation properties of the collision operator constraint the \textit{test} and \textit{field} components. In fact, while particle conservation is satisfied by both components, such that

 \begin{align} \label{eq:particleconservation}
      \int d \vi C^T  = \int d \vi  C^F  = 0,
 \end{align}
 \\
 the momentum and energy conservations require that 
 
 \begin{align} \label{eq:momentumconservation}
 \int d \vi \bm v C^T  = -  \int d \vi\bm v C^F , \\
  \int d \vi v^2  C^T = - \int d \vi v^2 C^F ,\label{eq:energyconservation}
 \end{align}
 \\
 respectively. From the conservation properties of the linearized collision operator $C$, \cref{eq:particleconservation,eq:momentumconservation,eq:energyconservation}, constraints on the coefficients $\C^{pj}$ can be derived in the zero gyroradius limit, when the gyrocenter and particle position correspond. In fact, using \cref{eq:Cpj} combined with \cref{eq:particleconservation,eq:momentumconservation,eq:energyconservation} yields the relations

\begin{align} \label{eq:conservaltionCpj}
    \C^{00} = 0, \quad \C^{10}= 0, \quad \C^{20} = \sqrt{2} \C^{01},
\end{align}
\\
which express the conservation of gyrocenters, momentum and energy. The constraints in \cref{eq:conservaltionCpj} are satisfied by all linearized collision operator models considered in the present work in the long wavelength limit. We have checked that, indeed, \cref{eq:conservaltionCpj} are fulfilled up to machine precision when numerically implementing the gyro-moment expansions of the collision operators considered in this work. Finally, we note that the relations given in \cref{eq:conservaltionCpj} are used to derive the high-collisional limit of the gyro-moment hierarchy in \cref{sec:64GM}

\section{Collisionless Limit}
 \label{sec:CollisionlessHermiteLaguerreTheory}
  
  In this section, we derive the collisionless limit of the gyro-moment hierarchy equation given in \cref{eq:dNipjdt}, showing that the ITG collisionless results \citep[see, e.g.,][]{Romanelli1989,Hahm1989,Romanelli1990,Chen1991} can be retrieved as an asymptotic limit of the gyro-moment hierarchy. We first derive the collisionless ITG dispersion relation in \cref{subsec:collsionlessITGdisprel} by solving directly the GK equation, \cref{eq:LinGK}. Then, we obtain closed semi-analytical expressions of the collisionless linear gyro-moment response in \cref{subsec:collisonlessgyromomentresponse}. The expressions of the gyro-moments we obtain are valid at arbitrary order in the normalized magnetic frequency, $\omega_{\grad B}/\omega$, and are written in terms of linear combinations of definite integrals of velocity-independent hypergeometric functions, which reduce to linear combinations of derivatives of plasma dispersion functions in the slab limit. Finally, in \cref{subsec:inifinitemomentlimit} and thanks to the expressions of the gyro-moments developed in \cref{subsec:collisonlessgyromomentresponse}, we show the equivalence between the gyro-moment and GK dispersion relations by considering the infinite gyro-moment limit focusing on the slab branch. Ultimately, the results of the present section allow us to determine the scaling of the collisionless gyro-moments spectrum that will be used for the convergence studies in \Cref{sec:GMspectrum,sec:convergence}. 

\subsection{Collisionless GK ITG Dispersion Relation}
\label{subsec:collsionlessITGdisprel}

We first derive the ITG dispersion relation by solving directly the GK equation, \cref{eq:LinGK}. Fourier transforming  \cref{eq:LinToroidalGKB} in time with $\C =0$, such that $\partial_t \to - i \omega$, and introducing $\omega = \omega_r + i \gamma$, with $\omega_r$ and $\gamma$ the real mode frequency and the growth rate respectively, the normalized perturbed ion gyrocenter distribution function, $ \g/ \phi$, can be expressed as

\begin{align} \label{eq:fi}
  \frac{\g}{\phi}=  \sum_{l=1}^3 \hat{\g}_{l}
\end{align}
\\
with
\begin{subequations} \label{eq:ghatl}
\begin{align}
\hat{\g}_{1 }  & = - \frac{q}{\tau}  J_0(b \sqrt{x}) F_{M}, \\
\hat{\g}_{2 }  & = \frac{q}{\tau} \frac{  \omega }{ \omega - \omega_{\grad B}- z_\parallel s_\parallel } J_0(b \sqrt{x}) F_{M}, \\
\hat{\g}_{3}  & = -\frac{\omega_{T}^*}{\omega - \omega_{\grad B}- z_\parallel s_\parallel }J_0(b \sqrt{x}) J_0(b \sqrt{x}) F_{M},
\end{align}
\end{subequations}
\\
where we introduce $z_\parallel = \sqrt{2 \tau} k_\parallel$. The $l=1$ contribution to the sum in \cref{eq:fi} is the adiabatic response, while the $l=2$ and $l=3$ terms are associated with the Landau damping and equilibrium gradient drive. We remark the presence of the velocity-dependent magnetic frequency $\omega_B$ in the resonant denominator of the $l=2$ and $l=3$ contributions.

The GK ITG dispersion relation is deduced from the linearized GK quasineutrality condition given in \cref{eq:LinGKPoisson}, 

\begin{align} \label{eq:tdisperlationf}
  1 + \frac{q^2}{\tau} \left( 1 - \Gamma_0(a) \right) -  q  \hat{n} =0,
\end{align}
\\
where $\hat{n} = \sum_{l=1}^3  \hat{n}_l$ is the density perturbation with $ \hat{n}_l = \int d \vi J_0 \hat{\g}_l$ and $\hat{\g}_l$ given in \cref{eq:ghatl}. 
By considering an unstable mode, i.e. $\gamma > 0$, the velocity integrals appearing in $\hat{n}_l$ can be performed explicitly by writing the resonant term, $1/(\omega- \omega_{\grad B} - z_\parallel s_\parallel )$, in integral form \citep{Beer1996}

\begin{align} \label{eq:resonantintegral}
    \frac{1}{  \omega - \omega_{\grad B} -  z_\parallel  s_\parallel} = -  i \int_0^\infty d \tau e^{ i \tau (\omega - \omega_{\grad B} - z_\parallel s_\parallel)}.
\end{align}
\\
We remark that \cref{eq:resonantintegral} allows us to retain the effects of magnetic drift resonance in $\hat{N}_{l}^{pj}$ at arbitrary order in  $\omega_B / \omega$, since no expansion in the $\omega_B / \omega$ parameter is carried out to evaluate the velocity integrals, in contrast to previous toroidal collisionless ITG theories \citep{Jarmen1987,Kim1993}. Using \cref{eq:resonantintegral}, one derives

\begin{subequations}\label{eq:deltanl}
\begin{align}
 \hat{n}_{1} & = - \frac{q}{\tau} \Gamma_0(a), \\
\hat{n}_{2} & =  - \frac{ iq}{\tau} \omega \int_0^\infty d \tau e^{i \tau \omega} I_\perp(\tau) I_\parallel(\tau), \\
 \hat{n}_{3} & =  i k_\perp \int_0^\infty d \tau e^{i \tau \omega} \left[  I_\parallel (\tau)  I_\perp (\tau) \right. \nonumber \\ 
& \left.+ \eta \left(  I_{\parallel 2} (\tau)  I_\perp (\tau)  + I_{\parallel } (\tau)  I_{\perp 1} (\tau) - \frac{3}{2}  I_\parallel (\tau)  I_\perp (\tau) \right) \right],
\end{align}
\end{subequations}
\\
where we introduce

\begin{subequations} \label{eq:deltanlintegrals}
\begin{align} \label{eq:Itau1}
    I_\perp(\tau)& = \frac{1}{1 + i\alpha \tau} I_0\left(\frac{a
}{1 + i\alpha \tau} \right) e ^{-a /(1 + i \alpha\tau)}, \\
 I_\parallel(\tau)  & = \frac{1}{\sqrt{1 + 2i \alpha  \tau}} e^{ - z_\parallel^2 \tau^2 /[4 (1 + 2 i  \alpha \tau)]}, \\
 \label{eq:Itau2}
   I_{\perp 1}(\tau) & =    \frac{e^{- a /  (1 + i \alpha \tau)}}{2 (1 + i \alpha  \tau)^3} \nonumber \\
&  \times \left[ (2 (1 + i \alpha \tau) - 2 a) I_0 \left( \frac{a}{ (1 + i \alpha \tau)} \right)+ 2 a I_1 \left( \frac{a}{ (1 + i \alpha \tau)} \right) \right], \nonumber \\
   I_{\parallel 2}(\tau) & =   \frac{(2(1 + 2 i \tau \alpha ) - \tau^2 z_\parallel^2) }{4 (1 + 2 i \tau \alpha)^{5/2}}
    e^{- z_\parallel^2 \tau^2 / ( 4 (1 + 2 i \tau \alpha))},
\end{align}
\end{subequations}
\\
with $\alpha = q \omega_B / \tau$. \Cref{eq:tdisperlationf}, with the definitions in \cref{eq:deltanl} and the velocity integrals in \cref{eq:deltanlintegrals} constitute the GK ITG collisionless dispersion relation for the unstable modes $\gamma > 0$. We note that the exponential factors $e^{i \tau \omega}$, appearing in \cref{eq:deltanl}, ensure the convergence of the integrals at $\tau \to \infty$ for $\gamma > 0$. On the other hand, the inclusion of stable modes ($\gamma < 0$) in \cref{eq:tdisperlationf} can be obtained by expressing the velocity integrals in $\hat{n}$ using generalized plasma dispersion functions \citep{gultekin2019generalized} and efficient numerical algorithms can be used to solve the dispersion relation in the entire complex plane \citep{xie2017comparisons,gultekin2018stable}. However, here we focus on the upper half of the complex plane since one of the advantages of the transformation \cref{eq:resonantintegral} is that it yields one-dimensional integrals that can be easily evaluated numerically.

We finally note that the GK ITG dispersion relation can be simplified in the slab ITG (sITG) case. Neglecting $\omega _B$ in \cref{eq:ghatl}, the velocity integrals appearing in $\hat{n}$ and given in \cref{eq:deltanl}, can be expressed in terms of the plasma dispersion function. Then, the GK ITG  dispersion relation, \cref{eq:tdisperlationf}, becomes  

\begin{align} \label{eq:dispESITG}
& 1 +  \frac{1}{\tau}  + \frac{1}{\tau}   \xi  Z(\xi)  \Gamma_0(a) - \frac{k_\perp}{k_\parallel}  \frac{1}{\sqrt{2 \tau}} \left[  \Gamma_0(a) Z(\xi) \right. \nonumber \\
& \left.  + \eta \left( \Gamma_0(a) \xi \left( 1 + \xi Z(\xi)\right) - \frac{\Gamma_0(a)}{2}Z(\xi) + a Z(\xi)\left(\Gamma_1(a) - \Gamma_0(a)   \right)   \right)\right] =0,
\end{align}
\\
where  $\xi  = \omega \sqrt{2 \tau}/ k_\parallel$ is the normalized mode phase velocity, $Z(\xi) = \int_{-\infty}^{\infty} d s_\parallel e^{- s_\parallel^2}/(s_\parallel - \xi)/\sqrt{\pi}$ is the plasma dispersion function, and $\Gamma_1(a) = I_1(a) e^{- a}$.

  \subsection{Collisionless Linear Gyro-Moment Response}
  \label{subsec:collisonlessgyromomentresponse}

 We now obtain the collisionless expressions of the gyro-moments by projecting \cref{eq:fi} onto the Hermite-Laguerre basis. This yields

\begin{align} \label{eq:tNapj}
\frac{N^{pj}}{\phi} = \sum_{l=1}^3 \hat{N}_{l}^{pj},
\end{align}
\\
where we introduce 

\begin{align} \label{eq:defhatNlpj}
    \hat{N}_{l}^{pj} = 2 \pi \int d v_\parallel d \mu \frac{B}{m} \frac{H_p(s_\parallel ) L_j(x) \hat{g}_l }{ \sqrt{2^p p !}}.
\end{align}
\\ 
with $\hat{g}_l$ defined in \cref{eq:ghatl}. Performing the velocity integral in \cref{eq:defhatNlpj} using \cref{eq:resonantintegral} for unstable mode with $\gamma >0$ and identities involving hypergeometric functions and Hermite-Laguerre polynomials (described in \cref{sec:AppGeneralizedLinearMomentResponse}), we obtain

\begin{subequations}\label{eq:tdisperlation}
\begin{align}
    \hat{N}_1^{pj} & = - \frac{q}{\tau} \kernel{j} \delta_p^0, \\
    \hat{N}_{2}^{pj}& = - \frac{i q \omega}{\tau} \sum_{n=0}^\infty  \frac{ \kernel{n} }{\sqrt{2^p p!} }\mathcal{I}_{n}^{pj} , \\
        \hat{N}_{3}^{pj}& =  \frac{i k_\perp}{\sqrt{2^p p!}}\sum_{n=0}^{\infty}\kernel{n}  \left\{ \mathcal{I}_n^{pj}  + \eta \left[ (2j+1) \mathcal{I}_n^{pj}   - j  \mathcal{I}_n^{pj-1}- (j+1)  \mathcal{I}_n^{pj+1} \right.\right. \nonumber \\
    & \left. \left. + \frac{\mathcal{I}_{n}^{p+2j}}{4} + \left(p + \frac{1}{2} \right) \mathcal{I}_n^{pj} + p(p-1) \mathcal{I}_n^{p-2j}- \frac{3}{2}\mathcal{I}_n^{pj}\right] \right\}.
      \end{align}
 \end{subequations}
 \\
 where $\mathcal{I}_{n}^{pj}$ is given in \cref{eq:Inpj} as a finite sum of terms that involve integrals of velocity-independent hypergeometric functions with complex argument. Generalized plasma dispersion relations can also be used to express the collisionless gyro-moments $\hat{N}_{l}^{pj}$, given in \cref{eq:tdisperlation}, extending the formulas in \cref{sec:AppGeneralizedLinearMomentResponse} to the lower half complex plane of $\omega$.

In the case of the slab ITG branch, the velocity integrals containing the resonant term, $1 /(\omega - z_\parallel s_\parallel)$, being $\omega_B = 0$, can be performed without using \cref{eq:resonantintegral}. In fact, in this case, the collisionless gyro-moments becomes

 \begin{align} \label{eq:Nipjslab}
   \frac{N^{pj}}{\phi}& = -\frac{q}{\tau} \kernel{j} \delta_{p}^0  + \frac{1}{\sqrt{2 \tau}}\frac{k_\perp}{k_\parallel} \frac{1 }{\sqrt{2^p p!}}  \left( \kernel{j}  (-1)^p Z^{(p)} (\xi) \right.  \nonumber \\
&   \left. + \eta  \left[ \kernel{j} \left( \frac{(-1)^{p+2}}{4} Z ^{(p+2)}(\xi) + (p + 2j) (-1)^p Z ^{(p)}(\xi) + p(p-1) (-1)^{p-2} Z ^{(p-2)}(\xi)\right)  \right. \right.\nonumber  \\
&   \left. \left. - \left[ j \kernel{j-1} + (j+1) \kernel{j+1} \right] (-1)^p Z ^{(p)}(\xi) \right) \right]    - \frac{q}{\tau} \frac{ \xi \kernel{j}}{\sqrt{2^p p!}}    (-1)^p Z^{(p)}(\xi),
\end{align}
\\
where $Z^{(p)} (\xi)= d^{p} Z(\xi)/ d \xi^p$ is the $p$th-order derivative of the plasma dispersion relation, which can be defined in terms of the Hermite polynomials, $H_p$, by  $Z^{(p)} (\xi) = (-1)^p \int_{-\infty}^{\infty} d s_\parallel H_p(s_\parallel) e^{- s_\parallel^2}/(s_\parallel - \xi)/\sqrt{\pi}$.

As an aside, we also notie that \cref{eq:tNapj} can be related to the kinetic response of the monomial velocity-space moments,  $M_{j,k} = v_T^{j + 2k}\int d \vi \g s_\parallel^j x^k / 2^k$, used in \citet{Beer1996} to derive general closure expressions for toroidal gyrofluid equations, in the $k_\perp = 0$ limit, such as

 \begin{align} \label{eq:Npjtobeer}
     N^{pj} = \sum_{p_1=0}^{\lfloor p/2 \rfloor} \sum_{j_1=0}^j \frac{(-1)^{p_1} 2^{p-2p_1}}{p_1!(p-2p_1)!}  \frac{(-1)^{j_1} j !}{(j - j_1)!(j_1!)^2}  \frac{2^{j_1}}{v_T^{2j_1 + p - 2p_1}} \frac{M_{p-2p_1,j_1}}{\sqrt{2^p p!}}.
 \end{align}
\\
We remark that \cref{eq:Npjtobeer} provide a direct link between the gyro-moment hierarchy equation described here and the moment equations derived in \citet{Beer1996} and associated closures expressions.

\subsection{Gyro-Moment Dispersion Relation and Infinite Gyro-Moment Limit}
\label{subsec:inifinitemomentlimit}

 The ITG dispersion relation expressed in terms of $N^{pj}$ is obtained from the GK quasineutrality condition \cref{eq:GKPoissonNpj}, yielding

\begin{align} \label{eq:DispRelNpj}
1 + \frac{q^2}{\tau}\left( 1 - \sum_{j=0}^{\infty} \kernel{j}^2 \right) - \sum_{j=0}^{\infty} \kernel{j} \sum_{l=1}^{3 }\hat{N}^{0j}_l = 0,
\end{align}
\\
where $\hat{N}_l^{0j}$ is given by \cref{eq:tdisperlation} with  $p=0$. 

\begin{figure}
  \centering 
    \includegraphics[scale = 0.48]{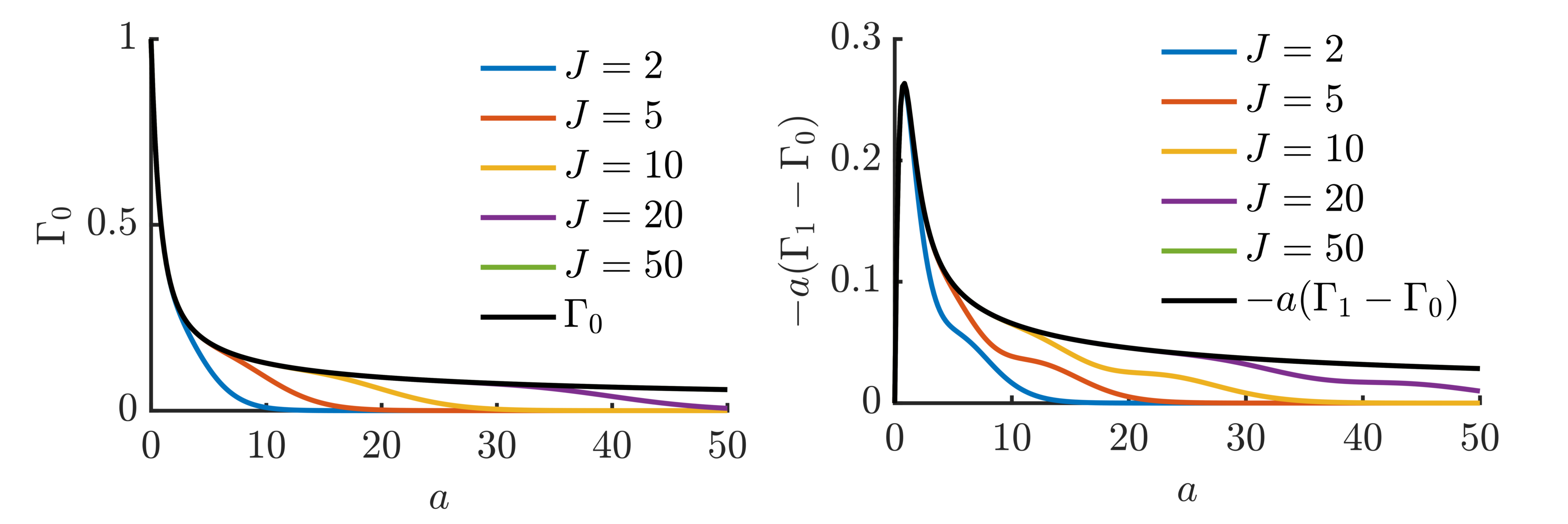}
   \caption{$\hat{\Gamma_0}(a,J)$ and $\hat{\Pi}_1(a,J) $ as a function of $a = b^2/2$ for increasing values of $J$. The solid black lines are their asymptotic limits, $\Gamma_0$ and $- a (\Gamma_1 - \Gamma_0)$, respectively. }
   \label{fig:Fig_sITG_FLROperator}
  \end{figure}

 Focusing on the sITG branch, we now demonstrate that, in the $P \to \infty$ and $J \to \infty$ limit, the ITG gyro-moment dispersion relation, \cref{eq:DispRelNpj}, is equivalent to the GK ITG dispersion relation given in \cref{eq:dispESITG}. In order to solve the  gyro-moment dispersion relation \cref{eq:DispRelNpj}, we use the expressions for the collisionless gyro-moments, $N^{pj} / \phi$ given in \cref{eq:Nipjslab} in the sITG case. Setting  $p =0$ in \cref{eq:Nipjslab} and plugging the resulting expression for $N^{p0} / \phi$ into  \cref{eq:DispRelNpj} yields

\begin{align} \label{eq:dispESITGmoment}
& 1 +  \frac{1}{\tau}  + \frac{1}{\tau}   \xi Z(\xi)  \hat{\Gamma_0}(a) - \frac{k_\perp}{k_\parallel}  \frac{1}{\sqrt{2 \tau}} \left[  \hat{\Gamma_0}(a,J) Z(\xi)   + \eta\left(\hat{\Gamma_0}(a,J) \xi \left( 1 + \xi Z(\xi)\right)    \right. \right. \nonumber \\
& \left.  \left.  - \frac{\hat{\Gamma_0}(a,J)}{2}Z(\xi) +Z(\xi) \hat{\Pi}_1(a,J)   \right)\right] =0,
\end{align}
\\
where we introduce the quantities

\begin{subequations} \label{eq:hatGamma0hatGamma1}
\begin{align}
 \hat{\Gamma_0}(a,J)  & = \sum_{j=0}^{J} \kernel{j}^2, \\
\hat{\Pi}_1(a,J)  & =  \sum_{j=0}^{J} \kernel{j} \left(2j  \kernel{j} - j \kernel{j-1} -(j+1)\kernel{j+1}  \right).
\end{align}
\end{subequations}
\\
which are associated with FLR effects. By comparing \cref{eq:dispESITGmoment} with \cref{eq:dispESITG}, one observes that they are equivalent in the infinite moment limits provided that $ \hat{\Gamma_0}(a,\infty) = \Gamma_0$ and $\hat{\Pi}_1(a,\infty) = a (\Gamma_1 - \Gamma_0)$. This can be verified explicitly by considering the $J \to \infty$ limit in \cref{eq:hatGamma0hatGamma1} and using the definition of the kernel function $\kernel{j}$ in \cref{eq:kerneldef}. In fact, we derive 

\begin{align} \label{eq:hatgamma0}
 \hat{\Gamma_0}(a,\infty)  & = \sum_{j=0}^\infty \frac{1}{(j!)^2} \left( \frac{b}{2}\right)^{4j} e^{- b^2 /2} \nonumber \\
  & = e^{- a} I_0\left( a\right) = \Gamma_0(a),
\end{align}
\\
and 

\begin{align} \label{eq:hatgamma1gamma0}
   \hat{\Pi}_1(a,\infty)  & = 2 e^{ - b^2 /2}\sum_{j=0}^\infty \left[ \frac{1 }{j!(j+1)!} \left(  \frac{b}{2} \right)^{4(j+1)} -   \frac{1}{(j!)^2} \left( \frac{b}{2}\right)^{4j + 2}\right] \nonumber \\
   & = a e^{- a} \left( I_1(a) - I_0(a)\right) = a \left( \Gamma_1(a)- \Gamma_0(a)\right).
\end{align}
\\
 \Cref{fig:Fig_sITG_FLROperator} displays the quantities, $\hat{\Gamma_0}(a,J)$ and $\hat{\Pi}_1(a,J)$ defined in \cref{eq:hatGamma0hatGamma1}, for different values of $J$, as well as their $J \to \infty$ limits. As observed, $\hat{\Gamma_0}(a,J)$ and $\hat{\Pi}_1(a,J)$ converges to $\Gamma_0$ and $a(\Gamma_1 - \Gamma_0)$ when $J\to \infty$. We notice that convergence is faster for low values of $a = b^2/2$, i.e. for long wavelength mode. This is in agreement with the fact that the truncation of the sums in \cref{eq:hatgamma0} and \cref{eq:hatgamma1gamma0} to include only $j \leq J$ terms yields an error $O( b^{4J})$.
 
\begin{figure}
  \centering 
    \includegraphics[scale = 0.48]{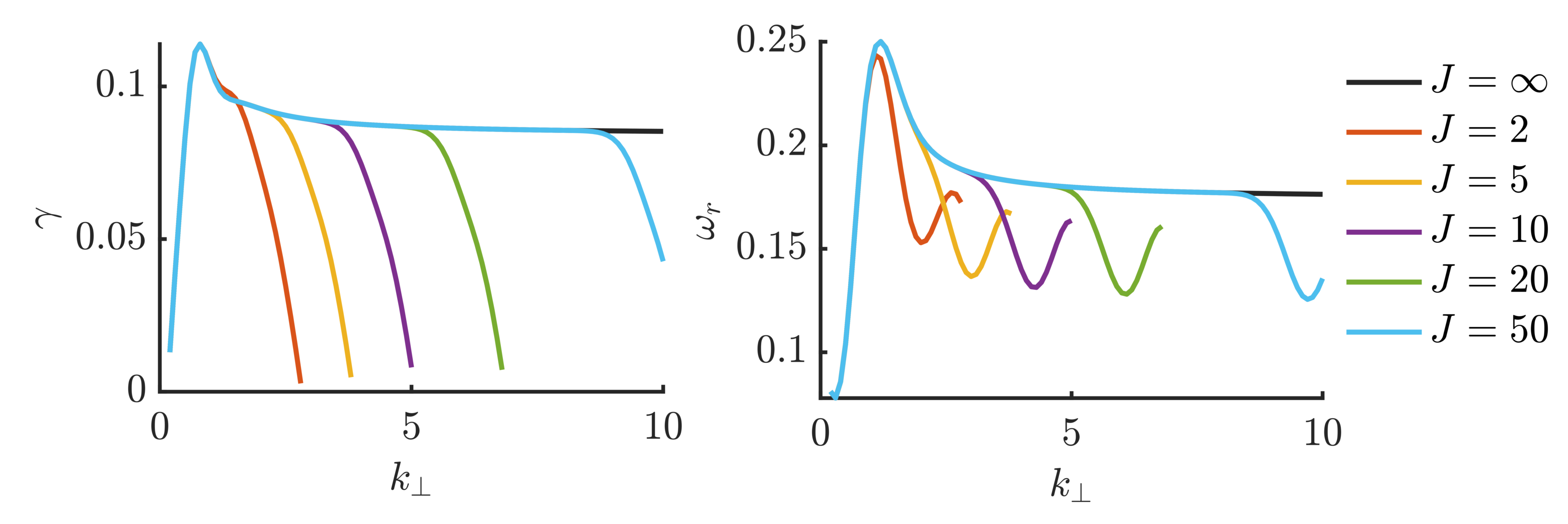}
   \caption{Collisionless sITG growth rate, $\gamma$, and frequency, $\omega_r$, as a function of the perpendicular wavenumber $k_\perp$ for different values of $J$. The color lines are the solution of the gyro-moment hierarchy dispersion relation, given in \cref{eq:dispESITGmoment}, while the black lines are solutions of \cref{eq:dispESITG}, i.e. the $J = \infty$ solution. Here, $\eta =3 $ and $k_\parallel = 0.1$.}
   \label{fig:sITGgammaowrcollisionless}
  \end{figure}

 The impact on the ITG mode due to approximating $\hat{\Gamma_0}(a,\infty)$ and $\hat{\Pi}_1(a,\infty)$ with $\hat{\Gamma_0}(a,J)$ and $\hat{\Pi}_1(a,J)$ can be illustrated by solving the ITG dispersion relation, \cref{eq:dispESITGmoment}, for the mode complex frequency $\omega = \omega_r + i \gamma$, as a function of $J$, and compare its prediction with the one from the GK ITG dispersion relation, given in \cref{eq:dispESITG}. \Cref{fig:sITGgammaowrcollisionless} shows (left panel) the maximum growth rate, $\gamma$, and (right panel) the corresponding real frequency, $\omega_r$, solution of the gyro-moment sITG dispersion relation, given in \cref{eq:dispESITGmoment}, plotted as a function of the perpendicular wavenumber $k_\perp$, when $\eta =3$ and $k_\parallel = 0.1$, for increasing values of $J$. For comparison, the collisionless GK solution, obtained from \cref{eq:dispESITG}, is shown. Both $\gamma$ and $\omega_r$, obtained by the dispersion relation expressed in terms of gyro-moments, \cref{eq:dispESITGmoment}, approach the one predicted by \cref{eq:dispESITG}, as $J \to \infty$. Convergence is achieved with small $J$ at long wavelength (e.g., the $k_\perp \simeq 0.6$ case is well reproduced with $J \simeq 5$). In general and as a rule of thumb, $J \gtrsim k_\perp ^2$ must be retained in order to retrieve accurately the GK collisionless prediction. We remark that, if $J$ is not sufficiently large, the truncation of the quantities, \cref{eq:hatGamma0hatGamma1}, can yield significant deviations in the growth rate $\gamma$ and frequency $\omega_r$. The presence of collisions is expected to affect the convergence estimate reducing $J$ even for short wavelength modes, as numerically demonstrated in \cref{sec:convergence}.

\section{High-Collisional Limit}
\label{eq:highcollisionallimit}

We now investigate the high collisional limit of the gyro-moment hierarchy, \cref{eq:dNipjdt}, focusing on two reduced models that retain only a finite number of gyro-moments. In the high-collisionality limit, high-order gyro-moments are damped by the presence of collisions and the ITG mode can be accurately described by considering only the evolution of the lowest order gyro-moments. Indeed, the perturbed gyro-center distribution function is expected not to significantly deviate from a perturbed Maxwellian at high collisionalities. We consider, first, a reduced model based on $6$ gyro-moments ($6$GM), obtained by using a closure by truncation of the gyro-moment hierarchy. Second, we consider a reduced model based on $4$ gyro-moments ($4$GM), rigorously derived as an asymptotic limit of the $6$GM using the Chapman-Enskog closure to express the highest-order gyro-moments as a function of the lower order ones in the limit of $ \epsilon \sim k_\parallel \lambda_{mfp}  \ll 1  $. While the Chapman-Enskog procedure allows us to reduce the number of gyro-moments, closures for the FLR terms associated with the $\kernel{n}$ kernel functions and inherent to gyro-fluid models \citep{brizard1992nonlinear,Waltz1997,Dorland1993,Beer1996,Madsen2013, Held2020}, are still an open issue in the literature. The closure issue appears explicitly in the Fourier exponential form of the kernel function defined in \cref{eq:kerneldef}. \Cref{appendix:FLRclosures} is dedicated to explore possible FLR closures by using a Pad\'e approximation technique applied to the $4$GM model. To make analytical progress, we focus here on long wavelength modes. Consistently, we model collisions using the DK Coulomb operator \citep{frei2021}. This has the advantage to avoid infinite sums with FLR terms that need to be truncated depending on the value of the perpendicular wavenumber, $k_\perp$. 

The present section is structured as follows. In \cref{sec:64GM}, we derive the $6$GM and $4$GM reduced models. The high-collisional limits are also verified against the GK code GENE. In \cref{subsec:highcollisionaldisprel}, we explore the high-collisional limit by deriving a general algebraic collisional dispersion relation from which we discuss the limits of the slab and toroidal branch of the ITG mode.

\subsection{$6$GM and $4$GM Reduced Models}
\label{sec:64GM}

We start by deriving the 6GM model. We introduce the lowest order normalized fluid gyro-moments, i.e. the gyrocenter density $ N  = N^{00}$, the parallel velocity $V = N^{10}/\sqrt{2}$, the parallel and perpendicular temperatures, $T_\parallel  = \sqrt{2} N^{20} + N$ and $T_\perp  = N - N^{01}$ and, finally, the parallel and perpendicular heat fluxes, associated with the deviations of $\g$ from the Maxwellian distribution function, $Q_\parallel = N^{30}$ and $Q_\perp  = N^{11}$. The evolution equations of these gyro-moments are obtained by setting $(p,j) = (0,0)$, $(1,0)$, $(2,0)$, $(0,1)$, $(3,0)$, and $(1,1)$ in \cref{eq:dNipjdt}, respectively, and by neglecting all higher order gyro-moments. This yields  

\begin{subequations}\label{eq:ITG6GM}
\begin{align}
\frac{\partial}{\partial t } N  &+ i k_\parallel \sqrt{2 \tau} V  + i \tau \omega_{B} \left( T_\parallel - T_\perp \right)   + i \Gamma_1 \phi  = 0 \label{eq:eqN}\\ 
\sqrt{2} \frac{\partial}{\partial t }V  &+  \frac{i k_\parallel}{\sqrt{\tau}} \kernel{0} \phi + i k_\parallel \sqrt{\tau} T_\parallel + i \tau \omega_B \left( 4 \sqrt{2} V-  Q_\perp+ \sqrt{6} Q_\parallel \right)= 0, \label{eq:eqV}\\
\frac{\partial}{\partial t } T_\parallel &  + i k_\parallel \sqrt{6\tau} Q_\parallel  + i \tau \omega_B \left( 5 T_\parallel - 4 N + T_\perp  \right) \nonumber \\
& +  i (\Gamma_2 - \Gamma_1) \phi = - \nu \alpha   \left( T_\parallel     - T_\perp \right) \label{eq:eqTparallel}, \\
     \frac{\partial}{\partial t }T_\perp&+ i k_\parallel \sqrt{\tau} ( \sqrt{2 } V+Q_\perp )  -  i \tau \omega_B \left( 3 N    - T_\parallel -3 T_\perp  \right) \nonumber \\
     & - i (\Gamma_{3}- \Gamma_1) \phi  =  \nu \frac{\alpha}{2}\left(  T_\parallel - T_\perp\right), \label{eq:eqTperp} \\
\frac{\partial}{\partial t } Q_\parallel & + i k_\parallel \sqrt{\frac{3 \tau}{2}} (T_\parallel - N)  + i \tau \omega_B \left( 2 \sqrt{3} V + 8Q_\parallel\right)= - \nu \left(  \frac{8}{5}\sqrt{\frac{2}{\pi}} Q_\parallel  + \frac{8}{5 \sqrt{3 \pi}} Q_\perp \right), \label{eq:Qparallel} \\
\frac{\partial}{\partial t }Q_\perp & +  \frac{i k_\parallel }{\sqrt{\tau}} \kernel{1} \phi  + i k_\parallel \sqrt{\tau} ( N - T_\perp)    \\
&  - i \tau \omega_B \left( \sqrt{2} V - 6 Q_\perp \right)= - \nu \left( \frac{28}{15}\sqrt{\frac{2}{\pi}} Q_\perp +  \frac{8}{5 \sqrt{3 \pi}} Q_\parallel \right) .\label{eq:Qperp} 
\end{align}
\end{subequations}
\\
where we introduce the numerical coefficients $\alpha = 16/15 \sqrt{2 / \pi}$. We remark that the collisional terms in \cref{eq:ITG6GM} are obtained by using the lowest-order gyro-moments of the DK Coulomb collision operator \citep{frei2021} and the energy constraints expressed in \cref{eq:conservaltionCpj}. The quantities appearing in the evolution equations, \cref{eq:ITG6GM}, are defined by

\begin{subequations} \label{eq:Gammas}
\begin{align} 
\Gamma_1 &=  \kernel{0} \left(    2  \omega_B - \omega_*   \right)  +  \kernel{1} \left(   \omega_*   \eta -  \omega_B\right)   ,  \\
\Gamma_2 & = \left( 2 \omega_B- \omega_*  \eta \right) \kernel{0}, \\
\Gamma_3 & =   \kernel{0} \left( \omega_*   \eta   - \omega_B \right)  + \kernel{1} \left( 4 \omega_B- \omega_*  ( 1 + 2 \eta)\right)  + 2 \kernel{2} \left( \eta \omega_*    - \omega_B \right),
\end{align}
\end{subequations}
\\
Additionally, the electrostatic potential, $\phi$, is determined by the self-consistent quasi-neutrality condition obtained from \cref{eq:GKPoissonNpj}, i.e.

\begin{align} \label{eq:poissonGMs}
  \left[ 1 + \frac{q^2}{\tau} \left( 1 - \sum_{j=0}^{1} \kernel{j}^2 \right) \right]\phi = q(\kernel{0}  + \kernel{1} ) N - \kernel{1}T_\perp,
\end{align}
\\
where the infinite sum appearing in the polarization term in \cref{eq:GKPoissonNpj} is truncated to $j = 1$ consistently with the fact that the FLR terms are represented up to $\kernel{1}$ in the right-hand side of the same equation. \Cref{eq:ITG6GM} constitutes a closed set of fluid-like equations that we refer to as the $6$GM model.



 To derive the $4$GM model, further reducing the number of gyro-moments appearing in the $6$GM model, we apply the Chapman-Enskog asymptotic closure scheme \citep{chapman1941velocity,Jorge2017} to the $4$GM model. We introduce the dimensionless small parameter $\epsilon \sim \lambda_{mfp} k_\parallel  \sim \sqrt{\tau} k_\parallel / \nu \ll 1$, which quantifies the importance of the non-Mawellian part of $\g$. This allows us to express the heat fluxes $Q_\parallel$ and $Q_\perp$ as a function of the lowest order gyro-moments, by applying the ordering $Q_\parallel  \sim Q_\perp \sim \epsilon N $ (with $T_\parallel \sim T_\perp \sim V  \sim N$) and $\partial_t \sim \gamma \sim \epsilon \nu $. We neglecting the terms proportional to $\omega_B$, which are smaller than the terms proportional to $k_\parallel T_\parallel $ and $k_\parallel T_\perp$ since $\omega_B / k_\parallel \lesssim 1$ in the boundary region. With these assumptions, \cref{eq:Qparallel,eq:Qperp} can be solved at the leading order in $\epsilon$ yielding 
%
%

\begin{subequations} \label{eq:N30andN11}
\begin{align}
Q_{\parallel } = i k_\parallel   \frac{\chi_{\perp}^{\parallel}   }{\tau} \kernel{1} \phi + i k_\parallel \chi_{\perp}^{\parallel} ( N - T_\perp) - i k_\parallel \chi_{\parallel}^{\parallel} (T_\parallel - N ), \\
Q_\perp =-   i k_\parallel \frac{\chi_\perp^\perp }{\tau} \kernel{1} \phi - i k_\parallel \chi_\perp^\perp ( N - T_\perp)  + i k_\parallel  \chi^\perp_{\parallel} (T_\parallel - N ).
\end{align}
\end{subequations}
\\
 where we introduce the thermal conductivities,
 
 \begin{align} \label{eq:conductivities}
\chi_{\perp}^{\parallel} = \frac{5}{16}\sqrt{\frac{\pi}{3}} \frac{\sqrt{\tau}}{\nu}, \quad  \chi_{\parallel}^\parallel = \frac{35}{32} \sqrt{\frac{2 \pi}{3}}\frac{\sqrt{\tau}}{\nu}, \quad \chi_\perp^\perp = \frac{5\sqrt{2\pi}} {16}\frac{\sqrt{\tau}}{\nu}, \quad \chi_\parallel^\perp = \frac{5\sqrt{\pi}}{16} \frac{\sqrt{\tau}}{\nu}.
 \end{align}
 \\
 We remark that $\chi_{\parallel}^\parallel > \chi_{\perp}^\parallel$ and that $\chi_\perp^\perp > \chi_\parallel^\perp$. \Cref{eq:ITG6GM}, with the closure relations expressed in \cref{eq:N30andN11}, yield the $4$GM for the lowest-order gyro-moments fluid quantities $N$, $V$, $T_\parallel$ and $T_\perp$, providing the rigorous $\nu \gg 1$ asymptotic limit of the gyro-moment hierarchy. We remark that an analytical procedure to close the gyro-moment hierarchy similar to the one considered here for the ITG mode, can be applied in more the general cases that include variation of equilibrium quantities along the magnetic field line and electron dynamics.

\begin{figure}
    \centering
    \includegraphics[scale=0.6]{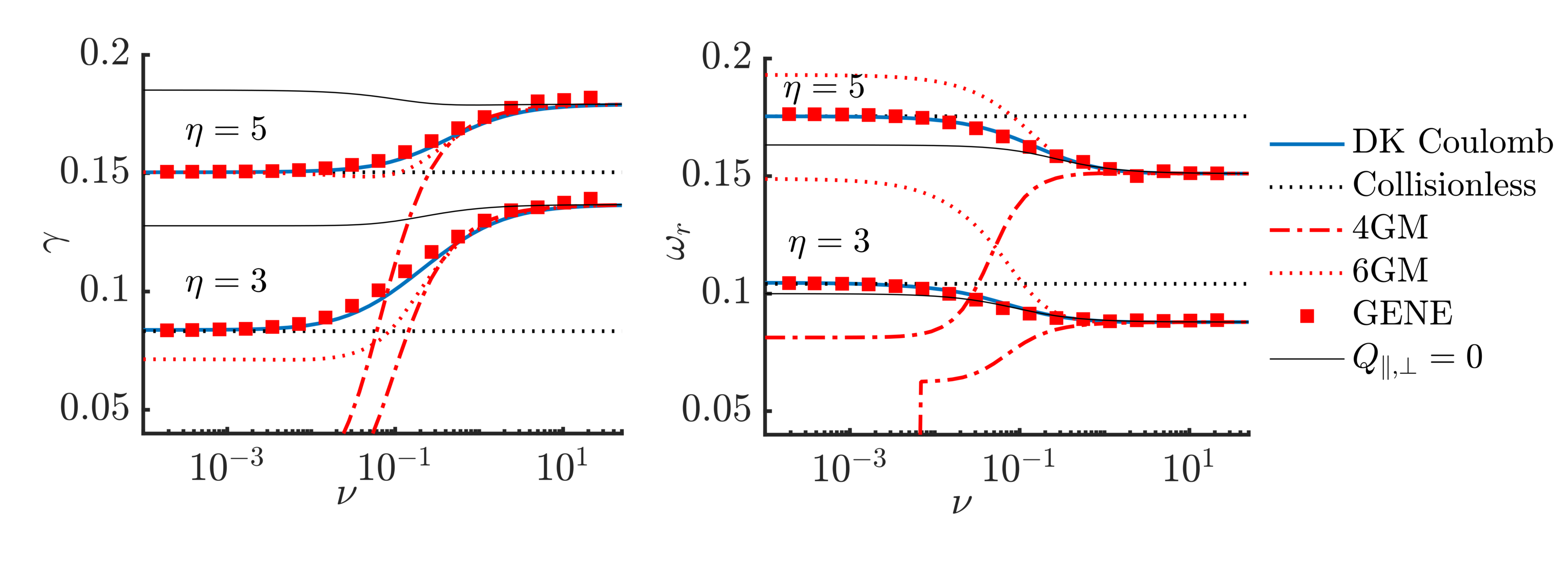}
    \caption{(Left) slab ITG growth rate $\gamma$ and (right) real frequency $\omega_r$ as a function of collisionality $\nu$ obtained using the gyro-moment approach with the DK Coulomb collision operator (solid blue line), GENE with the same operator (red markers), the collisionless GK dispersion relation (dotted black) and $4$GM and $6$GM high collisional limits  (red dashed-dotted and red dotted lines, respectively). The solution of the dispersion relation in the case of $Q_{\parallel, \perp} = 0$, given by \cref{eq:cubicsITG}, is also plotted for comparison by the solid thin black lines. Here, we consider $k_\parallel = 0.1$ and $k_\perp = 0.5$.}
    \label{fig:Fig_GENEVSMOLI_nuScan}
\end{figure}

 To study the limits of validity of the $6$GM and $4$GM models, we compare in \cref{fig:Fig_GENEVSMOLI_nuScan} the estimate of the ITG growth rate $\gamma$ and real frequency $\omega_r$ with the results of the gyro-moment hierarchy using the DK Coulomb collision operator \citep{frei2021} and the GK ITG collisionless dispersion relation, given in \cref{eq:tdisperlationf}. For verification purposes, the results of GENE simulations using the same operator \citep{Jenko2000} are also shown. Here, we consider $(P,J) = (18,6)$ (convergence studies are reported in \cref{sec:convergence}) and $k_\perp = 0.5$, which corresponds approximately to the fastest growing ITG mode. The results are shown as a function of the collisionality $\nu$, for two different normalized temperature gradients, $\eta=3$ and $\eta =5$. First, an excellent agreement is observed between the gyro-moment approach and the GENE simulations at arbitrary collisionality $\nu$ using the same DK Coulomb collision operator, confirming the preliminary study reported in \citet{frei2021}. Second, both GENE and the gyro-moment ITG growth rates and real frequencies agree well with the high-collisional ($\nu \gg 1$) predictions obtained from the $6$GM and $4
 $GM models, in addition to being in agreement between them in the collisionless limit, i.e. $\nu \ll 1 $. Overall, the $6$GM provide a better approximation to the low-collisionality solution than the $4$GM, while they both agree in the high-collisional limits. It is remarkable that the $6$GM retrieves surprisingly well the ITG growth rate $\gamma$ when $\eta =5$, but fails to predict correctly the mode frequency $\omega_r$. Finally, we remark that the $4$GM yields an unphysical $\omega_r < 0$, when the collisionality is below $\nu \lesssim 10^{-2}$ with $\eta =3$. This is mainly due to the fact that in the low-collisionality limit, the $1/\nu$ dependence of the thermal conductivities given in \cref{eq:conductivities} produces arbitrarily large heat fluxes. The good agreement observed in \cref{fig:Fig_GENEVSMOLI_nuScan} illustrates the multi-fidelity nature of the gyro-moment approach that allows the derivation of reduced models able to address both the collisionless and high-collisional limits of the GK equation.
 
\subsection{High-Collisional Dispersion Relation} 
\label{subsec:highcollisionaldisprel}

 We now derive an algebraic dispersion relation using the $4$GM equations and discuss the limits of the slab and toroidal branch of the ITG mode. As collisional effects enter through the parallel and perpendicular thermal conductivities such that $\chi^{\parallel}_{\perp} \sim \chi^{\parallel}_{\parallel} \sim \chi^{\perp}_{\parallel} \sim  \chi^{\perp}_{\perp} \sim 1/\nu$ (see \cref{eq:conductivities}),  we assume that the heat fluxes become negligible when $\nu \gg 1$, i.e. $Q_{\parallel} \simeq 0 $ and $Q_\perp \simeq 0 $ from \cref{eq:N30andN11}. Hence, a dispersion relation can be derived from the $4$GM equations, \cref{eq:ITG6GM}, yielding

   \begin{align} \label{eq:cubicsITG}
a \omega^4 + b \omega^3 + c \omega^2 + d \omega + e   = 0,
 \end{align}
 \\
 where

 \begin{subequations}
 \begin{align}
     a &=   \Gamma_\phi , \\     
     b & =  \frac{3 }{2} i \alpha \nu \Gamma_\phi     - \Gamma_1 \kernel{0} - \Gamma_3 \kernel{1}- 16 \Gamma_\phi \tau \omega_{B},  \\
     c &= \frac{ i \alpha  \nu}{2} \left( \kernel{1}\left( \Gamma_2 - 2 \Gamma_3\right) - 32  \tau \Gamma_\phi \omega_B \Gamma_1 \kernel{0}  \right) \nonumber \\
& - 3 k_\parallel^2 \tau \Gamma_\phi -  k_\parallel^2 \kernel{0}^2 + \tau \omega_B \left( \kernel{0} \left( 14 \Gamma_1 - \Gamma_2 + \Gamma_3 \right)  + \kernel{1} \left(\Gamma_1 + 12 \Gamma_3 \right) \right)  , \\
   d&  =  i \alpha \nu \left[  k_\parallel^2 \left(   \kernel{0} \kernel{1} - \frac{5}{2} \tau \Gamma_\phi - \frac{3}{2}  \kernel{0}^2 \right) + \tau \omega_B \left(  \Gamma_1 \left(13 \kernel{0} + 2 \kernel{1} \right) - (\Gamma_2 - 2 \Gamma_3)(\kernel{0} + 3 \kernel{1}) \right)  \right]  \nonumber \\
   &   +  \tau  k_\parallel^2 \left[ \omega _{B } \left(18
   \tau  \Gamma _{\phi }+\kernel{0} \left(8 \kernel{0} +\kernel{1} \right)\right)+ \left(2 \Gamma _1-\Gamma _2\right) \kernel{0} +3 \Gamma _3 \kernel{1}  \right], \\
   e & = i \alpha \nu \left[  5  k_\parallel^2 \tau \omega_B \left( \tau \Gamma _{\phi }  +  \kernel{0}^2 \right)
  +k_\parallel^2 \tau \left( \Gamma _1- \frac{\Gamma _2}{2}+ \Gamma _3\right) \left(\kernel{0}+ \kernel{1}\right] \right) \nonumber \\
   &  + k_\parallel^2 \tau^2 \omega_B \left(\kernel{1}  \left(\Gamma _2-6 \Gamma _3 -2 \Gamma _1 \right)  -2 \kernel{0}  \left(4 \Gamma _1-2 \Gamma _2+\Gamma _3\right) \right),
 \end{align}
 \end{subequations}
 \\
and where terms proportional to $\omega_B^2$ are neglected (being $\omega_B \ll 1$). We also define $\Gamma_\phi = 1 + q^2 (1 - \kernel{0}^2 - \kernel{1}^2) / \tau $. In \cref{eq:cubicsITG}, collisional effects are represented through the terms proportional to $ i\alpha \nu$. The solution of \cref{eq:cubicsITG} is plotted in \cref{fig:Fig_GENEVSMOLI_nuScan} and compared with the $6$GM and $4$GM models, showing a good agreement when $\nu \gg 1$, such that $Q_\parallel = Q_\perp \simeq 0$.

\begin{figure}
    \centering
    \includegraphics[scale = 0.57]{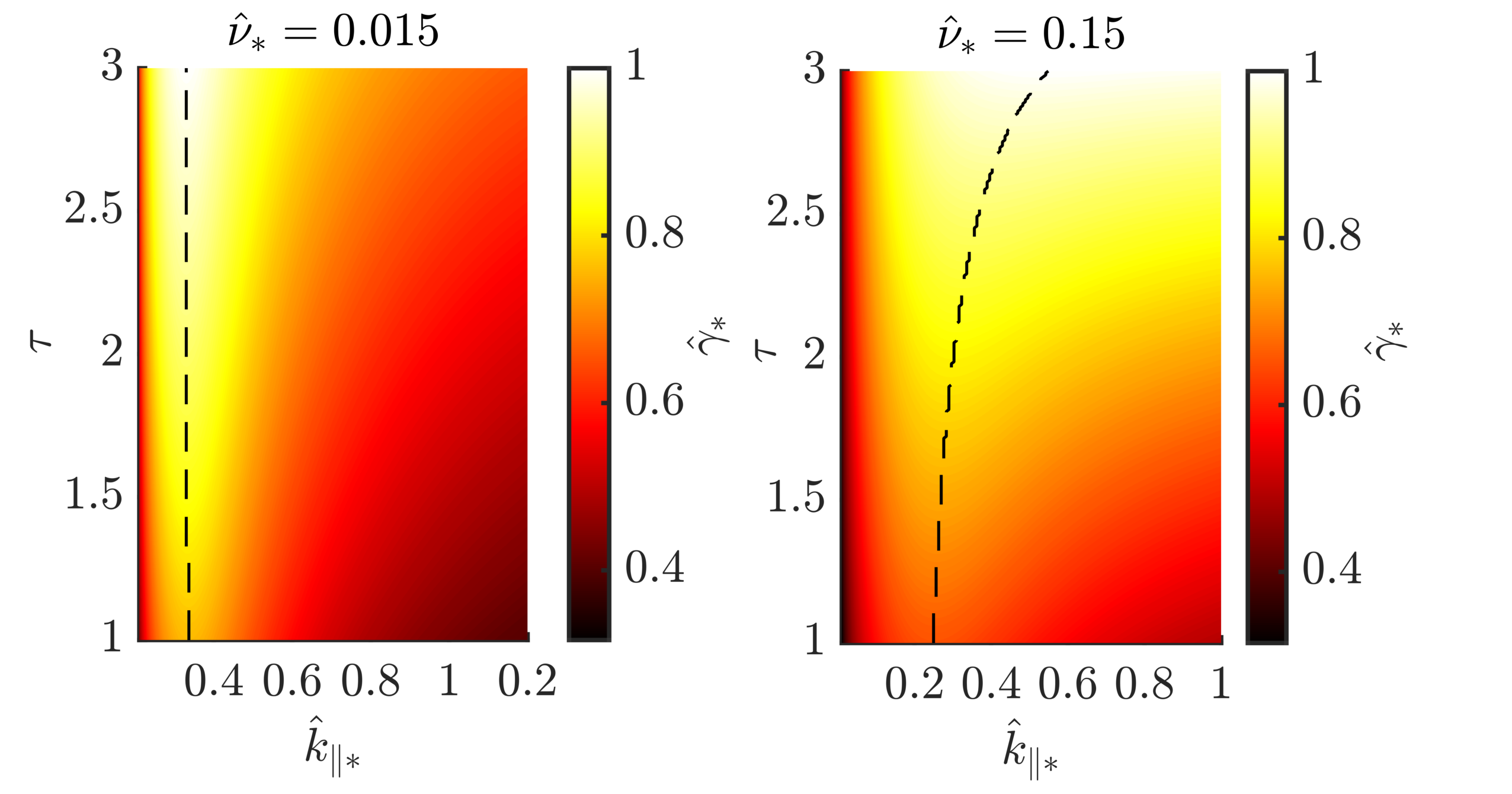}
    \caption{Estimates of the normalized slab ITG peak,  $\hat{\gamma}_*$, obtained from \cref{eq:slabITGmaxGamma}, for increasing values of $\hat{\nu}_* = 0.015$ (from left to right), as a function of $\hat{k}_{\parallel *}$ and temperature ratio $\tau$. The trajectory of the maximum growth rate is indicated by the dashed black line. The growth rates $\hat{\gamma}_*$ are normalized to their maximal values. }
    \label{fig:fig_ITG_peak_deltaT}
\end{figure}

We now consider the slab and toroidal limits of \cref{eq:cubicsITG} separately in details. An estimate the slab ITG peak can be derived from \cref{eq:cubicsITG} as a function of $k_\parallel$, $\nu$ and $\eta$. Besides neglecting the terms proportional to $\omega_B$ and FLR effects (i.e. imposing $\kernel{0} \simeq 1 $, $\kernel{1} \simeq 0$, $\kernel{2} \simeq 0 $, such that $\Gamma_\phi \simeq 1 $), we assume that the slab ITG mode is far from marginal stability limit $\eta \sim \nu \gg 1$. Then, the dispersion relation in \cref{eq:cubicsITG} reduces to
 
 \begin{align} \label{eq:slabITGmaxGamma}
     \hat{\omega}_*^4 + \frac{3}{2} i \alpha \hat{\nu}_*  \hat{\omega}_*^3 -   \hat{k}_{\parallel *}^2 \left( 1 + 3 \tau\right)  \hat{\omega}_*^2 +   \hat{k}_{\parallel *}^2 \left[ \tau - \frac{3}{2} i \alpha \hat{\nu}_*  \left( 1 + \frac{5}{2} \tau\right)\right]  \hat{\omega}_*  +  \frac{3}{2} i \alpha \hat{k}_{\parallel *} \hat{\nu}_* \tau =0.
 \end{align}
 \\
 where we introduce $ \hat{\omega}_*  = \omega / (\eta \omega_*)$,  $ \hat{\nu}_*  = \nu / (\eta \omega_*)$ and $\hat{k}_{\parallel *} = \hat{k}_\parallel / (\eta \omega_*)$. Given the normalization of \cref{eq:slabITGmaxGamma}, the maximum growth rate is expected to be proportional to $\gamma \simeq g(\tau, \hat{\nu}_*) \eta \omega_*$ and occur at $k_\parallel \simeq f(\tau, \hat{\nu}_*) \eta \omega_*$. In \cref{fig:fig_ITG_peak_deltaT}, we show the maximum ITG growth rate, $\hat{\gamma}_* = \gamma / (\eta \omega_*)$ (i.e., the imaginary part of $ \hat{\omega}_*$) obtained by solving \cref{eq:slabITGmaxGamma}, as a function of  $\hat{k}_{\parallel *} $ and $ \hat{\nu}_*$. The effects of increasing $\hat{\nu}_*$ is primarily to shift the growth rate peak to larger values of $\hat{k}_\parallel^*$ for all values of $\tau$. Additionally, while $g(\tau,\nu)$ increases with $\tau$ for all $\nu$, it is found that $f(\tau, \hat{\nu}_*)$ is an increasing function of $\tau$ at large values of $\hat{\nu}_*$.

 We remark that the collisionality dependence of the functions $f$ and $g$ is not present in previous fluid ITG models based on the drift-reduced fluid equations \citep{Mosetto2015}. This dependence arises here because the $4$GM model (and more generally the gyro-moment hierarchy equation in \cref{eq:dNipjdt}) allows for a different evolution of the perpendicular and parallel temperature fluctuations assuming, in general, $T_\parallel \neq T_\perp$. On the other hand, the drift-reduced Braginskii fluid model in, e.g., \citet{Zeiler1997} assumes that $T_\parallel = T_\perp$. In fact, the $\nu$ dependence in \cref{eq:slabITGmaxGamma} enters in the dispersion relation, given in \cref{eq:slabITGmaxGamma}, through terms proportional to $\nu \alpha(T_\parallel - T_\perp)$ which arise from the gyro-moment $\C^{20}$ and $\C^{01}$ components of the DK Coulomb collision operator.
 
To investigate the effect of anistotropic temperature fluctuations, we derive the evolution equation of the temperature, $T = (T_\parallel + 2 T_\perp) /3 - N$. Using \cref{eq:eqTparallel,eq:eqTperp} and the continuity equation \cref{eq:eqN}, this can be expressed as

 \begin{align} \label{eq:eqT}
     \frac{\partial }{\partial t} T - i k_\parallel \frac{ \sqrt{2 \tau}}{3}V  + \frac{i}{3} \left( \Gamma_2 - 2 \Gamma_3 - 2 \Gamma_1\right) \phi =0.
 \end{align}
 \\
 From \cref{eq:eqT} with \cref{eq:eqN} and \cref{eq:eqV}, and considering the limits used to derive \cref{eq:slabITGmaxGamma}, we obtain the estimate of the maximum sITG growth rate,
 
 \begin{align} \label{eq:reducedsITG}
 \hat{\omega}_*^3  -  \hat{k}_{\parallel *}^2 \left(   1   + \frac{2}{3}\tau  \right)  \hat{\omega}_* +  \hat{k}_{\parallel*}^2 \tau     = 0.
\end{align}
\\
From \cref{eq:reducedsITG}, one finds that the peak of the sITG growth rate is proportional to $\gamma \simeq g(\tau) \eta \omega_*$ and occurs at $k_{\parallel} \simeq f(\tau) \eta \omega_*$. It is found that $g(\tau)$ is an increasing function of $\tau$, while $f(\tau) $ decreases with $\tau$. We remark that \cref{eq:reducedsITG} agrees with \citet{Mosetto2015} by replacing the numerical factor $2/3$ by $5/3$, a difference due to the additional degree of freedom assumed in the $4$GM model.



\begin{figure}
    \centering
    \includegraphics[scale = 0.55]{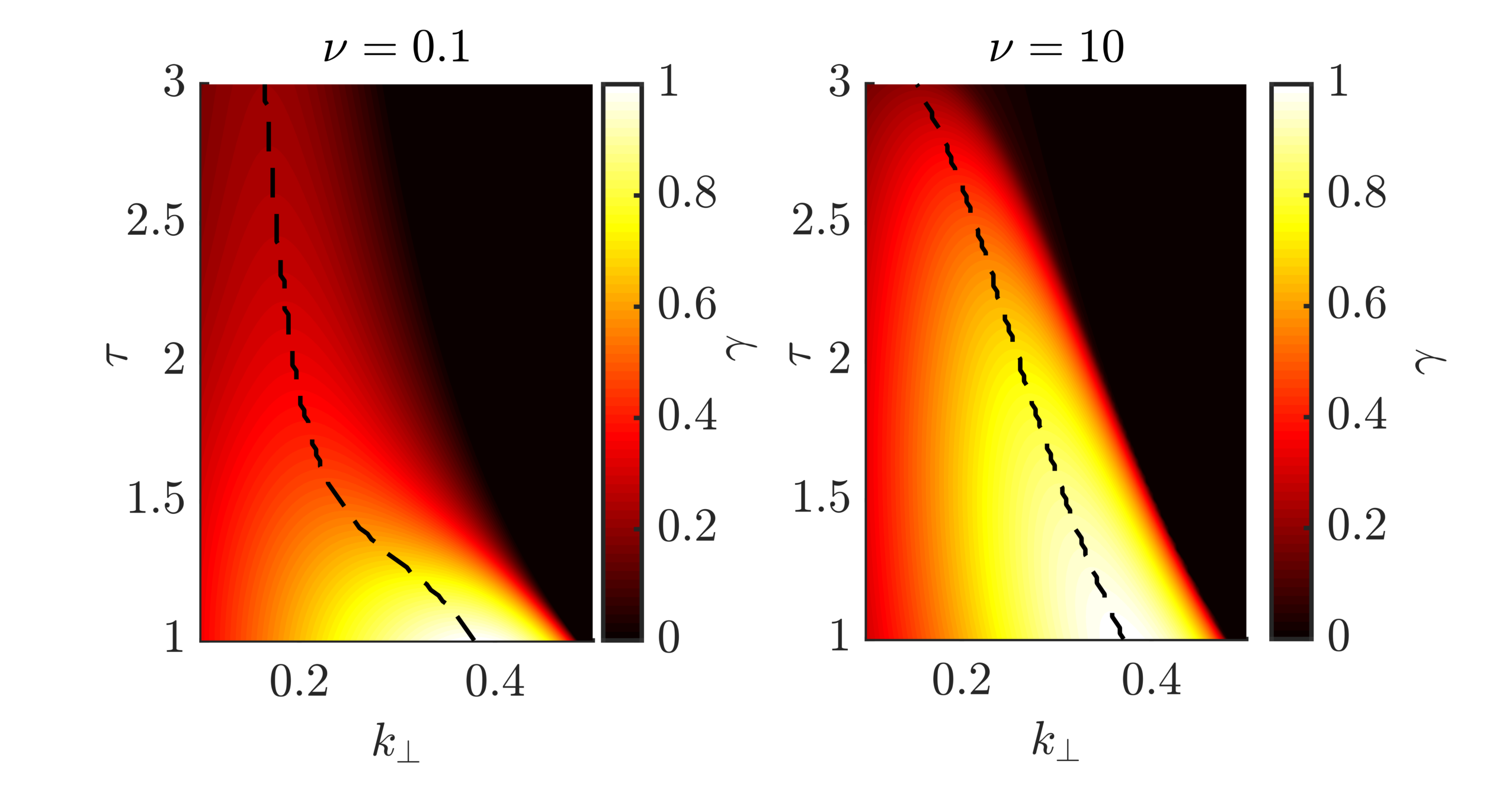}
    \caption{Toroidal ITG growth rate as a function of the perpendicular wavenumber, $k_\perp$ and temperature ratio $\tau$ obtained from \cref{eq:cubictITG}, for a low (left) and a high (right) value of collisionality. The maximum growth rate is indicated by the dashed black line. The growth rates are normalized to the maximum value. Here, $\eta = 7$ and $R_B = 0.1$.}
    \label{fig:fig_tITG_tau_vs_kperp}
\end{figure}

We now turn to the toroidal ITG mode, driven unstable by the magnetic drifts and by the presence of a finite temperature gradient. We derive a simple dispersion relation for the toroidal ITG from \cref{eq:cubicsITG} indicative of its properties. Since, in contrast to the sITG, the toroidal ITG subsists at $k_\parallel = 0$, we set $k_\parallel =0$ removing the coupling with sound waves, and adopt the long perpendicular wavelength limit, keeping terms up to  $O(k_\perp^2)$, such that $\kernel{0} \simeq 1  - \tau k_\perp^2/2 $, $\kernel{1} \simeq \tau k_\perp ^2/2$, and $\kernel{2} \simeq 0$. Since $\omega_B \ll \omega_* $ in the boundary region, we obtain

   \begin{align} \label{eq:cubictITG}
\omega^3  + b_t \omega^2 + c_t \omega + d_t   = 0,
 \end{align}
\\
where 

\begin{subequations} \label{eq:btcddt}
\begin{align}
b_t & = \frac{3}{2} i \alpha \nu + \omega_* - 2 \omega_B \left( 1 + 8 \tau \right) + k_\perp^2 \left[ 2 \omega_B \left( 1 + \frac{3}{2} \tau  \right)  - \omega_* \left( 1 + \tau + \eta \tau \right) \right],  \label{eq:bt}\\
c_t & = \frac{3}{2 } i \alpha \nu \left[ \omega_* - 2 \omega_B - \frac{32}{3} \tau \omega_B + k_\perp^2 \left(  2 \omega_B \left( 1 + \frac{19}{12} \tau \right)  -\omega_* (1 + \tau + \tau \eta)  \right) \right] \nonumber \\
& + \omega_B \omega_* \left[ 2   \tau  ( \eta -7)  + k_\perp^2 \left(10 \eta  \tau \left(\tau - \frac{1}{5}\right) 
   +\tau\left(13 \tau  +14\right) \right) \right], \\
   d_t & =i \alpha \nu  \omega _{B } \omega _* \tau \left[  \left(3  \eta  -13  \right) + k_\perp^2  \left(3 \eta \left(2   \tau-1 \right)
     +11  \tau +13   \right)\right].
\end{align}
\end{subequations}
\\
An estimate of the value of $k_\perp$ that yields the largest growing mode can be derived from \cref{eq:cubictITG} by neglecting terms proportional to $\nu$, for simplicity. Then, the dispersion relation in \cref{{eq:cubictITG}} becomes a second order equation for $\omega$ presenting the largest solution $\omega$ when the coefficient $b_t$ is minimized. Hence, by minimizing $b_t$ and considering $\omega_B  \ll \omega_*$, the growth rate is maximized when $k_\perp^2 \simeq 1 / [ 1 + \tau (1 + \eta)]$. \Cref{fig:fig_tITG_tau_vs_kperp} shows the toroidal ITG growth rate obtained by solving \cref{eq:cubictITG}, as a function of the perpendicular wavenumber $k_\perp$ and temperature ratio $\tau$. We observe that the decrease of the perpendicular wavenumber of the fastest growing mode with $\tau$ is present also at finite collisionality, $\nu$. We  also remark that collisions tend to destabilize long wavelength modes.

\section{ITG Mode with GK Coulomb Collision Operator}
\label{sec:ITGwithGKCoulomb}
\begin{figure}
    \centering
    \includegraphics[scale = 0.65]{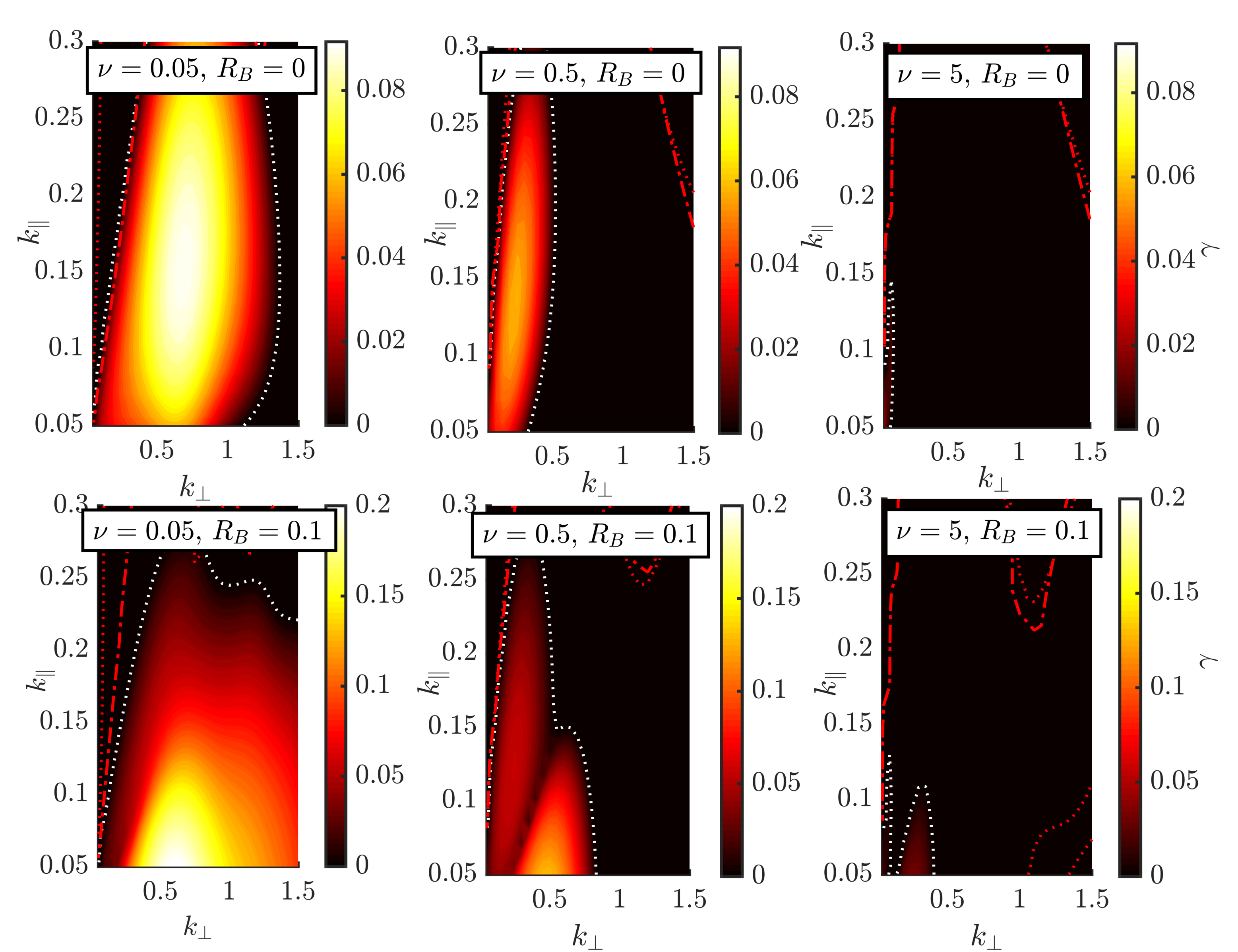}
    \caption{ITG linear growth rate, $\gamma$, in the $(k_\perp, k_\parallel)$ parameter space, for increasing collisionality, $\nu$ (from left to right) and magnetic gradient strength $R_B$ (from top to bottom) at a temperature gradient strength of $\eta =3$. The stability boundary of the gyro-moment hierarchy (dotted white line) and $6$GM and $4$GM (dotted and dot-dashed lines, respectively) are shown for comparison. Here, we fix $(P,J) = (18,6)$.}
    \label{fig:Fig_kperpVSkpars_GKCoulomb_RTi_3}
\end{figure}

We now perform parameter scans to explore the dependence of the ITG linear growth rate on the perpendicular and parallel wavenumbers, on the level of collisionality and on the temperature and magnetic gradient strengths. Collisional effects are modelled with the GK Coulomb collision operator \citep{frei2021}. As reference value, we consider $k_\parallel \sim 0.1 $. This is because of the strong ballooning character of the ITG mode in boundary region such that $ k_\parallel$ is of the order of $ L_N / q R_0$, and typical H-mode edge values are $q \sim 4$ and $L_N / R_0  \lesssim 1/50$. In addition, we note that $ 1 \lesssim \tau \lesssim 4 $ in the boundary. Therefore, we consider a temperature ratio of $\tau = 1$ for simplicity, and note that the effect of increasing $\tau$ can be inferred from \cref{fig:fig_ITG_peak_deltaT,fig:fig_tITG_tau_vs_kperp} for the slab and toroidal cases, respectively. For the present numerical calculations, we fix $(P,J) = (18,6)$. As we show in \cref{sec:convergence}, this choice ensures good convergence of the numerical results in the range of parameters explored in this Section.

\Cref{fig:Fig_kperpVSkpars_GKCoulomb_RTi_3} displays the ITG growth rate in the $(k_\parallel,k_\perp)$ parameter space when $\eta=3$ for increasing collisionality $\nu$ (from left to right) and magnetic gradient $R_B$ (slab at the top and toroidal at bottom). For comparison purposes, we show the stability boundaries obtained numerically from the gyro-moment hierarchy  (dotted white) and from the $6$GM and $4$GM reduced models (red dotted and dashed-dotted). First, we observe that, as collisionality increases, FLR collisional stabilization effects become important and eventually suppress the ITG modes when $k_\perp \gtrsim 1$ and $\nu \gtrsim 0.5$. In fact, we notice that, at intermediate and high collisionalities, the gyro-moment hierarchy stability boundary follows closely the ones predicted by the $6$GM and $4$GM models for long parallel ($k_\parallel \lesssim 0.15$) and perpendicular ($k_\perp \lesssim 0.2$) wavelengths where the FLR collisional effects are negligible and kinetic effects are small. However, while the ITG mode is completely suppressed for $k_\perp \gtrsim 0.5$ when $\nu \gtrsim 0.5$ according to the full gyro-moment calculation, the $6$GM and $4$GM stability boundaries extend to $k_\perp \gtrsim 1$. This highlights the importance of FLR terms in the Coulomb collision operator model since the $6$GM and $4$GM express the $\nu \gg 1$ limits of the gyro-moment hierarchy when collisions are modelled with the DK Coulomb (see \cref{subsec:SWITG}), i.e. neglecting FLR terms in the collision operator. Second, at a given collisionality, the toroidal ITG mode extends to shorter perpendicular wavelength with respect to the slab case. At the same time, the peak of the growth rate of the toroidal ITG, which occurs near $k_\perp \sim 0.5$, is reported at longer wavelength in the parallel direction, compared to the slab branch, consistently with the fact that the toroidal mode subsists in the $k_\parallel = 0$ limit. In all cases, the stabilization of the growth rate at large $k_\parallel$ is caused by Landau damping. Third, the peak growth rate of the toroidal ITG is about twice as large as the slab one at all collisionalities. 

    \begin{figure}
        \centering
        \includegraphics[scale =0.55]{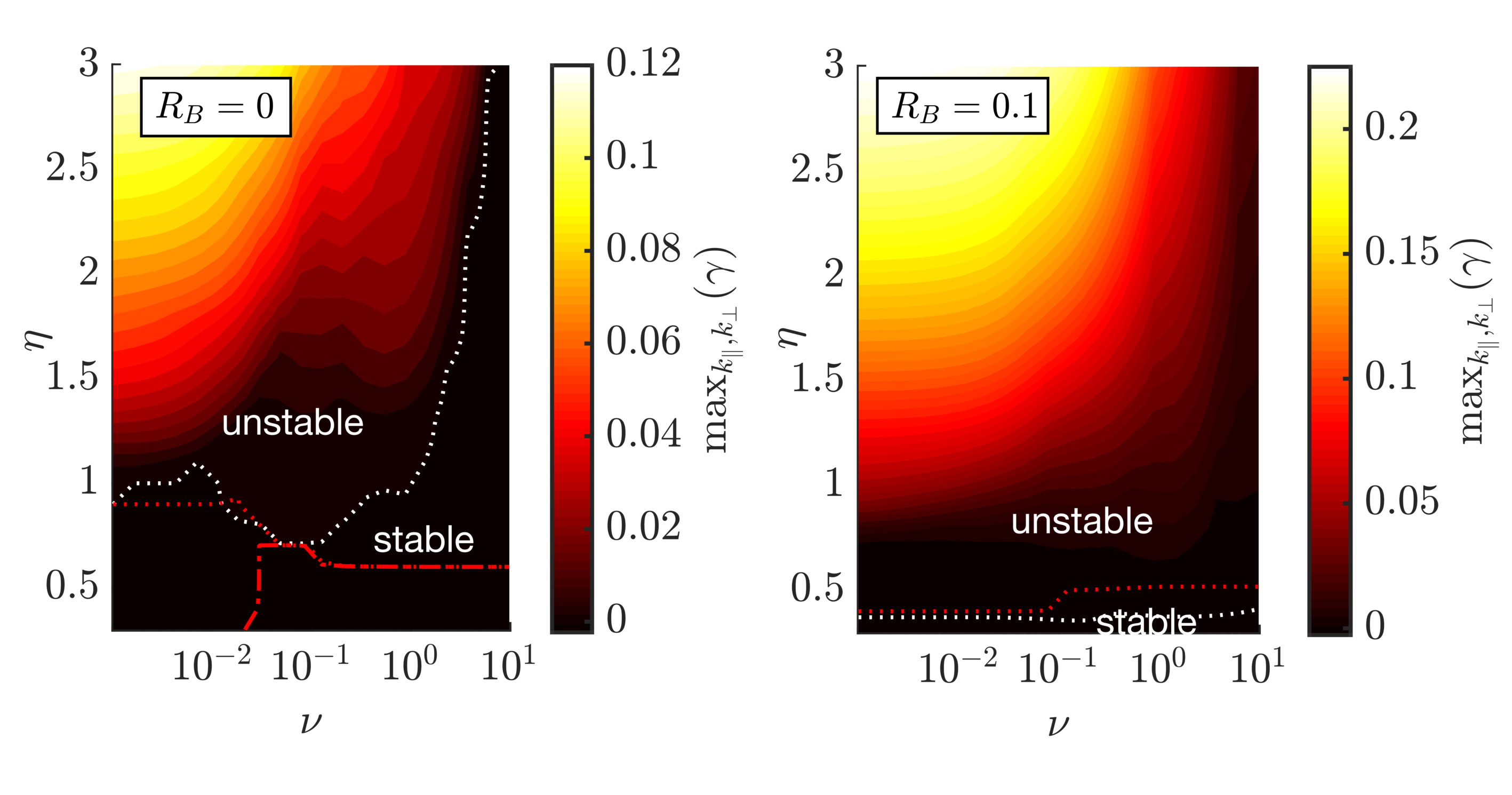}
        \caption{ITG growth rate  maximized over $k_\parallel$ and $k_\perp$, denoted by $\max_{k_\parallel, k_\perp}(\gamma)$, a a function of the collisionality $\nu$ and normalized temperature gradient $\eta$ for the slab (left) and toroidal (right) ITG modes. The stability boundaries predicted by the gyro-moment approach using the GK Coulomb collision operator (which separates the unstable and stable modes) and the $6$GM and $4$GM models are plotted by the white dotted, red dotted and red dotted-dashed lines, respectively. The collisionless temperature gradient critical value is retrieved when $\nu \ll 1$. }
        \label{fig:fig_etathreshold_GKCoulomb}
    \end{figure}
    
We now assess the effects of collisions on the critical value of the temperature gradient above which the ITG develops. For this purpose, we compute the ITG growth rate, maximized over $k_\parallel$ and $k_\perp$, as a function of $\nu$ and $\eta$. This allows us to evaluate the stability boundary shown in \cref{fig:fig_etathreshold_GKCoulomb} for the slab (left) and toroidal (right) ITG modes and compared with the predictions of the $6$GM and $4$GM models. The slab and toroidal ITG mode are destabilized by increasing $\eta$ at all collisionality and are damped (and/or suppressed) by FLR collisional effects. We remark that the gyro-moment stability boundary matches the collisionless one when $\nu \ll 1$. The sITG stability boundary is highly sensitive to the level of collisionality, in particular for $\nu \gtrsim 0.1$. In fact, the critical temperature gradient for the sITG mode onset is increased (from $\eta \simeq 1$ when $\nu \lesssim 1$ to $\eta \gtrsim 3$ when $\nu \gtrsim 1$ with an abrupt transition) in order to overcome the strong FLR collisional stabilization effects that suppress all unstable modes. On the other hand, the toroidal ITG mode stability boundary is not significantly affected by collisions and remains near the collisionless threshold. The reason of this difference can be inferred from \cref{fig:Fig_kperpVSkpars_GKCoulomb_RTi_3} which shows a peak near $k_\perp \simeq 0.3$ when $\nu = 5$ in the toroidal case (bottom row), while the mode is completely suppressed in the slab case (top row). Finally, we notice that the $6$GM and $4$GM stability boundaries converge to a constant value near $\eta \simeq 0.5$ in the high collisional limit. At low collisionality, $\nu \lesssim 0.1$, the ITG mode is completely suppressed according to the $4$GM model because of kinetic effects associated with the parallel streaming.


\section{Comparisons Between Collision Operator Models}
\label{sec:comparisonsbetweencollisionoperatormodels}

In this section, we investigate the differences between the collision operator models introduced in \cref{sec:CollisionOperatorModels}. In particular, we compare them as a function of collisionality and perpendicular wavenumber and measure their relative deviation with respect to the GK Coulomb collision operator in \cref{subsec:comparisonsModelsandGKCoulomb}. We consider different collisionality regimes and temperature gradient strengths. We demonstrate that energy diffusion and FLR effects are important in the collisional damping of ITG modes, and find that the deviations, when compared to the GK Coulomb collision operator, decrease with temperature gradient strength at all collisionalities. One of the main result of this comparison is that the smallest relative deviation ($\lesssim 15 \%$) is produced by the GK Sugama operator, while the largest ($\lesssim 50 \%$) is obtained from the GK Dougherty operator. The importance of FLR effects is highlighted in \cref{subsec:SWITG} by investigating a short wavelength ITG branch (SWITG) typically peaking near $k_\perp \sim 1.5$ that can be destabilized if collisional FLR terms are neglected. The present results are particular important since DK collision operators are often used in the continuum GK codes to simulate the plasma boundary. In this section, we truncate the gyro-moment hierarchy, \cref{eq:dNipjdt}, at $(P,J) = (18,6)$ for all cases. We have checked that convergence is achieved in all cases, in agreement with the convergence studies performed in \cref{sec:convergence}.

\subsection{Comparisons between the GK Coulomb and Collision Operators Models}
\label{subsec:comparisonsModelsandGKCoulomb}

We focus on the differences between the GK Coulomb collision operator, the GK/DK Sugama collision operator \citep{Sugama2009}, the GK/DK momentum-conserving pitch-angle scattering (pitch) operator \citep{Helander2002}, the zeroth-order DK Hirshman-Sigmar-Clarke (HSC) collision operator \citep{Hirshman1976} and the GK/DK Dougherty collision operator \citep{Dougherty1964}. The gyro-moment expansion of the Sugama collsion operator are detailed in \citet{frei2021}, while the gyro-moment expansion of the other collision operators are reported in \cref{sec:pitchangle,sec:HSC,sec:dougherty}.

\begin{figure}
    \centering
    \includegraphics[scale=0.5]{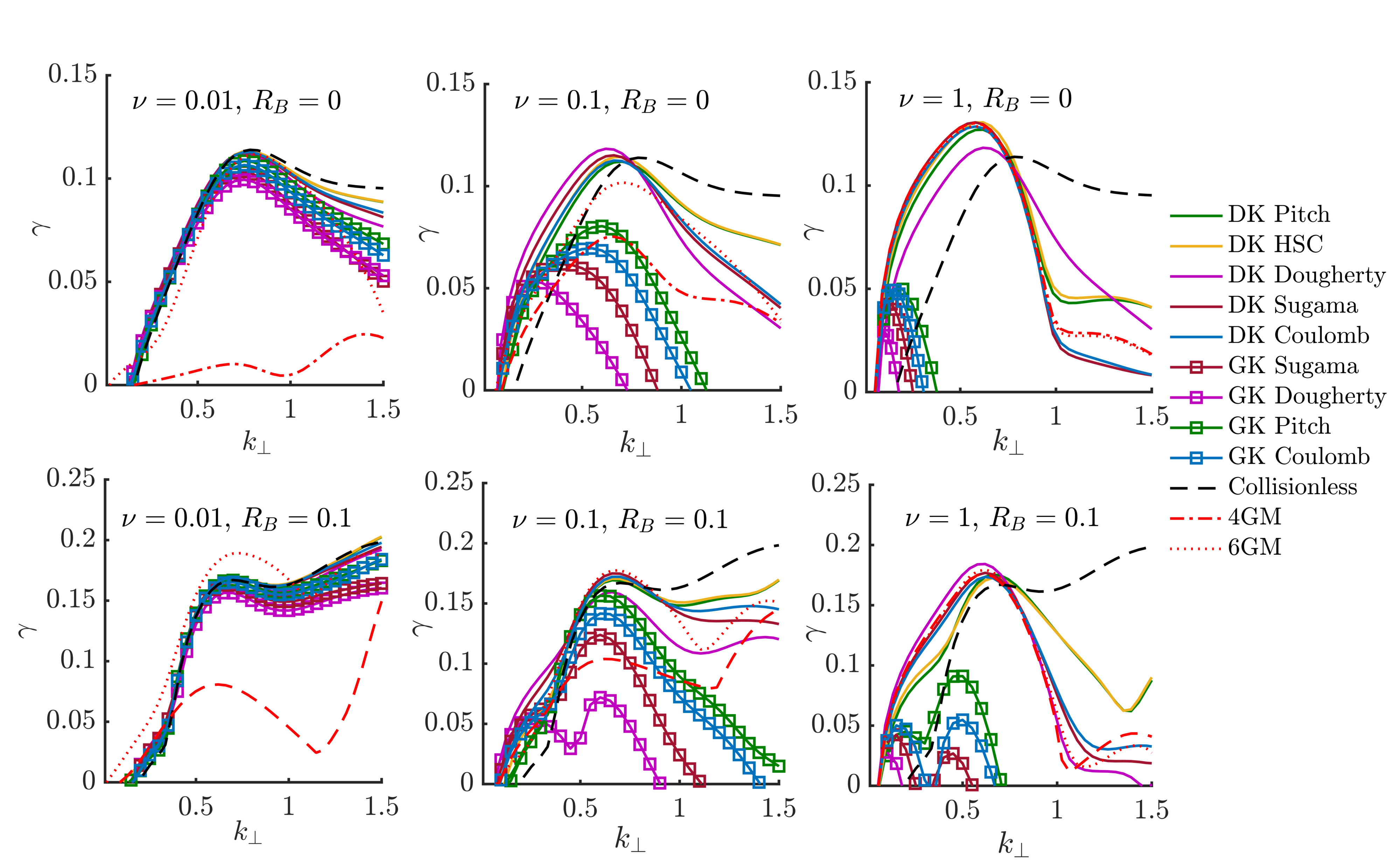}
    \caption{Comparisons between DK (solid lines) and GK (square markers) collision operator models in the case of the slab (top) and toroidal (bottom) ITG growth rate as a function of the perpendicular wavenumber $k_\perp$, from the low (left) to high (right) collisionality regime. The collisionless (dashed black) and $6$GM and $4$GM (red dotted and dashed-dotted lines, respectively) growth rates are plotted for comparisons. Here, $\eta = 3$, $k_\parallel  = 0.1$.}       
      \label{fig:Fig_ITG_kperpscan_collision_ops_eta_3}
      \end{figure}

\Cref{fig:Fig_ITG_kperpscan_collision_ops_eta_3} shows the slab (top) and the toroidal (bottom) ITG linear growth rate at a relative temperature gradient strength of $\eta =3$ from low (left) to high (right) collisionality, as a function of the perpendicular wavenumber $k_\perp$. The DK collision operators are displayed by the colored solid lines, while their GK operators by the colored square markers. The collisionless and high-collisional limits are plotted for comparison. Is is remarkable that the ITG growth rate is sensitive to both the collision operator models and to the presence of FLR related terms. The GK Dougherty collision operator produces the results that most differ with respect to the other GK operator models, and the GK pitch-angle operator yields a systematic overestimation of the ITG growth rate, an effect of the absence of energy diffusion in the latter operator. On the other hand, the DK collision operators lead to a larger ITG growth rate than the GK operators, and can eventually destabilize a SWITG peaking near $k_\perp \simeq 1.5$. The presence of this SWITG can be inferred in \cref{fig:Fig_ITG_kperpscan_collision_ops_eta_3} from the increase of the ITG growth rate occurring for $k_\perp \gtrsim 1$ when $\nu \gtrsim 0.1$ (see \cref{subsec:SWITG}). This behaviour is particularly pronounced in the pitch-angle and HSC collision operators at short wavelengths, and it is related to the absence of energy diffusion in these operators \citep{Barnes2009,Ricci2006}, which is retained in the Coulomb, Sugama, and Dougherty collision operators. We also notice that the absence of energy diffusion in the pitch-angle and HSC collision operators affects the high collisional limit. While the deviations between the DK Coulomb, Sugama, and Dougherty collision operators are reduced as the collisionality increases and, ultimately, becomes negligible when $\nu \gg 1$, converging to the $6$GM and $4$GM solutions, a difference subsists in the pitch-angle and HSC collision operators. Finally, among the DK collision operators considered, the DK Sugama approaches better the DK Coulomb collision operator.

\begin{figure}
    \centering
    \includegraphics[scale = 0.55]{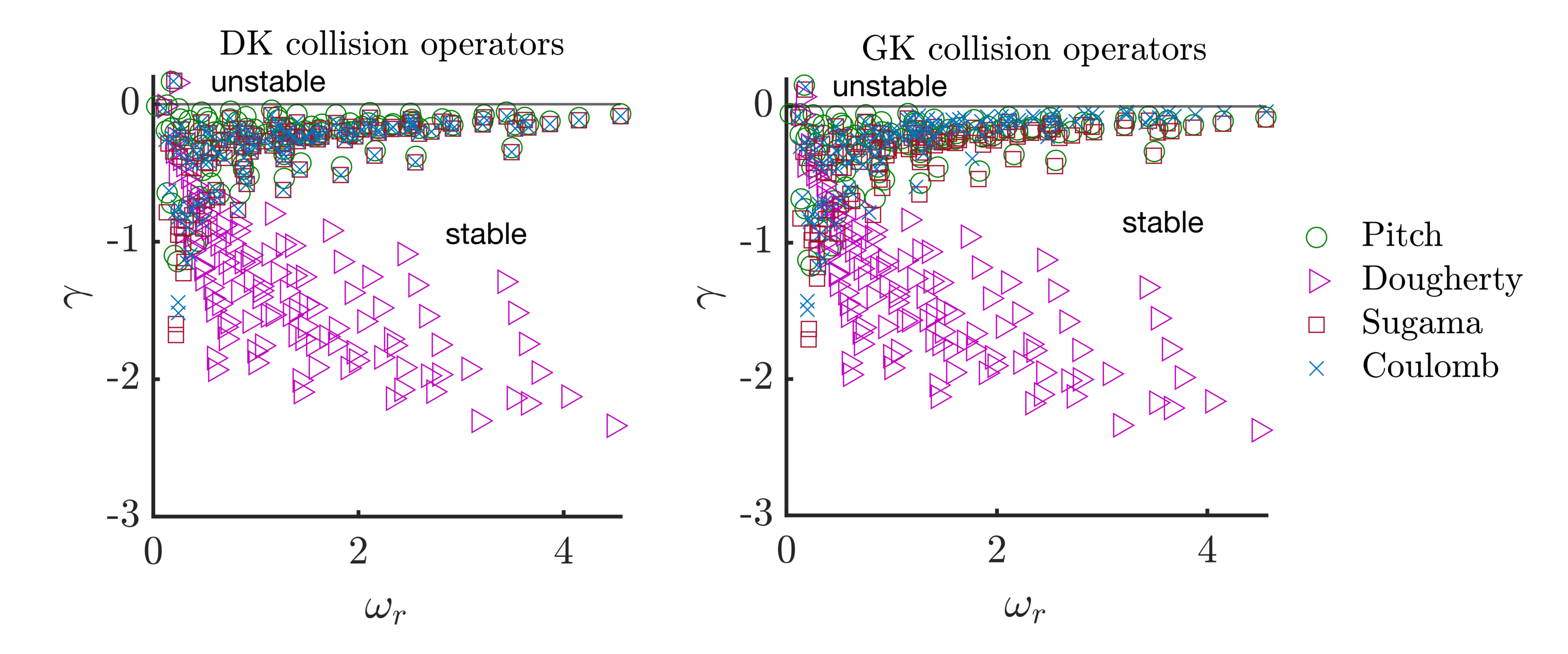}
    \caption{Toroidal ITG eigenvalue spectrum when $\nu = 0.1$ and $k_\perp = 0.6$ using the DK (left) and GK (right) collision operator models. Here, $R_B = 0.1$ and $\eta = 3$.}
    \label{fig:fig_eig_spectrum}
\end{figure}

\begin{figure}
\centering
\includegraphics[scale=0.45]{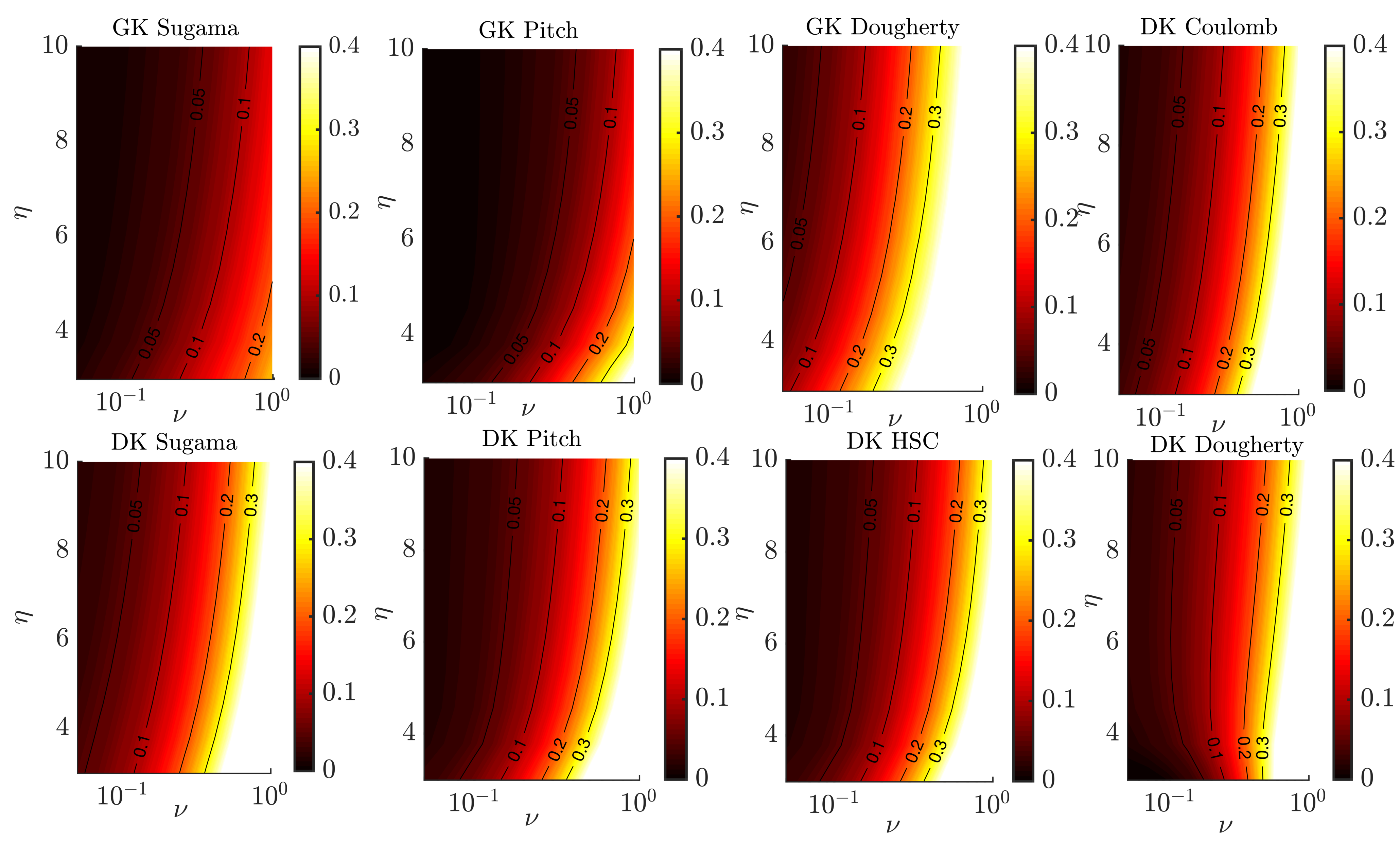}
\caption{Relative deviations in the growth rate obtained using the collision operator models, used in \cref{fig:Fig_ITG_kperpscan_collision_ops_eta_3}, with respect to the GK Coulomb results, as a function of collisionality $\nu$ and temperature gradient strength $\eta$. The colorbars are saturated at a maximal relative deviation of $0.4$. Here, $k_\parallel = 0.1$, $k_\perp = 0.5$ with $R_B = 0.1$ are considered, corresponding approximately to the peak ITG growth rate.}
\label{fig:Fig_diff_ITG}
\end{figure}

The deviations of the collision operator models can be quantified by computing their relative difference, $|\gamma_C - \gamma| / \gamma_C$ where $\gamma_C$ is the growth rate using the GK Coulomb and $\gamma$ is the growth rate obtained using the other collision operator models. The results are shown in \cref{fig:Fig_diff_ITG} where we plot  $|\gamma_C - \gamma| / \gamma_C$ as a function of collisionality, $\nu$, and of temperature gradient strength, $\eta$, for the toroidal branch ($R_B = 0.1$). The GK Sugama collision operator provides the results that best approach the GK Coulomb collision operator. Among the other GK operators considered, the largest deviation is produced by the GK Dougherty where the relative difference is larger than $30 \%$ when $\nu \gtrsim 0.5$. Surprisingly, while the GK pitch-angle operator systematically overestimates the ITG growth rate, it yields an error smaller than $ 30 \% $ in all the parameter space considered. In particular, the relative difference between the GK Coulomb and the GK pitch-angle collision operators is similar to the one of GK Sugama at low $\eta$ and high collisionality. For all operators, this deviation decreases at steeper temperature gradients while it increases with collisionality.

The gyro-moment approach allows the investigation of ITG eigenvalue spectrum at finite collisionality. An example is shown in \cref{fig:fig_eig_spectrum} for the toroidal case when $\nu = 0.1$ using the DK (left) and GK (right) pitch-angle, Dougherty, Sugama and Coulomb collision operators near to the peak growth rate occurring at $k_\perp = 0.6$. First, focusing on the spectrum of the DK collision operators, we note that, while the DK Sugama reproduces qualitatively well the eigenvalues of the DK Coulomb, the DK Dougherty collision operator displays an eigenvalue spectrum with strongly damped subdominant modes characterized by $\gamma < 0$. Similar differences between the DK Dougherty and the DK Coulomb collision operators in the eigenvalue spectrum has been reported in the study of electron plasma waves at arbitrary collisionality \citep{jorge2019linear}. In addition, despite the absence of energy diffusion, the DK pitch-angle operator features similar eigenvalues than the DK Coulomb. Focusing now on the spectrum of the GK collision operators, we remark that a large difference exists between the GK Dougherty operator and the other GK collision models. It is noticeable that the GK Coulomb collision operator produces subdominant modes that are less damped than the ones predicted by the GK Sugama and pitch-angle operators but, nevertheless, yield a similar eigenvalue spectrum. The differences observed in the ITG spectrum shown in \cref{fig:fig_eig_spectrum} can potentially yield large deviations in nonlinear ITG simulations where subdominant mode can interact nonlinearly. We remark that the eigenvalues have all a positive real frequency, $\omega_r > 0 $, corresponding to the ion diamagnetic direction.

\subsection{Destabilization of the Short Wavelength ITG Mode}
\label{subsec:SWITG}

 \begin{figure}
    \centering
    \includegraphics[scale=0.55]{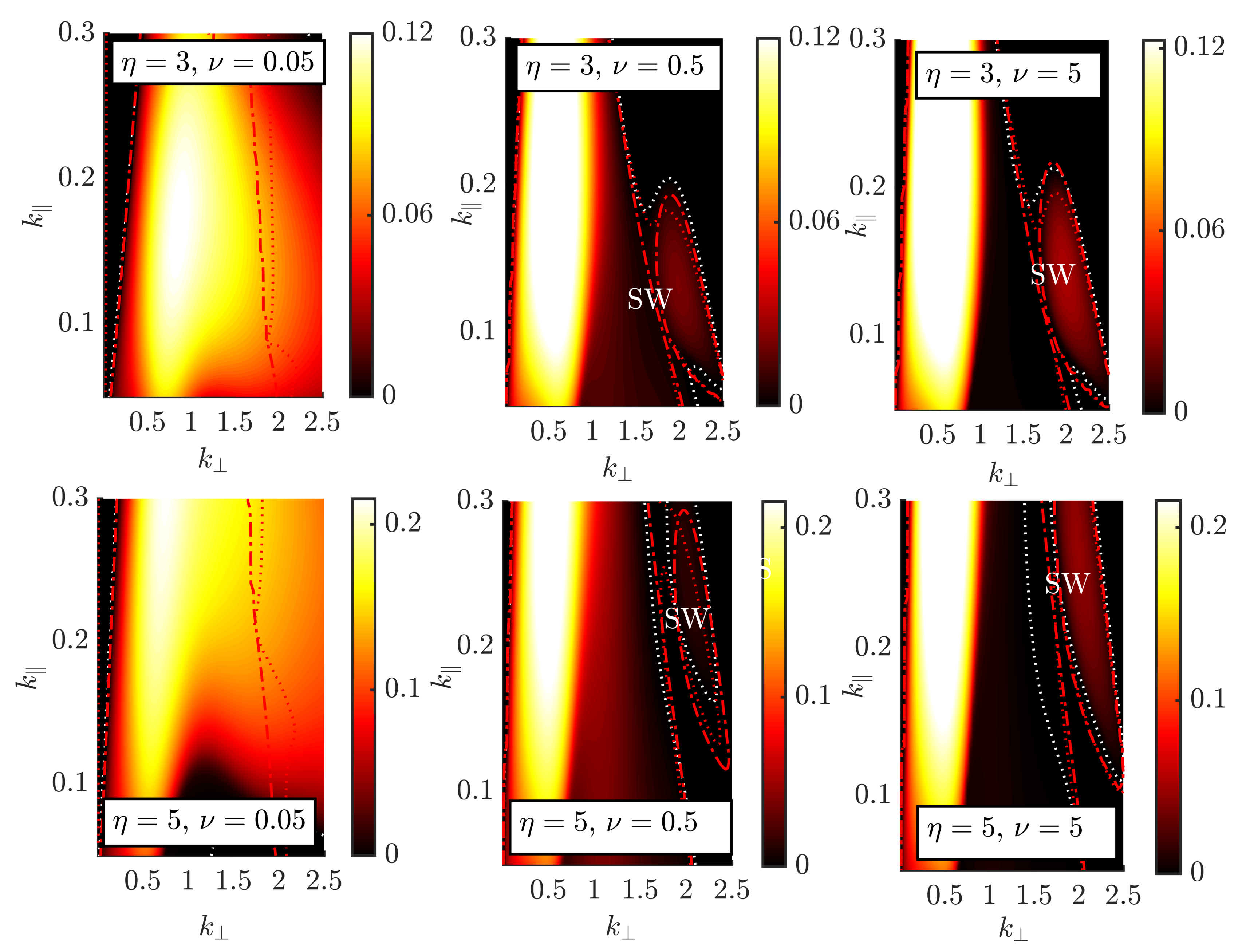}
    \caption{Slab SWITG branch driven unstable by the DK Coulomb model when $\eta =3$ (top) and $\eta=5$ (bottom). The stability boundaries of the gyro-moment hierarchy (dotted white line), $6$GM and $4$GM (red dotted and dot-dashed lines, respectively) are plotted for comparison. The main ITG branch is identified near $k_\perp \simeq 0.5$ and the SWITG near $k_\perp \simeq 2$, appearing as the collisionality increases when $\nu \gtrsim 0.5$. The colobar is saturated at the maximum of $\gamma$ when $\nu = 0.05$.}
    \label{fig:Fig_SWsITG_kperpVSkpar}
\end{figure}

An initial analysis of the effects of FLR terms in the collision operators can be carried out by observing, first, the discrepancies between the GK Coulomb and the $6$GM and $4$GM stability boundaries in the $\nu =1$ case (see \cref{fig:Fig_kperpVSkpars_GKCoulomb_RTi_3}) and, second, from the comparison between GK and DK collision operators (see \cref{fig:Fig_ITG_kperpscan_collision_ops_eta_3}). A clear deviation between GK and DK Coulomb operators occurs at sufficiently steep temperature gradient, $\eta \gtrsim 3$, and large collisionality, $\nu \gtrsim 0.1$. For a more detailed analysis, we present in \cref{fig:Fig_SWsITG_kperpVSkpar} the ITG growth rate evaluated with the DK Coulomb in the slab case for two different values of temperature gradient strength.

The comparison between \cref{fig:Fig_kperpVSkpars_GKCoulomb_RTi_3} and \cref{fig:Fig_SWsITG_kperpVSkpar} reveal that a short wavelength branch of the ITG, referred to as the SWITG mode and typically peaking at $k_\perp \sim 2$, is present when $\nu \gtrsim 0.5$ if the DK Coulomb operator is used. We remark that the SWITG is also captured by the $6$GM and $4$GM models. This mode has been identified in the collisionless limit as a continuous extension of the main branch of the ITG mode in the $k_\perp > 1$ region. In fact, the collisionless ITG growth rate reveals a typical "double-humped" behaviour when plotted as a function of $k_\perp$ \citep{Smolyakov2002,Gao2003,Gao2005}. Additionally, \cref{fig:Fig_SWsITG_kperpVSkpar} shows the presence of a stability region that increases with collisionality, isolating the conventional ITG and the SWITG modes. We note that, in the toroidal case, the SWITG is stabilized (or even suppressed) by increasing magnetic gradients, a similar feature observed also in the collisionless limit \citep{Gao2005}.

The SWITG mode is also predicted by the other DK collision operators considered in this work, while it is suppressed by all the other GK collision operators. This questions the applicably of DK collision operator models for the prediction of ITG driven turbulence level in edge conditions, since the SWITG can produce significant, yet spurious, level of transport in addition to the main ITG mode peaking at larger $k_\perp$ \citep{Gao2005}. However, further investigations are required to go beyond the local approximation used in the present work.

\section{Gyro-Moment Spectrum}
\label{sec:GMspectrum}

Using the collisionless results obtained in \cref{sec:CollisionlessHermiteLaguerreTheory}, we now study the structure of the ITG gyro-moment spectrum. In particular, we compute numerically the collisionless gyro-moment spectrum using the semi-analytical expressions obtained in \cref{subsec:collisonlessgyromomentresponse} for the slab and toroidal branches. We find that magnetic drift resonance effects broaden the collisionless spectrum compared to the slab case. The semi-analytical expression of the collisionless gyro-moment, given in \cref{eq:tNapj}, allows us to evaluate numerically the collisionless gyro-moment spectrum of the normalized ion perturbed distribution function, $\g / \phi$. We first focus on the slab limit case and, then, on the purely toroidal ($k_\parallel = 0$) case. Finally, we consider the effect of collisions by deriving an approximate expression of the cut-off of the gyro-moment spectrum.

\begin{figure}
    \centering
    \includegraphics[scale = 0.6]{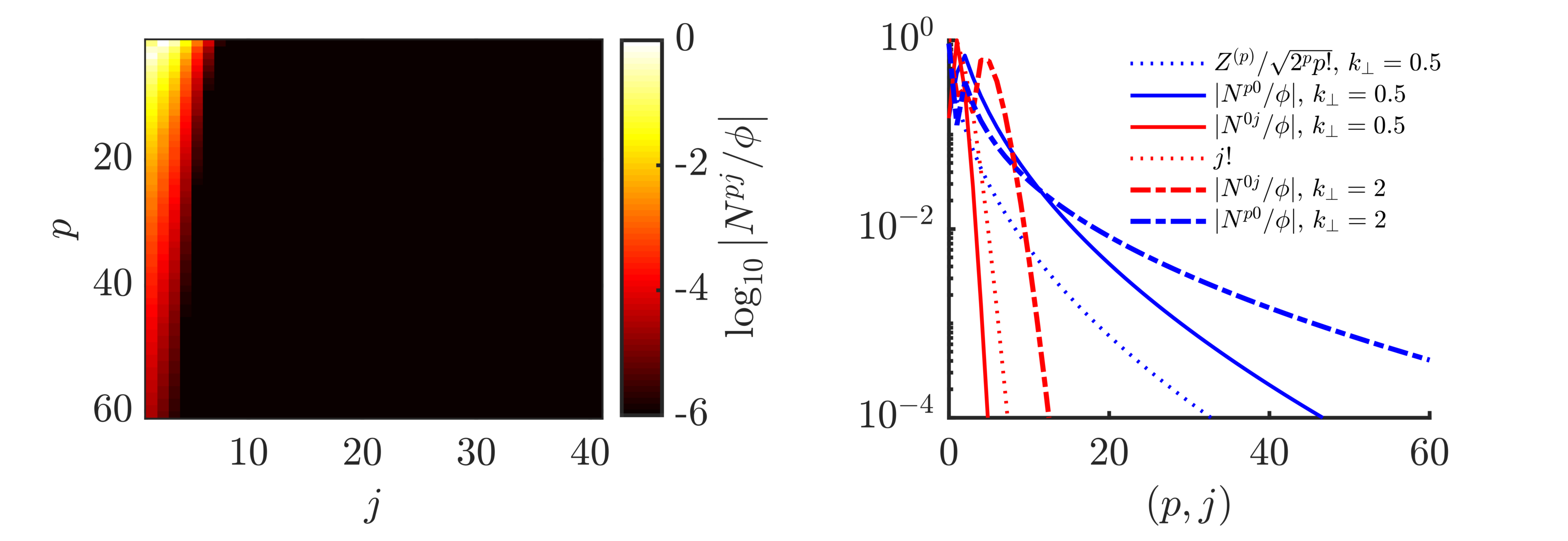}
    \caption{The modulus of the normalized collisionless gyro-moments spectrum, $N^{pj} / \phi$, in the case of slab ITG, as a function of $p$ and $j$, is shown on the left panel (plotted on a logarithmic scale and artificially saturated at $10^{-6}$). The spectrum $N^{p0} / \phi$ and  $N^{0j} / \phi$ are shown on the right panel (solid blue and red lines, respectively). Here, $k_\parallel = 0.1$, $ \eta =5$ and $k_\perp = 0.5$. Dashed lines in the right plot correspond to the case of $k_\perp = 2$.}
    \label{fig:sITG_HL_spectrum}
\end{figure}

Considering the slab ITG mode, we focus on the parameters $k_\parallel = 0.1$, $\eta =5$, and $k_\perp = 0.5$ (i.e., near the growth rate peak). We first note that the solution of the GK dispersion relation, given in \cref{eq:tdisperlationf},  yields the mode complex frequency $\omega = 0.1754 +  0.1504 i$. Given $\omega$, the normalized collisionless gyro-moments spectrum, $N^{pj} / \phi$, can be evaluated using \cref{eq:Nipjslab} (or equivalently \cref{eq:tdisperlationf} with $R_B = 0$). This is shown in \cref{fig:sITG_HL_spectrum} as a function of $(p,j)$ on a logarithmic scale. The collisionless gyro-moment spectrum decreases rapidly with $j$, while the decrease of the magnitude of the gyro-moments is slower along the $p$ direction, as highlighted on the right panel of \cref{fig:sITG_HL_spectrum} that shows $N^{p0} / \phi$ and $N^{0j} / \phi$. The finite amplitude of the gyro-moments $N^{pj} / \phi$ at $j \neq 0 $ is due to the presence of FLR effects, yielding terms proportional to $\kernel{j}$ in the collisionless gyro-moment and, to a smaller extent, to the presence of temperature gradients, yielding terms proportional to $\eta \kernel{j \pm 1}$ (see \cref{eq:Nipjslab}). The rapid decay of the gyro-moment amplitude with $j$ stems from the behaviour of the FLR kernel $\kernel{j}$ for large $j$, such that $\kernel{j} \sim (b/2)^{2j}/j! $. In fact, as \cref{fig:sITG_HL_spectrum} shows, the decrease of the $\hat{N}^{pj}$ coefficients  with $j$ is slower at large $b$ and it follows, in general, $1/j!$. On the other hand, the gyro-moments of $p \neq 0 $ arise because of the parallel resonance terms proportional to the $p$th derivative of the plasma dispersion functions, $Z^{(p)}(\xi) / \sqrt{2^p p!}$, as illustrated by the black solid and dashed dotted lines in \cref{fig:sITG_HL_spectrum}.

\begin{figure}
    \centering
    \includegraphics[scale = 0.6]{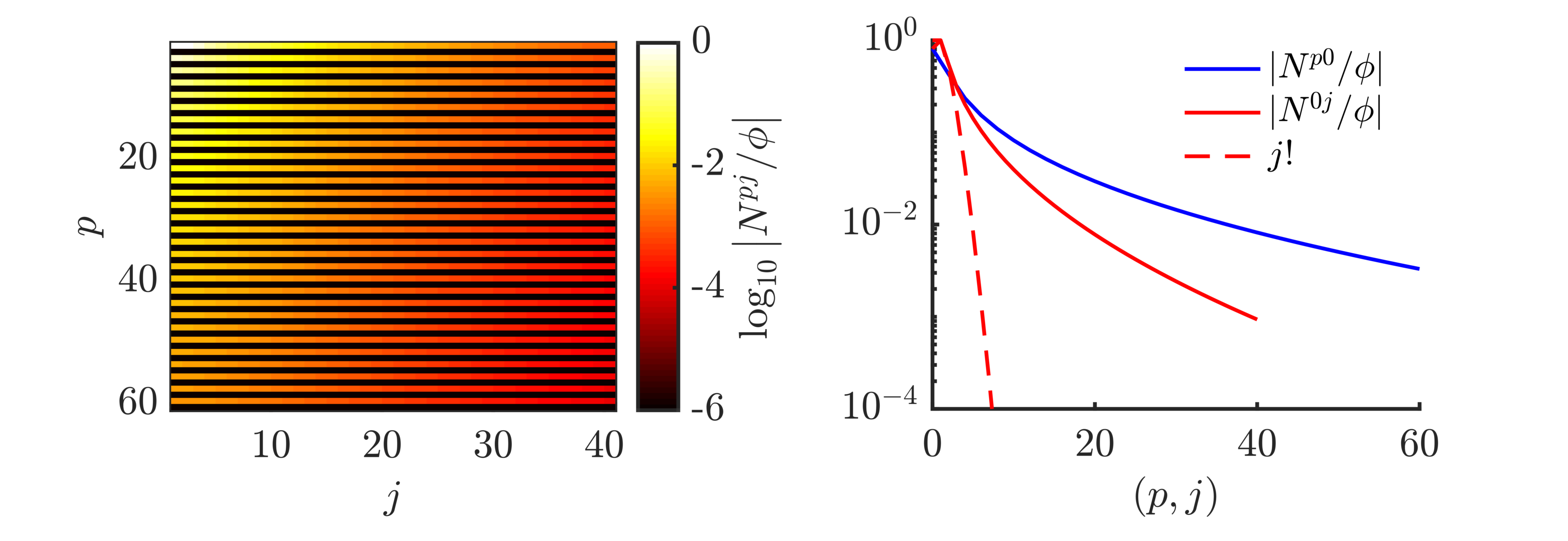}
    \caption{The modulus of the normalized collisionless gyro-moments spectrum, $N^{pj} / \phi$, in the case of toroidal ITG ($R_B = 0.5$), as a function of $p$ and $j$, is shown on the left panel (plotted on a logarithmic scale and artificially saturated at $10^{-6}$). The spectrum $N^{p0} / \phi$ and  $N^{0j} / \phi$ are shown on the right panel (solid blue and red lines, respectively). Here, $k_\perp = 0.5$, $k_\parallel = 0$, $ \eta =5$.}
    \label{fig:tITG_HL_spectrum}
\end{figure}

We now consider the collisionless gyro-moment spectrum in the presence of magnetic gradients. As in the slab case, we focus on $k_\perp = 0.5$, $k_\parallel = 0.1$ and $\eta = 5$, but we choose $R_B = 0.5$. The solution of the GK dispersion relation \cref{eq:tdisperlationf}, for the complex mode frequency, yields $\omega = 0.7972 + 0.4364 i$. \Cref{eq:tdisperlation} is used to evaluate the definite integrals in $\mathcal{I}_n^{pj}$, given in \cref{eq:curlyInpj} numerically, while the infinite sums in \cref{eq:tdisperlation} are truncated when machine precision is reached. Similarly to \cref{fig:sITG_HL_spectrum}, we plot the toroidal collisionless gyro-moment spectrum, $N^{pj}  / \phi$, on a logarithmic scale in \cref{fig:tITG_HL_spectrum}. Compared with the slab case, the gyro-moment spectrum is significantly broader. This is attributed to the velocity-dependence of the magnetic drift resonant coefficients, proportional to $1/[\omega -  \omega_{\grad B}]$, that drives high-order gyro-moments. In fact, while the high $j$ dependence of the gyro-moments stems mostly from terms proportional to $\kernel{j}$ in the slab case, the definite integrals of Gauss hypergeometric functions, appearing in $\mathcal{I}_{n}^{pj}$ (see \cref{eq:curlyInpj}), extend the gyro-moment spectrum to higher $j$, producing a considerably slower decay than $1/j!$. We also notice that, since $k_\parallel =0$, the gyro-moment spectrum vanishes for odd values of $p$, where $\mathcal{I}_n^{pj} = 0$ (see \cref{eq:curlyInpj}). Finite $k_\parallel$ yields non vanishing odd gyro-moments in $p$ due to the term proportional to $(1 + (-1)^{p-2p_1 +1})$ in \cref{eq:Inpj}. 

\begin{figure}
    \centering
    \includegraphics[scale= 0.6]{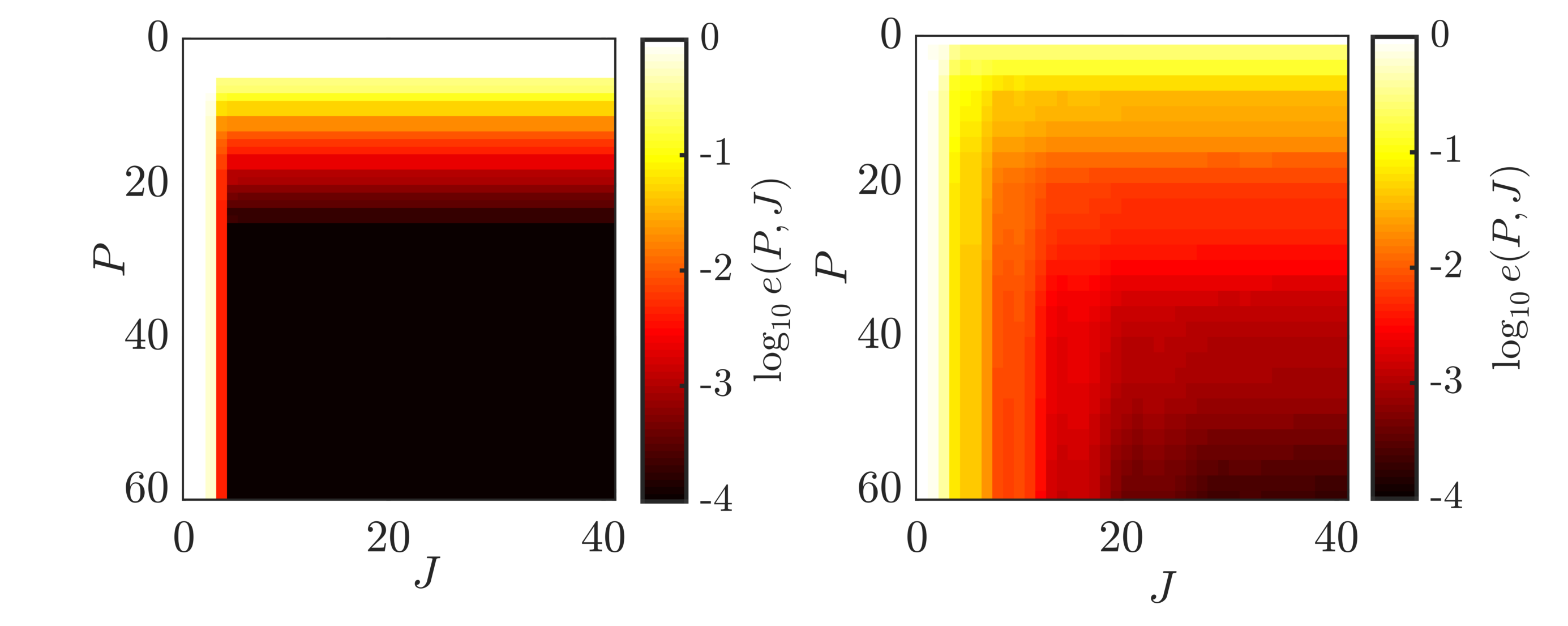}
    \caption{Error norm $e(P,J)$, \cref{eq:errornorm}, on a logarithmic scale for the slab (left) and toroidal (right) ITG cases. The same parameters as in \cref{fig:sITG_HL_spectrum} and \cref{fig:tITG_HL_spectrum} are used respectively. The colorbars are saturated artificially at $\log_{10} e(P,J) = -6$.}
    \label{fig:ITG_fanderror}
\end{figure}

Despite the fact that the gyro-moment spectrum is significantly broader with respect to the slab limit when the magnetic drifts are important (large $R_B$), we numerically demonstrate the convergence of the gyro-moment representation of the perturbed distribution function, $\g /\phi$. Given the complex frequency $\omega$, a truncated distribution function $g^{PJ}$ is defined from the Hermite-Laguerre expansion in \cref{eq:fHL} according to

\begin{align} \label{eq:fpjtruncated}
   \g^{PJ} = \sum_{p=0}^P \sum_{j=0}^J \frac{\hat{N}^{pj}}{\phi} \frac{H_p(s_{\parallel}) L_j(x)}{\sqrt{2^p p!}} F_{M}.
\end{align}
\\
Then, from $g^{PJ}$ and $\g /\phi$ obtained from \cref{eq:fi}, the error norm $e(P,J)$ can be defined by

\begin{align} \label{eq:errornorm}
e(P,J) = \int_{- \infty}^{\infty} d s_\parallel \int_0^{\infty} x \left| \g^{PJ} -  \frac{\g}{\phi}  \right|^2,
\end{align}
\\
and can be computed as a function of $(P,J)$. The results are shown in \cref{fig:ITG_fanderror} for the slab and toroidal cases. Convergence is observed in both cases. However, while the magnitude of  $e(P,J)$ becomes negligibly small when $(P,J) \gtrsim
(20,10)$ (corresponding to $e(20,10) \lesssim 10^{-3}$) in the slab case, a larger number of gyro-moments is necessary to achieve a similar error, i.e. $(P,J ) \gtrsim (50,20)$.

  We now turn to the collisional effects, which provide a natural cutoff of the gyro-moment hierarchy. We aim to derive an \textit{ad-hoc} proxy of the scaling of $(P,J)$ at which the gyro-moment hierarchy can be truncated as a function of the collisionality. In particular, we consider an equation for the generalized gyro-moment amplitude $G
^{pj} =  \sqrt{p} (g^{pj})^2 /2$ (where $g^{pj} = (-1)^j i^{ p +j } \text{sgn}(k_\parallel)^p  N^{pj}$ is introduced to remove the oscillatory behaviour of $N^{pj}$) using the gyro-moment hierarchy equation, \cref{eq:dNipjdt} in the limit $p,j \gg 1$ \citep{Zocco2011,Loureiro2013,Jorge2018}. In this limit, the amplitude $g^{pj}$ can be considered as a continuous complex function of $p$ and $j$. This allows us to approximate $g^{p \pm 1j} \simeq g^{pj} \pm \partial_p g^{pj}$ (and $g^{p  j \pm 1} \simeq g^{pj} \pm \partial_j g^{pj}$). Assuming that an unstable mode is considered, $\partial_t G^{pj} = 2 \gamma G^{pj}$, we derive the equation for $G^{pj}$, up to first order in $1/p$ and $1/j$, i.e.

\begin{align}\label{eq:GLDEqGpj}
    |k_\parallel |\sqrt{\tau} \partial_{s} G^{pj} +   \frac{   \tau}{q} \omega_{B}  ( 2 j +1) \partial_j G^{pj}+ \frac{2 \tau}{q} \omega_B  \left( 1 + i \left( 2j +1\right)\right) G^{pj}  = 2 \nu \left(  f_{p,j} + b^{\alpha} h_{p,j} -  \gamma \right) G^{pj},
\end{align}
\\
with $s = \sqrt{p}$. In deriving \cref{eq:GLDEqGpj}, collisional effects are modelled using the \textit{ansatz} $\C^{pi} \simeq \nu \left[ f_{p,j} + b^\alpha h_{p,j}\right] N^{pj}$, where the functions $f_{p,j} = f(p,j)$ and $h_{p,j} = h(p,j)$, and the constant $\alpha$ depend on the choice of the collision operator. We remark that the coupling along the Hermite direction associated with the curvature drifts vanishes at the leading order in $1/p$ in \cref{eq:GLDEqGpj}. We solve \cref{eq:GLDEqGpj} for the slab and toroidal $k_\parallel = 0$ ITG cases using the GK Dougherty and GK Coulomb collision operators. While the functions $f_{p,j}$ and $h_{p,j}$ are evaluated numerically for the GK Coulomb collision operator, the sparse gyro-moment expansion of the GK Dougherty collision operator, given in \cref{sec:dougherty}, allows us to solve analytically \cref{eq:GLDEqGpj}.

Using the gyro-moment expansion of the GK Dougherty collision operator, we solve first \cref{eq:GLDEqGpj} in the case of the slab ITG mode. With $f_{p,j} = - p - 2j $, $h_{p,j} = -1/2 $ and $\alpha = 2$, the solution of \cref{eq:GLDEqGpj} is

\begin{align} \label{eq:slabmodNpjDougherty}
\left| N^{pj}\right|    \sim \frac{1}{p^{1/4}} \exp\left\{  - \left( \frac{p}{p_\gamma}\right)^{1/2} - \left( \frac{p}{p_\perp}\right)^{1/2}- \left(\frac{p}{p_{D\nu}} \right)^{3/2} \right\},
\end{align}
\\
where we introduce $p_\gamma =  \tau / \gamma_\parallel^2$ (with $\gamma_\parallel = \gamma / k_\parallel$), $p_{D \nu } = (9 \tau / \nu_\parallel^2)^{1/3}$ ($\nu_\parallel = \nu / k_\parallel$) and $p_\perp =  \tau /[ \nu_\parallel (2 j + a)]^2$. The $1 / \nu_\parallel$ of $p_{D \nu}$ indicates that the exponential decay of \cref{eq:slabmodNpjDougherty} is slower at high $k_\parallel$ and faster at high $\nu$. At high $k_\parallel$, higher-order $p$ gyro-moments are excited due to parallel resonance effects. Indeed, in the high-collisional limit, the Chapman-Enskog closure procedure ensures that $\nu_\parallel \sim \sqrt{\tau} / \epsilon \gg 1$, such that $p_{D\nu} \sim \nu_\parallel^{-2/3} \sim \epsilon^{2/3} / \tau^{1/3}$ and, therefore, that the ITG dynamics can be described by the lowest order gyro-moments at high collisionality (see \cref{sec:64GM}). Moreover, when $\nu \gg \gamma$ (such that $p_{D\nu}$ is small that $p_\gamma$) and in the DK limit, the collisional damping provide a truncation at $P \gtrsim p_{D \nu}$. We remark that FLR terms have a small effect on the scaling of \cref{eq:slabmodNpjDougherty} since, for ITG modes, $p_\perp \gtrsim p_{ D\nu}$ and the term with the exponent $3/2$ is dominant at large $p$. Since \cref{eq:GLDEqGpj} does not couple the Laguerre gyro-moment, the evolution at large $j$ is dominated by collisions, and, thus, are damped in the high-collisional regime.

We now derive the solution of \cref{eq:GLDEqGpj} for the toroidal $k_\parallel = 0$ ITG case, still considering the GK Dougherty collision operator. It yields 

\begin{align}  \label{eq:toroidalmodNpjDougherty}
    \left| N^{pj}\right|    \sim \frac{1}{p^{1/4}} \exp\left\{  - \nu_B j- \frac{1}{2}\ln\left( 1 + 2j \right) \left[ \gamma_B + 1  + \nu_B \left(   p - 1+ a \right) \right]\right\}.
\end{align}
\\
where we introduce $\gamma_B =  q \gamma  /( \tau \omega_B)$ and $\nu_B = q \nu / (\tau \omega_B)$. The $\nu_B$ dependence highlights the broadening of the gyro-moment spectrum associated with the presence of magnetic gradient drifts (see \cref{fig:tITG_HL_spectrum}), an effect reduced at high collisionality. Similar to the slab case, FLR terms do not affect the scaling of \cref{eq:slabmodNpjDougherty}, since the linear term dominates over the second term at large $j$ values, and $p-1 \gg a \sim k_\perp^2 $ for ITG modes. We remark that, in the case of toroidal ITG with $k_\parallel = 0$, the Hermite gyro-moment evolution is dominated by collisional damping.

\begin{figure}
    \centering
    \includegraphics[scale =0.55]{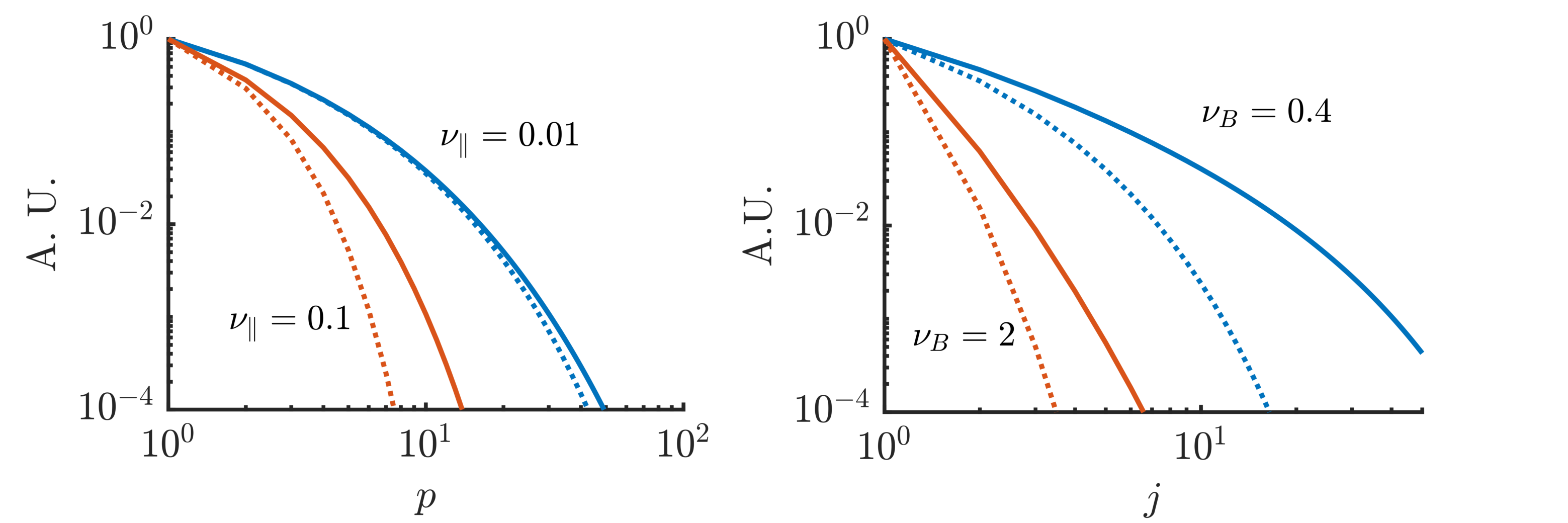}
    \caption{Normalized slab (left) and toroidal (right) gyro-moment spectra as a function of $p$ and $j$, respectively. The solid lines are the gyro-moment spectrum obtained using the GK Coulomb collision operator, given in \cref{eq:slabmodNpjCoulomb} and \cref{eq:toroidalmodNpjCoulomb}, and the dotted lines the ones obtained from the GK Dougherty collision operator, given in \cref{eq:slabmodNpjDougherty} and \cref{eq:toroidalmodNpjDougherty}. The slab and toroidal growth rate are estimated at $\gamma \simeq 0.1$ and $\gamma \simeq 0.4$, respectively near the ITG peaks ($k_\perp \simeq 0.5$).} 
    \label{fig:fig_Npj_spectrum}
\end{figure}

Because the Coulomb collision operator has a non-trivial block matrix structure when projected onto the Hermite-Laguerre basis \citep{frei2021}, an approximate expression of $\C^{pj}$ needs to be determined numerically. Based on the results derived from the GK Dougherty collision operator, we assume that $\C^{pi} \simeq \nu  f_{p,j}  N^{pj}$, thus neglecting the FLR effects in the spectrum scaling. Fitting numerically the values of $\C^{pj}$ using the GK Coulomb collision operator yields $f_{p,j} \simeq  - A j - B \sqrt{p} $ ($A \simeq 0.3$ and $B \simeq 0.8$). Similar values (within a $10 \%$ accuracy) are typically found in the case of the DK Coulomb and show that FLR terms have a small effects on the ITG spectrum scaling when $k_\perp \sim 1$. We remark that the $\sqrt{p}$ dependence of $f_{p,j}$ is consistent with previous DK moment hierarchy using the DK Coulomb collision operator \citep{Jorge2018}. Given $f_{p,j}$, we solve again \cref{eq:GLDEqGpj} for the case of the slab and toroidal $k_\parallel =0$ ITG modes yielding, respectively,

\begin{align} \label{eq:slabmodNpjCoulomb}
\left| N^{pj} \right|    \sim \frac{1}{p^{1/4}} \exp\left\{ - \left(  \frac{p}{p_\gamma}\right)^{1/2} -  \left( \frac{p}{p_j} \right)^{1/2} - \frac{p}{p_\nu} \right\},
\end{align}
\\
with $p_j = \tau / \nu_\parallel A j^2$ and $p_\nu = 2 \sqrt{\tau} / (B \nu_\parallel)$, and 

\begin{align} \label{eq:toroidalmodNpjCoulomb}
\left| N^{pj} \right|   \sim \frac{1}{p^{1/4}} \exp\left\{ - \frac{ A}{2} \nu_B j  - \frac{1}{2}\ln\left( 1 + 2j \right) \left[ \gamma_B + 1  + \nu_B \left( B \sqrt{p} - A /2  \right)  \right] \right\},
\end{align}
\\
The main differences between the results of the GK Coulomb collision operator, given in \cref{eq:slabmodNpjCoulomb} and \cref{eq:toroidalmodNpjCoulomb} and the results of the GK Dougherty operator, in \cref{eq:slabmodNpjDougherty,eq:toroidalmodNpjDougherty}, are that the scaling in $p$ predicted by the GK Coulomb collision operator is weaker in the slab case (compared to a $3/2$ exponent in \cref{eq:slabmodNpjDougherty}), and that the numerical factor in the linear term in $j$ in the toroidal case,  $A / 2 \simeq 0.15$, is smaller.

We compare the scaling of the gyro-moment spectra obtained using the GK Dougherty and GK Coulomb collision operators in \cref{fig:fig_Npj_spectrum} for the slab (left) and toroidal (right) ITG modes for different values of $\nu_\parallel$ (at $k_\parallel = 0.1$) and $\nu_B$ (at $\nu = 0.1$). The differences between the $p$ exponent and the numerical factors between GK Dougherty (dotted lines) and GK Coulomb (solid lines), do not significantly impact the gyro-moment spectra that feature the same qualitative behaviour. The larger deviation appear in the toroidal case when $\nu_B$ is small (large $R_B$ and/or small $\nu$). We compare the predictions of the values of $P$ and $J$, at which the gyro-moment hierarchy can be truncated deduced from \cref{fig:fig_Npj_spectrum}, with the numerical convergences results obtained in \cref{sec:convergence}.

\section{Convergence Studies}
\label{sec:convergence}
  
  Following the theoretical analysis of the gyro-moment spectrum presented in \cref{sec:GMspectrum}, we perform convergence studies of the ITG growth rate as a function of collisionality, perpendicular wavenumber and magnetic gradient. Collisions are modelled using the GK and DK Coulomb collision operator \citep{frei2021}. Similar convergence properties are observed with other collision operator models considered in this work. We show that the convergence of the gyro-moment approach improves with collisionality. 

\begin{figure}
    \centering
    \includegraphics[scale=0.5]{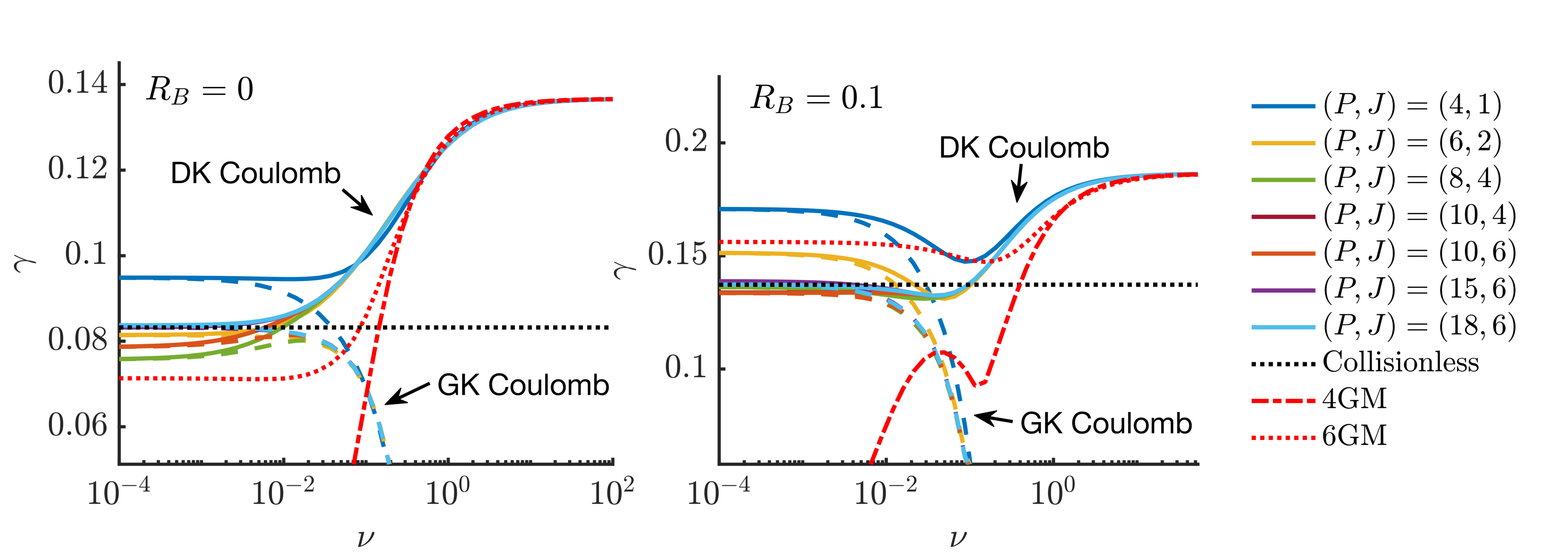}
    \caption{Convergence of the ITG linear growth rate, $\gamma$, as a function of the collisionality $\nu$ for different values of $(P,J)$, using the GK (dashed lines) and DK (solid lines) Coulomb operators, in the case of slab (left) and toroidal (right), with $R_B = 0.1$ branches. The collisionless limit (dotted black lines) and the high collisional limits, $6$GM and $4$GM models (red dotted and dash-dot lines, respectively) are shown for comparisons. Similar plots are obtained for the other GK and DK operator models. The parameters are $k_\perp  = 0.5$, $k_\parallel = 0.1$, $\eta = 3$.}
    \label{fig:ITG_nuconvergence}
\end{figure}

First, convergence on the ITG linear growth mode using the GK and DK Coulomb collision operators are performed numerically as function of the collisionality $\nu$. The collisionless and high-collisional limits are considered for comparison by solving the collisionless dispersion relation, \cref{eq:tdisperlationf}, and the $6$GM and $4$GM models. The results are shown in \cref{fig:ITG_nuconvergence} for the slab (left) and for the toroidal (right) ITG modes. Different number of gyro-moments, $(P,J)$, are used. 
The collisionless solution is retrieved for $(P,J) \simeq (18,6)$ with the GK (dashed) and DK (solid) Coulomb collision operators. In the slab case, for $\nu \simeq 0.1$ and $\nu \lesssim 10^3$, convergence is achieved with $P \lesssim 10$ and $P \gtrsim 18$. This is in qualitative agreement with the estimates deduced from \cref{fig:fig_Npj_spectrum} with the corresponding values of $\nu_\parallel =  \nu / k_\parallel \simeq 1$ and $\nu_\parallel \lesssim 0.01$. Also consistent with the observations made in \cref{sec:GMspectrum}, only a few number of Laguerre gyro-moments are necessary to resolve the slab ITG, as illustrated by the overlapping between the $(10,4)$ and $(10,6)$ lines in \cref{fig:ITG_nuconvergence} as the slab ITG mode is mainly driven by the parallel dynamics which is resolved by the Hermite gyro-moments. This remains also true for the toroidal ITG case considered in \cref{fig:ITG_nuconvergence}. Larger values of $R_B$ deteriorates the convergence, as discussed below. While the GK Coulomb yields a strong FLR collisional damping as $\nu$ increases, the gyro-moment hierarchy, with DK Coulomb, agrees well with the $6$GM and $4$GM when $\nu \gg 1$. Nevertheless, we notice that, because of finite magnetic drifts (see \cref{sec:CollisionlessHermiteLaguerreTheory}), agreements with the $6$GM and the 4GM models occur at higher collisionality in the toroidal case. In fact, in the slab case, the high-collisional limits are retrieved when $\nu \gtrsim 1$, while for the toroidal case ($R_B = 0.1$) when $\nu \gtrsim 10^2$. 

\begin{figure}
    \centering
    \includegraphics[scale = 0.48]{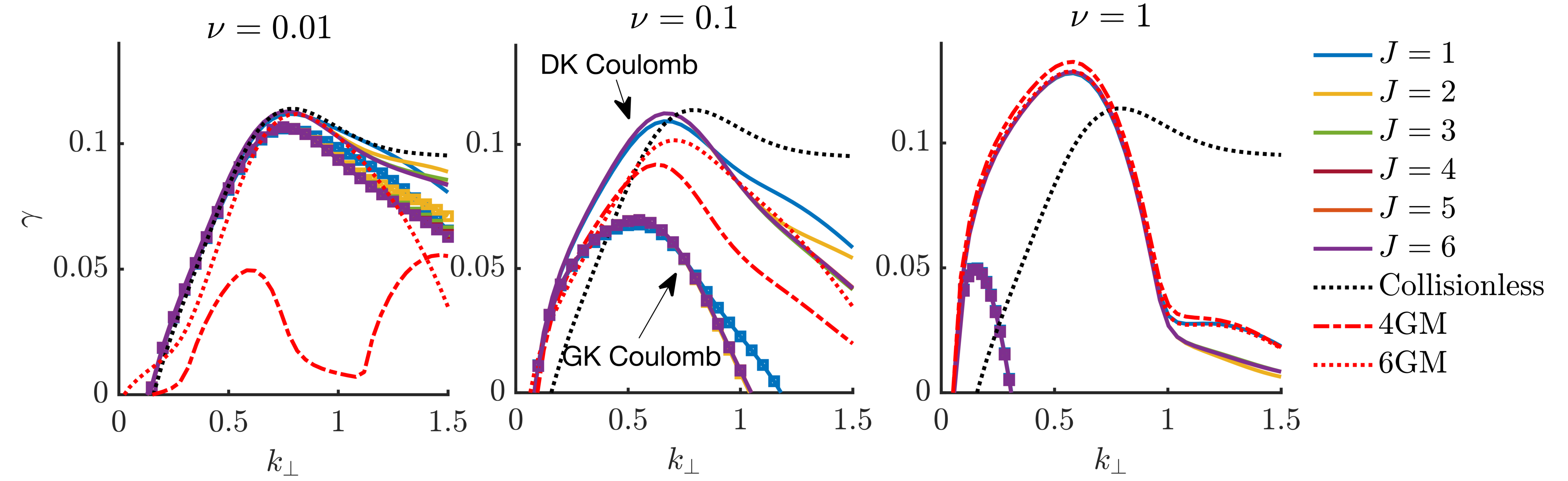}
    \caption{Slab ITG growth rate $\gamma$ as a function of $k_\perp$ for different values of $J$, in the low (left), intermediate (center), and high (right) collisionality regimes, with the GK (square markers) and DK (solid lines) Coulomb collision operators. Similar plots are obtained for the other GK and DK operator models.}
    \label{fig:sITG_FLRscan_GKDKCoulomb}
\end{figure}

We investigate the convergence associated with the perpendicular dynamics in velocity space, i.e. with the Laguerre gyro-moments and associated with the FLR effects due to the gyro-average of the electrostatic potential. The coupling between the Laguerre gyro-moments driven by magnetic gradients are neglected by focusing on the slab ITG branch. We scan the growth rate $\gamma$ as a function of the perpendicular wavenumber, $k_\perp$, for different values of $J$ and collisionality $\nu$, while we fix $P = 18$. The results are shown in \cref{fig:sITG_FLRscan_GKDKCoulomb}. Convergence occurs at lower $J$ for high collisionalities, but in all cases the values of $J$ for convergence increases with $k_\perp$. This is in agreement with the unfavorable scaling of the FLR kernel with $b$, i.e. $\kernel{j} \sim (b/2)^{2j}/j!$ (see \cref{sec:GMspectrum}).

\begin{figure}
    \centering
    \includegraphics[scale = 0.48]{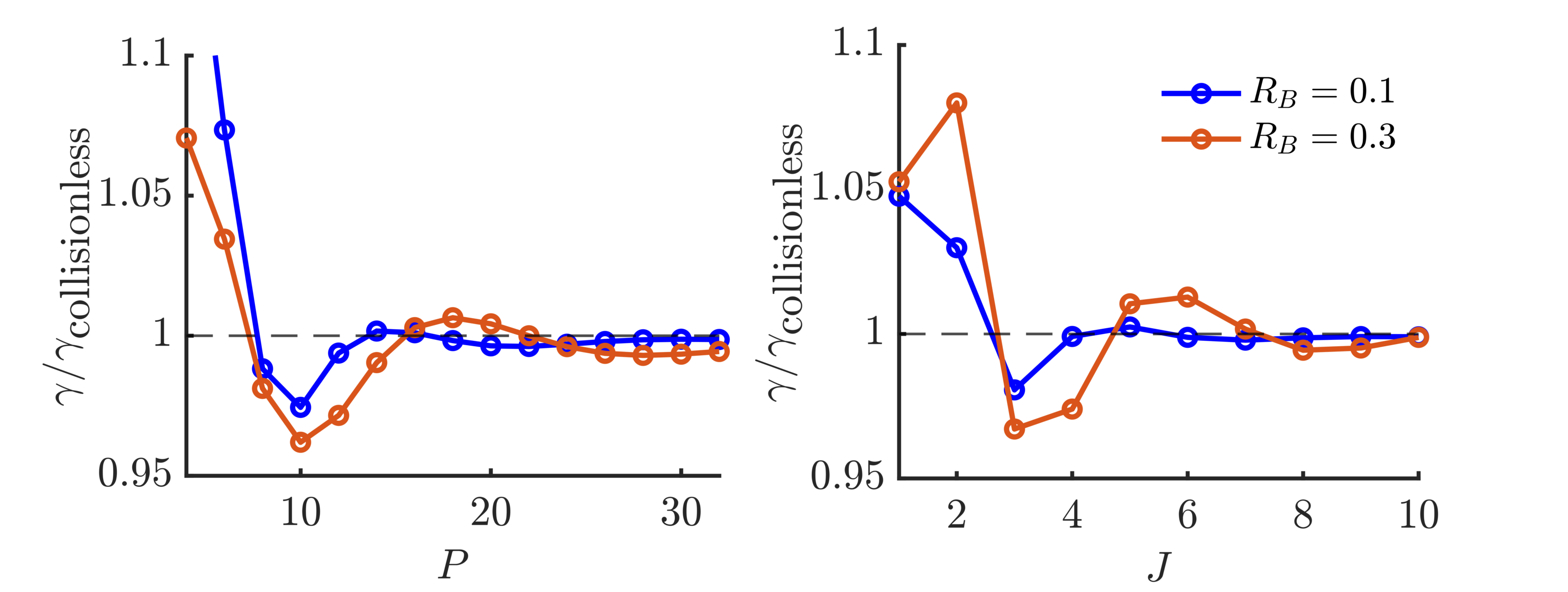}
    \caption{Ratio between the ITG growth rate $\gamma$ and the collisionless solution $\gamma_{\textrm{collisionless}}$ as a function of $P$ with $J = 10$ (left) and as a function of $J$ with $P=32$ (right) for two different values of $R_B$. Here, $k_\perp =0.5$, $k_\parallel = 0.1$ and $\nu = 0.001$.}
    \label{fig:fig_ITG_gammacollisionless_VSPJ}
\end{figure}

\begin{figure}
    \centering
    \includegraphics[scale = 0.48]{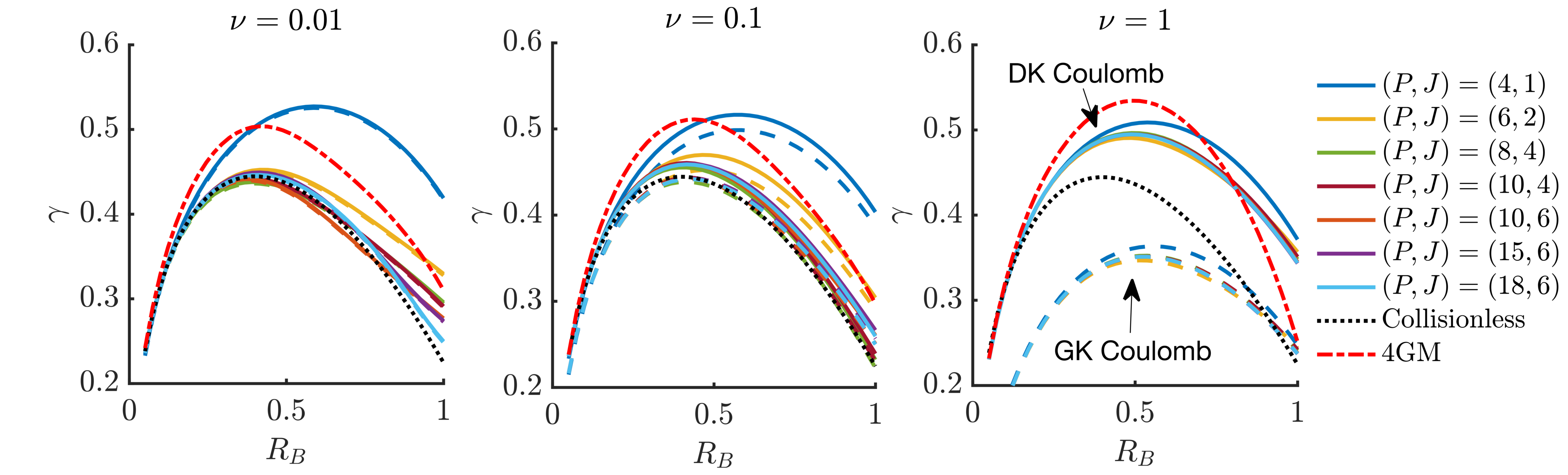}
    \caption{Toroidal ($k_\parallel = 0$) ITG growth rate $\gamma$ as a function of the normalized magnetic gradient, $R_B$, for different $(P,J)$ and increasing collisionality (from left to right). Here, the GK (dashed lines) and DK (solid lines) Coulomb collision operators are shown, with the collisionless (black dotted) and high collisional $4$GM limit (red dotted-dashed lines).} 
    \label{fig:TG_RBconvergence}
\end{figure}

Finally, we consider the convergence of the gyro-moment hierarchy as a function of the normalized magnetic gradient, $R_B$. As illustrated in \cref{sec:GMspectrum}, the presence of finite magnetic drifts broadens significantly the collisionless gyro-moment spectrum compared to the slab case. However, in all cases, the gyro-moment approach converge correctly to the collisionless solution of \cref{eq:tdisperlationf}, as shown in \cref{fig:fig_ITG_gammacollisionless_VSPJ}, where the ratio of the ITG growth rate to the collisionless value is plotted as a function of $P$ and $J$. It is remarkable that a larger number of gyro-moment is needed as $R_B$ increases. Collisions can compete against this strong kinetic drive by damping higher order gyro-moments (see \cref{fig:fig_Npj_spectrum}). To investigate the convergence of the gyro-moment approach at finite $R_B$ in the presence of collisions, we focus on the toroidal ITG with $k_\parallel =0$, i.e. neglecting the coupling between the Hermite gyro-moments associated with the parallel streaming. We scan the toroidal ITG linear growth rate $\gamma$ as a function of $R_B$, and increase the collisionality. The results are reported in \cref{fig:TG_RBconvergence}. At all collisionalities, the convergence deteriorates as $R_B$ increases. Nevertheless, convergence is observed with the increase of the collisionality, showing the competition between the magnetic gradient kinetic and collisional effects. We also remark that the $4$GM limit is retrieved at higher collisionality when $R_B$ increases.

\section{Conclusion}
\label{sec:conclusion}

In this paper, the properties of the ITG mode are studied using the GK Coulomb collision operator for the first time. The investigation is based on a Hermite-Laguerre moment expansion of the perturbed gyrocenter distribution function, which we refer to as gyro-moment approach. By projecting the GK Boltzmann equation onto the Hermite-Laguerre basis, a gyro-moment hierarchy equation is deduced which retains the effects of like-species collisions between ions thanks to the numerical implementation of the GK Coulomb collision operator described in \citet{frei2021}. Using the gyro-moment hierarchy equation, the collisionless and high-collisional limits are derived analytically. In the collisionless limit, we find that the magnetic drift resonance effects significantly broaden the gyro-moment spectrum compared to the slab case. In the high-collisional limit, the gyro-moment hierarchy can be reduced to a fluid model retaining only a finite number of gyro-moments using the Chapman-Enskog asymptotic closure scheme. Using these analytical results, we show that the gyro-moment hierarchy can retrieve the collisionless limit, where kinetic features are essential and, at the same time, the high collisional limit using a reduced number of gyro-moments.

Numerical experiments are performed to study the ITG linear growth rate using the GK Coulomb collision operator and investigate the importance of the FLR collisional terms. We compare the GK Coulomb collision operator with other collision operator models such as the Sugama \citep{Sugama2009}, the momentum-conserving pitch-angle scattering operator \citep{Helander2002}, the zeroth-order DK Hirshman-Sigmar-Clarke collision operator \citep{Hirshman1976}, and the Dougherty collision operator \citep{Dougherty1964}. The gyro-moment expansions of some of these operators are also derived for the first time here. We find that the ITG mode is strongly damped (or even suppressed) as the collisionality increases and that a steeper temperature gradient for the mode onset is necessary to overcome the FLR collisional stabilization when using the GK Coulomb collision operator. We reveal the importance of FLR terms in the Coulomb collision operator by demonstrating that neglecting these FLR terms destabilizes a short wavelength branch of the ITG mode, peaking near $k_\perp \rho_s \sim 1.5$. These observations are also found when using the other collision operator models considered in this work. In addition to these findings, the gyro-moment expansion developed in this work allows us to perform a systematic comparison between collision operator models and assess their accuracy. The main outcome of this comparison is that the GK Sugama slightly underestimates the ITG growth rate, compared to the GK Coulomb operator, and that the GK pitch-angle scattering operator systematically yields a larger ITG linear growth because of the absence of energy diffusion in the latter. The largest deviations with respect to the GK Coulomb collision operator are observed for the GK Dougherty operator. In general, we observe that energy diffusion is important at high collisionality and in the damping of small-scale modes.  

Complementary to previous works (see, e.g., \citet{Jorge2017,Jorge2018,Frei2020,frei2021}), the present study of the ITG mode at arbitrary collisionality shows the ability of the gyro-moment approach to capture the relevant physics in the kinetic and collisional regimes, with a fidelity cost that depends on the collisionality. Finally, we remark that the present study can be extended to more realistic conditions including the parallel direction, electron dynamics, and electromagnetic effects.

\section*{Acknowledgement}

The authors acknowledge helpful discussions with S. Brunner, P. Donnel, and M. Held.

This research was supported in part by the Swiss National Science Foundation, and has been carried out within the framework of the EUROfusion Consortium and has received funding from the Euratom research and training programme 2014 - 2018 and 2019 - 2020 under grant agreement No 633053. The views and opinions expressed herein do not necessarily reflect those of the European Commission. 

\appendix

\section{Pitch-Angle Scattering Collision Operator}
\label{sec:pitchangle}

The linearized pitch-angle scattering operator model is the sum of the pitch-angle scattering operator for the test component coupled and an \textit{ad hoc} field component such that the momentum conservation is satisfied. This operator neglects energy diffusion, contrary to the Sugama and Coulomb collision operators, but conserves particle, momentum and energy. The test component of the GK pitch-angle scattering operator is given by \citep{Helander2002}

\begin{align}\label{eq:Tpitchangle}
    \C^T =  \nu^D(v) \left\{-  \mathcal{L}^2 h - \frac{b^2}{2 v_T^2} (2 v_\parallel^2 + v_\perp^2) h \right\},
\end{align}
\\
and the field component is defined by

\begin{align}\label{eq:Fpitchangle}
 \C^F =   2 F_M\nu^D(v)  \left( v_\perp J_1 \overline{U}_\perp^D[h] + v_\parallel J_0 \overline{U}_\parallel^D[h]  \right),
\end{align}
\\
where we introduce 

\begin{align} \label{eq:barUD}
  \overline{U}_\perp^D[h] =  \frac{3\int d \vi v_\perp J_1 \nu^D(v) h}{\int d \vi v^2 \nu^D(v) F_M}, \quad  \overline{U}_\parallel^D[h] =  \frac{3 \int d \vi v_\parallel J_0 \nu^D(v) h}{\int d \vi v^2 \nu^D(v) F_M}.
\end{align}
\\
with $h = \g + q J_0 \phi F_M / \tau$ the nonadiabatic part of the perturbed gyrocenter distribution function $\g$. In \cref{eq:Tpitchangle}, $\mathcal{L}^2$ is the spherical angular operator defined by $v^2 \grad_{\bm v}^2 f= \partial_{v}(v^2 \partial_v f) - \mathcal{L}^2 f$ and $\nu^D(v) =  \nu  \left[ \erf(s) -( \erf(s)  - s \erf'(s)) / 2 s^2\right]/ s^3$ (with  $s= v  / v_T$) is the pitch-angle scattering (or deflection) velocity-dependent frequency.

Following \citet{frei2021}, we now project \cref{eq:Tpitchangle,eq:Fpitchangle} onto the Hermite-Laguerre basis in terms of the nonadiabatic moments, $n^{pj}$, defined by 

\begin{align} \label{eq:npjdef}
n^{pj} = \frac{1}{N} \int d \mu  d \vparallel  d \theta\frac{B}{m} h \frac{H_p(s_{\parallel}) L_j(x)}{\sqrt{2^p p!}},
\end{align}
\\
and expressed in terms of the gyro-moments, $N^{pj}$, by $n^{pj} = N^{pj} + q \kernel{j} \phi \delta_p^0 / \tau$. The gyro-moment expansion of the test and field components, \cref{eq:Tpitchangle} and  \cref{eq:Fpitchangle}, are given by

\begin{align} \label{eq:pitchTlk}
    \C^{Tlk} = \C^{Dlk} + \overline{R}^{Dlk} ,
\end{align}
\\
and
 \begin{align} \label{eq:pitchFlk}
    \C^{Flk} =  \overline{R}_\parallel^{Dlk} + \overline{R}_{\perp}^{Dlk},
\end{align}
\\
respectively. In \cref{eq:pitchTlk,eq:pitchFlk}, we introduce

\begin{align} \label{eq:CDlk}
\C^{Dlk} =- \sum_{j=0}^{\infty} \sum_{p=0}^{l+2k} \sum_{h=0}^{k + \lfloor  l/2\rfloor} \sum_{g=0}^{p+2j}\sum_{t=0} ^{j + \lfloor l/2\rfloor} \sum_{d=0}^h  \frac{2^{p} (p!)^2\sqrt{2^g g!}}{(2p)!} \frac{ \left(T^{-1} \right)^{ph}_{lk} }{\sigma_p^j \sqrt{2^l l!}} \frac{  p(p+1)}{(2p +1)}T _{pj}^{gt}L_{hd}^p \overline{\nu}_{*}^{Dpjd}n^{gt},
\end{align}
\\
with

\begin{align} \label{eq:barnuDpjd}
\overline{\nu}_{*}^{Dpjd}   =  \frac{4 }{\sqrt{\pi}} \sum_{j_1=0}^j L_{jj_1}^p \overline{\nu}^{Dp+d+j_1+1},
\end{align}
\\
and 

\begin{subequations} \label{eq:Pitchterms}
\begin{align} 
\overline{R}^{D  lk} & = - \frac{b^2}{\sqrt{\pi}} \sum_{p,j} \sum_{g=0}^{l + 2 k} \sum_{h=0}^{k + \lfloor  l/2\rfloor} \sum_{r=0}^{ p + 2j} \sum_{s=0}^{j + \lfloor  p/2\rfloor} \sum_{s_1 =0}^s\sum_{h_1 =0}^h  \frac{\left( T^{-1} \right)^{gh}_{lk}}{\sqrt{2^l l!}}\frac{\left( T^{-1} \right)^{rs}_{pj}}{\sqrt{2^p p!}}  L^{r}_{ss_1} L^{g}_{hh_1} \mathcal{\alpha}_{\perp}^{grs_1 h_1}   n^{pj}, \label{eq:RlkS}\\
    \overline{R}_\perp^{lk} & = \overline{U}_\perp[h] \frac{4 }{\sqrt{\pi}} \sum_{n=0}^\infty \sum_{s =0}^{n + k + 1} \sum_{h =0}^{s + \lfloor l/2\rfloor}  \sum_{h_1=0}^h L_{hh_1}^0 \frac{ d_{nks}^1 b \kernel{n}}{ (n+1)} \frac{\left( T^{-1} \right)_{ls}^{gh}}{\sqrt{2^l l!}} \overline{\nu}^{Dh_1 +1}, \label{eq:RperplkS} \\
 \overline{R}_\parallel^{lk} & =  \overline{U}_\parallel[h] \frac{8 }{3 \sqrt{\pi}}\sum_{n=0}^\infty \sum_{s =0}^{n+k} \sum_{g =0}^{l + 2s} \sum_{h =0}^{s + \lfloor l/2\rfloor} \sum_{h_1 =0}^h d_{nks}^0 \kernel{n}  \frac{L_{hh_1}^1 \left( T^{-1} \right)_{ls}^{gh}}{\sqrt{2^l l!}} \overline{\nu}^{Dh_1 + 2} [l > 0 \cup s > 0], \label{eq:RparalkS} 
  \end{align}
\end{subequations}
\\
where $[\cdot]$ denotes the Iverson bracket. In \cref{eq:Pitchterms}, we define

\begin{align} \label{eq:alphaperp}
\mathcal{\alpha}_{\perp}^{grs_1 h_1}& =  \frac{2\delta_g^r}{(2g+1)}   \overline{\nu}^{D g  + 2 + s_1 + h_1 }+ d_{g}^r  \overline{\nu}^{D(g+r)/2
+ 2 + s_1 + h_1 }  , 
\end{align}
\\
with the numerical coefficient

\begin{align}
d_g^r =  \begin{cases}
\dfrac{2 (r + 1)(r+2)}{(2 r+1)(2r +3)(2r+5)}  & \mbox{for } g = r+2\\
\dfrac{2(2r^2 + 2r -1)}{(2r-1)(2r+1)(2r+3)} & \hbox{for } g=r \\
\dfrac{2 r(r-1)}{(2r -3)(2r -1)(2r +1)} & \hbox{for } g=r-2,
\end{cases}
\end{align}
\\
and

\begin{subequations} \label{eq:Pitchterms2}
 \begin{align} 
     \overline{U}_\parallel[h]     & = \frac{1}{ \overline{\nu}^{D2}} \sum_{p,j} \sum_{n=0}^\infty \sum_{s =0}^{j + n }\sum_{g =0}^{p + 2s} \sum_{h =0}^{s + \lfloor p/2\rfloor} \sum_{h_1=0}^h d_{njs}^0 \kernel{n}  \frac{ L_{hh_1}^1 \left( T^{-1} \right)_{ps}^{gh} }{\sqrt{2^p p!}}   \overline{\nu}^{Dh_1 +2} n^{pj}  [p > 0 \cup s > 1],  \\
    \overline{U}_\perp[h] & = \frac{3 }{2 \overline{\nu}^{D2}}  \sum_{p,j} \sum_{n=0}^\infty \sum_{s =0}^{n + j + 1}  \sum_{h =0}^{s + \lfloor p/2\rfloor} \sum_{h_1=0}^h L_{hh_1}^0\frac{b \kernel{n}}{ (n+1)} \frac{ d_{njs}^1  \left( T^{-1} \right)_{ps}^{0h}}{\sqrt{2^p p !}} \overline{\nu}^{Dh_1 +1 }  n^{pj},  \\
   \end{align}
\end{subequations}
\\
In \cref{eq:Pitchterms,eq:Pitchterms2}, we introduce the speed integrated deflection frequency by $\overline{\nu}^{Dk} = \int d s  e^{- s^2} s^{2k} \nu^D(v) $. Using the definitions of $\nu^D(v)$, we derive

\begin{align}
\overline{\nu}^{Dk} & = \frac{1}{2} \nu \left( 2 E^{k-2} -  E^{k-3} +  e^{k-2}\right), 
\end{align}
 \\
where the definitions of $E^{k}$ and $e^{k}$ can be found in \citep{frei2021}.

The DK limit of the pitch-angle scattering operator is obtained from \cref{eq:pitchTlk,eq:pitchFlk} by taking the zero gyroradius limit, $b \to 0$. Noticing that $n^{pj} \simeq N^{pj}$, we derive the DK test and the field components given by

\begin{align} \label{eq:Pitchlk}
\C_{}^{Tlk}& = \C^{Dlk}
\end{align}
\\
\begin{align} \label{eq:Pitchlk}
\C_{}^{Flk}& =   U_\parallel[\g] \frac{8 }{3 \sqrt{\pi}} \sum_{g =0}^{l + 2k} \sum_{h =0}^{k + \lfloor l/2\rfloor} \sum_{h_1 =0}^h   \frac{L_{hh_1}^1 \left( T^{-1} \right)_{lk}^{gh}}{\sqrt{2^l l!}} \overline{\nu}^{Dh_1 + 2} [ l > 0 \cup k > 0 ],
\end{align}
\\
respectively, where $C^{Dlk}$ is given in \cref{eq:CDlk} and with 

\begin{align}
    U_\parallel^D[\g]      & = \frac{1}{ \overline{\nu}^{D2}}  \sum_{p,j} \sum_{g =0}^{p + 2j} \sum_{h =0}^{j + \lfloor p/2\rfloor} \sum_{h_1=0}^h   N^{pj} \frac{ L_{hh_1}^1 \left( T^{-1} \right)_{pj}^{gh} }{\sqrt{2^p p!}}     \overline{\nu}^{Dh_1 +2} [ p > 0 \cup k > j ]. 
\end{align}

\section{Hirshman-Sigmar-Clarke Collision Operator}
\label{sec:HSC}
 We consider the gyro-moment expansion of the Hirshman-Sigmar-Clarke (HSC) collision operator \citep{Hirshman1976a}, also referred to as the zeroth-order Hirshman-Sigmar collision operator \citep{Hirshman1976b}. The HSC collision operator model has been widely used for neoclassical transport calculations \citep{Hirshman1976a,Belli2008}. This operator is only considered in the DK limit. Additionally, it conserves particle, momentum and energy. While the test component of the HSC operator is the pitch-angle scattering operator, i.e. 
 
 \begin{align}
 \C^T = - \nu^{D}(v) \mathcal{L}^2 \g, 
 \end{align}
\\
 the field component of the HSC operator is defined by 
 
\begin{align} \label{eq:HSC}
 \C^F =  v_\parallel \left[ \nu^S(v) \frac{2r[\g]}{v_T^2} + \frac{2 u[\g]}{v^2} \left( \nu^D(v)- \nu^S(v) \right)\right]F_{M},
\end{align}
\\
where 

\begin{align}
    r[\g] & =  \dfrac{3 \int d \vi v_\parallel \nu^S(v) \g}{\int d \vi s^2 \nu^S(v) F_{M}}, \\
    u[\g] & = \frac{3}{4 \pi}\int d \Omega v_\parallel \frac{\g}{F_{M}},
\end{align}
\\ 
with $ \int d \Omega = \int_0^{2 \pi} d \theta \int_{-1}^1 d \xi$ the integral over the velocity-space solid angle, and $\nu^S (v)  =  2 \nu ( \erf(s) - s \erf'(s) ) /s^3$.

We project the HSC collision operator, given in \cref{eq:HSC}, onto the Hermite-Laguerre basis. As the Hermite-Laguerre projection of $- \nu^{D}(v) \mathcal{L}^2 g$ is given by \cref{eq:CDlk}, we focus on the projection of \cref{eq:HSC}. Performing the Hermite-Laguerre projection yields

\begin{align} \label{eq:HSClk}
    \C^{Flk} =  R^{Slk}_1 + R^{Dlk},
\end{align}
\\
where we introduce 

\begin{subequations}
\begin{align} \label{eq:HSCterms}
     R_{1}^{Slk}  & = \sum_{g = 0}^{l + 2k} \sum_{h = 0}^{k + \lfloor l/2 \rfloor} \sum_{h_1=0}^{h} L_{hh_1}^1\frac{\left( T ^{-1} \right)^{gh}_{lk}}{\sqrt{2^ll!}} \frac{8}{3 \sqrt{\pi}}   \frac{r[\g]}{v_T} \overline{\nu}^{Sh_1 +2} [ l > 0 \cup k > 0], \\
R^{SDlk} & =   \frac{8}{3 \sqrt{\pi}} \sum_{p,j}  \sum_{g=0}^{p + 2j} \sum_{h=0}^{j + \lfloor p/2\rfloor} \sum_{s=0}^{l + 2k} \sum_{t=0}^{k + \rfloor l /2 \lfloor} \sum_{h_1 =0}^h \sum_{t_1 = 0}^t \frac{\left( T ^{-1} \right)^{st}_{lk}}{\sqrt{2^l l! }} \frac{\left( T ^{-1} \right)^{gh}_{pj}}{\sqrt{2^p p! }}  \nonumber \\
& \times L_{tt_1}^1 L_{hh_1}^1 \left( \overline{\nu}^{D2 + t_1 + h_1} - \overline{\nu}^{S2 + t_1 + h_1}\right) N^{pj} [ p > 0 \cup j > 0] [ l > 0 \cup k > 0],
\end{align}
\end{subequations}
\\
and 

\begin{align} \label{eq:rglk}
\frac{ r[\g] }{v_T} & =    \frac{1}{\overline{\nu}^{S2}} \sum_{p,j}  \sum_{s=0}^{p + 2j} \sum_{t=0}^{j + \lfloor p/2\rfloor} \sum_{t_1=0}^t  \left( T ^{-1} \right)^{st}_{pj}  L_{tt_1}^1  \frac{N^{pj}}{\sqrt{2^p p!}} \overline{\nu}^{St_1 +2} [ p > 0 \cup j > 0 ].
\end{align}
\\
In, \cref{eq:HSCterms,eq:rglk}, we introduce the speed integrated frequency by $\overline{\nu}^{Sk} = \int d s  e^{- s^2} s^{2k} \nu^S(v) $. Using the definitions of $\nu^S(v)$, we derive

\begin{align}
  \overline{\nu}^{Sk} & = 2 \nu   \left( E^{k-2} - e^{k-1}\right).
  \end{align}

\section{Dougherty Collision Operator}
\label{sec:dougherty}
The linearized GK Dougherty collision operator is given by the test and field components \citep{Dougherty1964}

 \begin{align} \label{eq:TDougherty}
 \C^T & = \nu \left[ \frac{T}{m} \frac{\partial^2}{\partial v_\parallel^2} h + 2 \frac{T}{B}  \frac{\partial}{\partial \mu} \left( \mu \frac{\partial}{\partial \mu} h \right)- \frac{1}{2} b^2 h + 3 h+ v_\parallel  \frac{\partial}{\partial v_\parallel } h + 2 \mu \frac{\partial}{\partial \mu} h  \right],
 \end{align}
 \\
 and 
  \begin{align} \label{eq:FDougherty}
 \C^F & = \nu \left[ 2   J_0  T[h]\left(  x + s_{\parallel }^2 - \frac{3}{2}  \right)  + 2 u_{\parallel}[h] s_\parallel J_0(b) + J_1(b) 2  u_\perp[h] \sqrt{x}   \right] F_{M},
 \end{align}
 \\
respectively.  Here, the particle fluid quantities are defined by $ T[h] = ( T_{\parallel}[h]+ 2 T_{\perp  }[h])/3 - n[h]$ where $T_{\parallel } [h]= \int d^3 \vi m v_\parallel^2  J_0 h / T$, $T_{\perp }[h] = \int d^3 \vi J_0 \mu B h/T $, $n[h] = \int d^3 \vi J_0 h/N$, and $u_\parallel[h] = \int d^3 \vi J_0 s_\parallel h $, $u_\perp[h] =  \int d^3 \vi  \sqrt{x} J_1  h $.

Projecting the GK Dougherty collision operator, \cref{eq:TDougherty,eq:FDougherty}, onto the Hermite-Laguerre basis yields,

 \begin{align} \label{eq:CDTlk}
 \C^{Tlk} =  -\nu \left( 2k + l+ \frac{b^2}{2}  \right) n^{lk
}
\end{align}
 \\
 and
 
 \begin{align}\label{eq:CDFlk}
 \C^{Flk}   = & \nu \left( T[h] \left[2 \delta_{l}^0 \left(   2 k  \kernel{k}-(k+1) \kernel{k+1} - k \kernel{k-1}\right) +  \kernel{k} \sqrt{2} \delta_l^2  \right] \right. \nonumber \\
&  \left.+ u_{\parallel }[h] \kernel{k} \delta_l^1 + u_{\perp}[h]  b \left( \kernel{k} -  \kernel{k-1} \right)\delta_l^0   \right) ,
 \end{align}
 \\
  where the particle fluid quantities are expressed as 
 
 \begin{subequations}
 \begin{align}
 n[h]& = \sum_{j=0}^{\infty} \kernel{j} n^{0j}, \\
 T_{\parallel }[h]&=   \sum_{j=0}^\infty \kernel{j} \left[ \sqrt{2} n^{2j} + n^{0j}\right], \\
T_{\perp } [h]& = \sum_{j=0}^\infty n^{0j} \left[  (2 j+1) \kernel{j} - (j+1) \kernel{j+1}  - j \kernel{j-1}\right], \\
u_{\parallel }[h] &= \sum_{j=0}^\infty \kernel{j} n^{1j}, \\
u_{\perp} [h]&=  b  \sum_{j=0}^\infty \frac{1}{2 } \left[ \kernel{j} - \kernel{j-1}\right]  n^{0j} .
 \end{align}
 \end{subequations}
\\
We remark that the GK Dougherty collision operator, in \cref{eq:TDougherty} and \cref{eq:FDougherty}, is used in a previous Hermite-Laguerre velocity-space pseudo-spectral GK formulation because of its sparse representation on this basis \citep{Mandell2018}. In fact,  $\C^{Tlk} = -\nu \left( 2k + l+ b^2 / 2  \right) n^{lk}$ and $\C^{Flk}=0$ for $l > 2$.

The DK Dougherty collision operator is obtained in the zero gyroradius limit of \cref{eq:CDTlk,eq:CDFlk}, and is given by

 \begin{align} \label{eq:DKDougherty}
 \C^{lk} =  \nu \left[  - (2 k + l ) N^{lk}  -  (  \sqrt{2} N^{20} - 2 N^{01})\left( \frac{2}{3} \delta_l^0 \delta_k^1- \frac{\sqrt{2}}{3} \delta_l^2 \delta_k^0 \right) +  N^{10}\delta_l^1 \delta_k^0 \right].
 \end{align}
 \\
 We remark that the DK Dougherty operator, given in \cref{eq:DKDougherty}, is equivalent to the one used in \citet{Jorge2018}.
 
 \section{Collisionless Gyro-Moment Expressions}
\label{sec:AppGeneralizedLinearMomentResponse}

This appendix reports on the derivations of the collisionless gyro-moment expressions, defined in \cref{eq:tNapj}. We start from \cref{eq:fi} that we multiply by the Hermite-Laguerre basis yielding

\begin{align} \label{eq:tNapjappendix}
\frac{N^{pj}}{\phi } = \sum_{l=1}^3 \hat{N}_{l}^{pj},
\end{align}
\\
where we introduce the velocity integrals,

\begin{subequations} \label{eq:linResponNpj}
\begin{align}
    \hat{N}_{1}^{pj} & = - \frac{q}{\tau}\int_0^\infty d x \int_{- \infty}^\infty  \frac{d s_\parallel}{\sqrt{\pi}}  \frac{H_p(s_\parallel ) L_j(x)}{\sqrt{2^p p!}}  J_0(b\sqrt{x}) e^{ - s_\parallel^2 -x}, \\
      \hat{N}_{2}^{pj}&  =      \frac{q}{\tau} \int_0^\infty d x \int_{- \infty}^\infty  \frac{d s_\parallel}{\sqrt{\pi}} \frac{\omega J_0(b \sqrt{x})e ^{-x - s_\parallel^2}} {     \omega - \omega_{\grad B}- \sqrt{2 \tau} k_\parallel s_\parallel } \frac{H_p(s_\parallel ) L_j(x)}{\sqrt{2^p p!}},  \label{eq:Nhat2pj}\\
          \hat{N}_{3}^{pj}&  =      - \int_0^\infty d x \int_{- \infty}^\infty  \frac{d s_\parallel}{\sqrt{\pi}} \omega_{T}^*\frac{ J_0(b \sqrt{x})e ^{-x - s_\parallel^2}} {     \omega - \omega_{\grad B }- \sqrt{2 \tau} k_\parallel s_\parallel } \frac{H_p(s_\parallel ) L_j(x)}{\sqrt{2^p p!}}. \label{eq:Nhat3pj}
\end{align}
\end{subequations}
\\
We now perform analytically the velocity integrals that contain the resonant term proportional to $1 / ( \omega-  \omega_{\grad B }- \sqrt{2 \tau} k_\parallel s_\parallel)$ using the transformation \cref{eq:resonantintegral} valid for unstable modes. The adiabatic part of the collisionless gyro-moment response is deduced from the orthogonality relations, \cref{eq:hermiteLaguerreorthogonality,eq:J02Laguerre}, such that

\begin{align} \label{eq:tNpj1}
\hat{N}_1^{pj} = - \frac{q}{\tau} \kernel{j} \delta_p^0.
\end{align}
\\
Expanding the Bessel function in terms of Laguerre polynomials using \cref{eq:J02Laguerre}, $\hat{N}^{pj}_2$, defined in \cref{eq:Nhat2pj}, can be written as

\begin{align} \label{eq:hatNpj2integral}
    \hat{N}_{2}^{pj}& = -i  \frac{q}{\tau} \sum_{n=0}^\infty \kernel{n} \int_0^{\infty} d \tau \omega e^{i \tau \omega} \int_{- \infty}^\infty  \frac{d s_\parallel}{\sqrt{\pi}} e ^{- (1 + i \tau 2 \alpha )s_\parallel^2- i \tau z_\parallel s_\parallel}\frac{H_p(s_\parallel ) }{\sqrt{2^p p!}}  \nonumber \\
  & \times  \int_0^\infty d xL_n(x)L_j(x) e ^{-(1 + i \tau \alpha)x},
\end{align}
\\
where $\alpha = \tau \omega_{ B}/q$. The $x$-integration can be performed using the identity \citep{gradshteyn}

 \begin{align} \label{eq:intxsLnLm}
     \int_0^\infty d x e^{- \beta x} L_n( x) L_m( x) & = \frac{(m + n ) !}{m! n!} \frac{(\beta - 1)^{n+m}}{\beta^{m + n  + 1}} \mathcal{F}^m_n\left[\frac{\beta(\beta - 2)}{(\beta - 1)^2}\right], 
 \end{align}
 \\
 for $\text{Re } \beta > 0$ and where $\mathcal{F}^a_b[z] = F[-a,-b;-a-b,z]$ is the Gauss hypergeometric function. We remark that, in the slab limit corresponding to $\alpha =0 $ in \cref{eq:hatNpj2integral} (and thus $\beta = 1$ in \cref{eq:intxsLnLm})), \cref{eq:intxsLnLm}  reduces to the orthogonality relation between Laguerre polynomials, \cref{eq:Laguerreorthogonality}. The $s_\parallel$-integration in \cref{eq:hatNpj2integral} is evaluated by expanding $H_p(s_\parallel)$ such that
 
 \begin{align} \label{eq:Hermiteseries}
    H_p (s_\parallel) = \sum_{p_1=0}^{\lfloor p/2\rfloor} \frac{(-1)^{p_1}2^{p - 2p_1}p! }{p_1!(p -2 p_1)!}  s_\parallel^{p - 2p_1}.
 \end{align}
\\
Using the above identities, we derive

\begin{align} \label{eq:tN2pj}
    \hat{N}_{2}^{pj}& = - \frac{i q \omega}{\tau} \sum_{n=0}^\infty  \frac{ \kernel{n} }{\sqrt{2^p p!} }\mathcal{I}_{n}^{pj} ,
      \end{align}
      \\
      where we introduce the one-parameter integral
 
 \begin{align} \label{eq:Inpj}
 \mathcal{I}_{n}^{pj} &=  \sum_{p_1=0}^{\lfloor p/2\rfloor} \frac{(-1)^{p_1 }2^{p - 2p_1 -1}p! }{p_1!(p -2 p_1)!}  \int_0^{\infty} \frac{d \tau}{\sqrt{\pi}} e^{i \tau \omega}     (1 + 2 i \tau \alpha)
   ^{p_1-\frac{p}{2}-\frac{1}{2}} \nonumber \\
  &\times  \frac{(n+j ) !}{n! j!} \frac{(i \alpha \tau)^{n+j}}{(1 +  i \alpha \tau)^{j + n  + 1}}\mathcal{F}^j_n\left[\frac{(1 +  i \alpha \tau)(  i \tau \alpha-1 )}{(  i \alpha \tau)^2}\right] \nonumber   \\
   & \times \left\{\left(1+(-1)^{p - 2p_1}\right)  \left(
   \frac{p}{2} - p_1 - \frac{1}{2}\right)! \mathcal{M}^{{p/2 - p_1 +1/2}}_{1/2}\left[- \frac{z_\parallel^2 \tau^2}{4(1 + 2 \alpha i \tau )} \right] \right. \nonumber  \\
&   \left. -2 i \frac{z_\parallel \tau  \left(1+(-1)^{p - 2 p_1+1}\right) }{2 \sqrt{(1 + 2 i \tau \alpha)}}  
  \left(\frac{p}{2} -p_1 \right)! \mathcal{M}^{\frac{p}{2}-p_1+1}_{3/2} \left[- \frac{z_\parallel^2 \tau^2}{4(1 + 2 \alpha i \tau )} \right]\right\}, 
\end{align}
\\
where $ \mathcal{M}^a_b[z] =  M(a;b;z)$ is the Kummer confluent hypergeometric function, which stems from the $s_\parallel$-integration. 

Finally, using the recursive properties of Hermite and Laguerre polynomials, we deduce that

\begin{align} \label{eq:tN3pj}
    \hat{N}_{3}^{pj}& =  \frac{i k_\perp}{\sqrt{2^p p!}}\sum_{n=0}^{\infty}\kernel{n}  \left\{ \mathcal{I}_n^{pj}  + \eta \left[ (2j+1) \mathcal{I}_n^{pj}   - j  \mathcal{I}_n^{pj-1}- (j+1)  \mathcal{I}_n^{pj+1} \right.\right. \nonumber \\
    & \left. \left. + \frac{\mathcal{I}_{n}^{p+2j}}{4} + (p + \frac{1}{2}) \mathcal{I}_n^{pj} + p(p-1) \mathcal{I}_n^{p-2j}- \frac{3}{2}\mathcal{I}_n^{pj}\right] \right\}.
      \end{align}
\\
With \cref{eq:tNpj1,eq:tN2pj,eq:tN3pj}, the closed semi-analytical expressions for the collisionless gyro-moment, \cref{eq:tNapj}, can be evaluated. 
 We remark that a simpler expression for  $\mathcal{I}_{n}^{pj} $ can be obtained in the purely toroidal limit, i.e. $k_\parallel =0 $, therefore, neglecting the resonance due to the parallel streaming. Hence, from \cref{eq:Inpj}, one obtains 

 \begin{align} \label{eq:curlyInpj}
\mathcal{I}_{n}^{pj} &=  \sum_{p_1=0}^{\lfloor p/2\rfloor} \frac{(-1)^{p_1 }2^{p - 2p_1}p! }{p_1!(p -2 p_1)!}\frac{(n+j ) !}{n! j! \sqrt{\pi}}  \left(
   \frac{p}{2} - p_1 - \frac{1}{2}\right)!  \nonumber \\
&   \times  \int_0^{\infty} d \tau e^{i \tau \omega}     (1 + 2 i \tau \alpha_s)
   ^{p_1-\frac{p}{2}-\frac{1}{2}}  \frac{(i \alpha_s \tau)^{n+j}}{(1 +  i \alpha_s \tau)^{j + n  + 1}} \mathcal{F}^{j}_n  \left[ \frac{1 +   \alpha_s^2 \tau^2}{ \alpha_s^2 \tau^2} \right],
   \end{align}
\\
if $p$ is even and $0$ otherwise.

\section{FLR Closures}
\label{appendix:FLRclosures}

We explore possible FLR closures for the terms, proportional to $\kernel{0}$, $\kernel{1}$ and $\kernel{2}$ appearing in the $6$GM and $4$GM using the Pad\'e approximation technique. In order to deduce Pad\'e approximation of the kernel functions, we rewrite the exact form of $\kernel{n}$ defined in \cref{eq:kerneldef} as 

\begin{align} \label{eq:kernelbasicFLRoperator}
\kernel{j} = \frac{(-1)^j}{j!}  \frac{\partial^j}{\partial \beta^j} \Upsilon( a \beta ) \bigg \rvert_{\beta =1},
\end{align}
\\
where we introduce the basic FLR operator $\Upsilon( a \beta )  = \sum_{l=0}^\infty (\beta  a)^l/ (l !2^l ) = e^{- \beta a/2} $. From \cref{eq:kernelbasicFLRoperator}, approximation to $\kernel{j}$ can be constructed by specifying the basic FLR operator $\Upsilon$. We remark that in contrast to the choice $\Upsilon(a \beta)  = e^{- \beta a/2} $ introduced in \citep{brizard1992nonlinear}, \citet{Dorland1993} proposed a modified basic FLR operator such as $\Upsilon(  a \beta)  = \sqrt{ \Gamma_0( a \beta)}$. The latter approximation is motivated by the fact that basic FLR operator, $\Upsilon( a \beta)  = e^{- \beta a/2} $, yields a large damping at short wavelength with respect to \citet{Dorland1993} when applied to, e.g., the collisionless ITG case. As collisions damp short wavelength modes (see \cref{fig:Fig_ITG_kperpscan_collision_ops_eta_3}), we focus on the basic FLR operator $\Upsilon( a \beta) = e^{- \beta a/2} $, since little difference with respect to $\sqrt{ \Gamma_0(a \beta )}$ is expected in the high collisional limits where short wavelength ITG modes are damped.

\begin{figure}
\centering
\includegraphics[scale = 0.55]{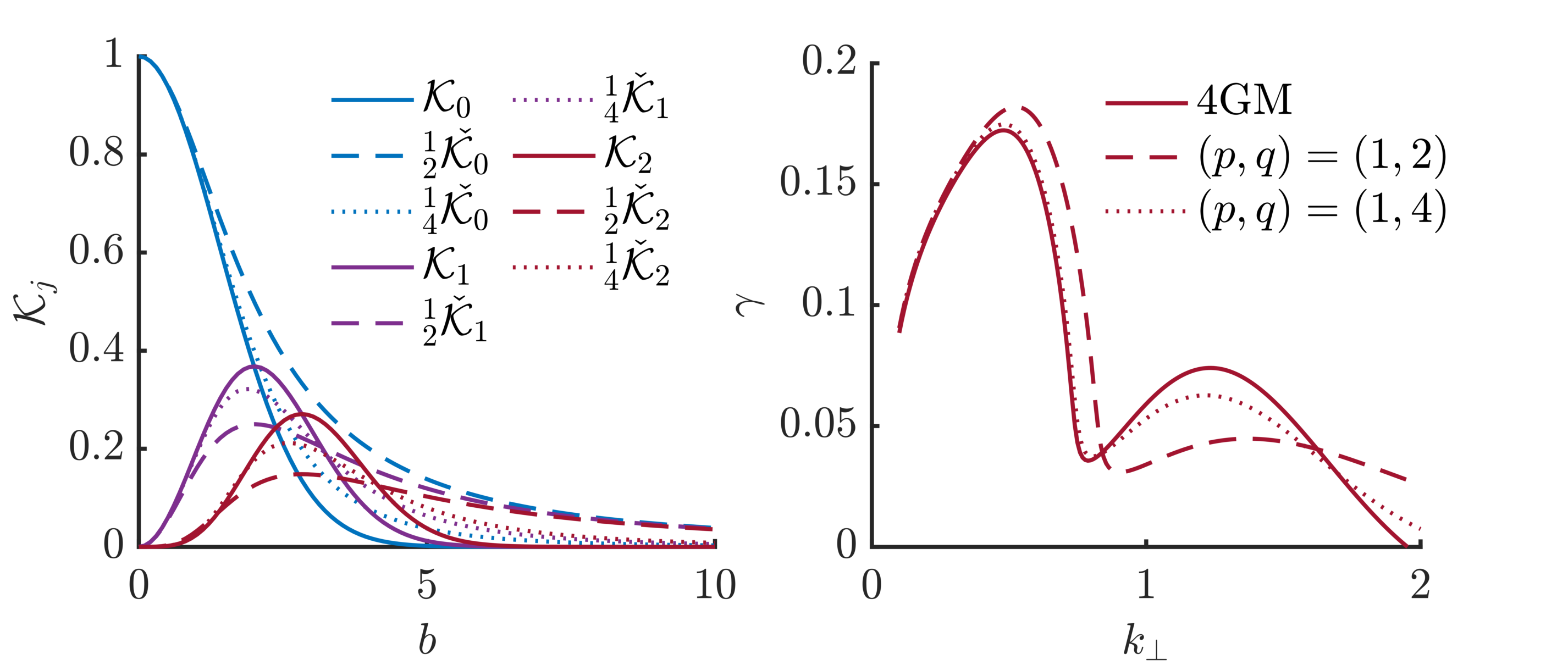}
\caption{(Left) lowest order kernel functions $\kernel{j}$ (solid line), and the correspinding Pad\'e approximants, $ {}_q^p\check{\kernel{j}}$ for $j =0,1,2$ when $(p,q) = (1,2)$ (dashed line) and $(p,q) = (1,4)$  (dotted line). (Right) ITG linear growth rate $\gamma$ as a function of the perpendicular wavenumber $k_\perp$, obtained by the $4$GM with the exact kernel functions $\kernel{j}$ (solid line) and with the Pad\'e approximants  ${}_q^p\check{\kernel{j}}$, with $(p,q) = (1,2)$ (dashed line) and $(p,q ) = (1,4)$ (dotted line), respectively.}
\label{fig:pade}
\end{figure}

We now aim to construct a Pad\'e approximant, of order $(p,q)$, of the basic FLR operator, that we denote by ${}_q^p\check{\Upsilon}(x) =  R^{(p)}(x) / S^{(q)}(x)$ where $R^{(p)}(x) $ and  $S^{(q)}(x)$ are polynomials in $x$ of order $p$ and $q$, respectively. The Pad\'e approximant $ {}_q^p\check{\Upsilon}(x) $ satisfies $ \Upsilon^{(p+q)}(0) = {}_q^p\check{\Upsilon}^{(p+q)}(0) $. Once ${}_q^p\check{\Upsilon}^{(p+q)}(0) $ is specified, the self-consistent Pad\'e approximant, of order $(p,q)$, of the $j$th order kernel function, denoted by ${}_q^p\check{\kernel{j}} \simeq\kernel{j} $, is expressed by

\begin{align} \label{eq:padekernelj}
 {}_q^p\check{\kernel{j}}  = \frac{(-1)^j}{j!}\frac{\partial^j}{\partial \beta^j} {}_q^p\check{\Upsilon}(\beta a)  \bigg \rvert_{\beta =1},
\end{align}
\\
 We consider Pad\'e approximant of order $(p,q) =(1,2)$ and $(1,4)$ for $j = 0,1,2$ to approximate $\kernel{0}$, $\kernel{1}$, and $\kernel{2}$ appearing in the FLR terms contained in both the $6$GM and $4$GM models \cref{eq:Gammas}. \Cref{eq:padekernelj} yields the  $(p,q) =(1,2)$ self-consistent Pad\'e approximants 

\begin{subequations} \label{eq:pade12}
\begin{align}
 {}_2^1\check{\kernel{0}} & = \frac{1}{1 + b^2/4}, \\
  {}_2^1\check{\kernel{1}} & = \frac{b^2}{4} \frac{1}{(1 + b^2/4)^2} , \\
    {}_2^1\check{\kernel{2}} & = \frac{b^4}{16}\frac{1}{(1 + b^2/4)^3},
\end{align}
\end{subequations}
\\
and the $(p,q) =(1,4)$ Pad\'e approximants

\begin{subequations} \label{eq:pade14}
\begin{align}
 {}_4^1\check{\kernel{0}} & = \frac{1}{1 + b^2/4 +b^4/32 }, \\
  {}_4^1\check{\kernel{1}} & =\frac{b^2/4 + b^4/16}{(1 + b^2/4 +b^4/32 )^2} , \\
    {}_4^1\check{\kernel{2}} & = \frac{1}{2} \left[ \frac{2(b^2/4 + b^4/16)^2}{(1 + b^2/4 +b^4/32 )^3}  - \frac{b^4}{16(1 + b^2/4 +b^4/32 )^2}  \right].
\end{align}
\end{subequations}
\\
The left panel of \cref{fig:pade} displays the lowest-order kernel functions, $\kernel{j}$ for $j =0$, $1$, $2$, and the Pad\'e approximants, given in \cref{eq:pade12} and \cref{eq:pade14}, respectively. As observed, the $(p,q) = (1,2)$ Pad\'e approximants decay slower in the larger $b$ limit, compared to the $(p,q) = (1,4)$ models, while they both agree with the kernel functions $\kernel{j}$ in the small $b$ limit. Also, $(p,q) = (1,4)$ case provides a better approximation near the $\kernel{1,2}$ maxima. Using the Pad\'e approximants, defined \cref{eq:pade12} and \cref{eq:pade14}, allows us to compute the slab ITG growth rate using the $4$GM. We consider the parameters $k_\parallel = 0.1$, $\eta = 5$, and $\nu = 1$ and show the results in the right panel of \cref{fig:pade} (similar results are obtained with the $6$GM). Notice that, while the Pad\'e approximants yield a good approximation when $k_\perp \lesssim 0.4$, the $(p,q) = (1,4)$ Pad\'e approximant models accurately describe the linear growth $\gamma$ rate near the peak and behaves qualitatively well when $k_\perp \gtrsim 1$. The second peak corresponds to the SWITG mode (see \cref{subsec:SWITG}). 
This simple application demonstrates the ability of Pad\'e approximant to accurately model FLR terms. Indeed, in conventional space, the kernel function, $\kernel{n}$ introduced in \cref{eq:kerneldef}, defines an infinite linear combination of differential operators in conventional space, i.e.

\begin{align} \label{eq:kernelgrad}
\kernel{j} =  \sum_{l =0}^{\infty} \frac{(-1)^{l +j}}{l!j!} \left(\frac{\rho}{4} \Delta_\perp\right)^{j+l},
\end{align}
\\
where $\Delta_\perp = \grad \cdot \grad_\perp$ is the perpendicular Laplacian. From a numerical point of view, \cref{eq:kernelgrad} is not practical due to the presence of the infinite sum over the $l$ index. A naive approach is by performing a truncation of the sum at a given order in $b$. But, it would yield an arbitrary lost of accuracy, in particular, for short wavelength mode, such as, e.g. ITG that peak at $k_\perp  \sim 0.6 / \rho_s$ (in physical units). A second option is to approximate the functional form of \cref{eq:kernelgrad} using Pad\'e approximant as discussed in this appendix. Such approach can be generalized to describe FLR terms in the GK collision operators and in the gyro-moment hierarchy for arbitrary number of gyro-moments.

 \bibliographystyle{jpp}
 \bibliography{biblio}

\end{document}